\documentclass[letterpaper,11pt]{article}
\pdfoutput=1
\usepackage{jheppub}
\usepackage{slashed}
\usepackage{graphicx}
\usepackage{subfigure}
\usepackage{soul}
\usepackage{amsmath}
\usepackage{multirow}
\usepackage{tablefootnote}
\usepackage{hyperref}
\usepackage{epstopdf}
\usepackage{comment}

\usepackage[dvipsnames]{xcolor}

\newcommand{\beq}{\begin{equation}}
\newcommand{\eeq}{\end{equation}}
\newcommand{\bea}{\begin{eqnarray}}
\newcommand{\eea}{\end{eqnarray}}
\newcommand{\beqs}{\begin{subequations}}
\newcommand{\eeqs}{\end{subequations}}

\newcommand{\ba}{\begin{array}}
\newcommand{\ea}{\end{array}}

\def\figureautorefname~#1\null{Fig.\,#1\null}
\def\tableautorefname~#1\null{Tab.\,#1\null}

\def\equationautorefname~#1\null{Eq.\,(#1)\null}

\def\m1{M_1}
\def\m2{M_2}
\def\m3{M_3}

\def\ch10{\tilde \chi^0_1}

\def\etc{{\it etc}}

\def\tev{\,{\rm TeV}}
\def\gev{\,{\rm GeV}}

\def\to{\rightarrow}

\newcommand{\lsim}{\mathrel{\mathop{\kern 0pt \rlap
  {\raise.2ex\hbox{$<$}}}
  \lower.9ex\hbox{\kern-.190em $\sim$}}}
\newcommand{\gsim}{\mathrel{\mathop{\kern 0pt \rlap
  {\raise.2ex\hbox{$>$}}}
  \lower.9ex\hbox{\kern-.190em $\sim$}}}

\definecolor{pink}{RGB}{255,105,180}

\def\cosba{\cos(\beta-\alpha)}

\newcommand{\ee}{{e^{+} e^{-}}}

\newcommand{\stu}{S,\ T\ \text{and}~ U}

\newcommand{\bpm}{\begin{pmatrix}}
\newcommand{\epm}{\end{pmatrix}}

\newcommand{\eehz}{e^+e^- \to hZ}

\newcommand{\eeww}{e^+e^- \to WW}
\newcommand{\tanb}{\tan \beta}
\newcommand{\lambvs}{\sqrt{\lambda v^2}}

\allowdisplaybreaks[4]

\title{Type-II 2HDM under the Precision Measurements \\ at the $Z$-pole and a Higgs Factory}

\author[1]{Ning Chen,}
\author[2,3]{Tao Han,}
\author[4]{Shufang Su,}
\author[4,5,6]{Wei Su,}
\author[7]{Yongcheng Wu}

\affiliation[1]{School of Physics, Nankai University, Tianjin 300071, China}
\affiliation[2]{Department of Physics and Astronomy, University of Pittsburgh,  Pittsburgh, PA 15260, USA}
\affiliation[3]{Department of Physics, Tsinghua University, and Collaborative Innovation Center of Quantum Matter, Beijing, 100086, China}
\affiliation[4]{Department of Physics, University of Arizona, Tucson, Arizona  85721, USA}
\affiliation[5]{CAS Key Laboratory of Theoretical Physics, Institute of Theoretical Physics, Chinese Academy of Sciences, Beijing 100190, China}
\affiliation[6]{School of Physics, University of Chinese Academy of Sciences, Beijing 100049, China}
\affiliation[7]{Ottawa-Carleton Institute for Physics, Carleton University, 1125 Colonel By Drive, Ottawa, Ontario K1S 5B6, Canada}

\preprint{PITT-PACC-1808}

\emailAdd{chenning$\_$symmetry@nankai.edu.cn}
\emailAdd{than@pitt.edu}
\emailAdd{shufang@email.arizona.edu}
\emailAdd{weisv@itp.ac.cn}
\emailAdd{ycwu@physics.carleton.ca}

\abstract
 {Future precision measurements of the Standard Model (SM) parameters at the proposed $Z$-factories and   Higgs factories may have significant impacts on new physics beyond the Standard Model in the electroweak sector. We illustrate this by focusing on the Type-II two Higgs doublet model (Type-II 2HDM). 
The contributions from the heavy Higgs bosons at the tree-level and at the one-loop level are included in a full model parameter space.
 We perform a multiple variable global fit
and study the extent to which the parameters of non-alignment and non-degenerate masses can be probed by the precision measurements.   We find that the allowed parameter ranges are tightly constrained by the future Higgs precision measurements, especially for small and large values of $\tan\beta$. 
Indirect limits on the masses of heavy Higgs can be obtained, which can be complementary to the direct searches of the heavy Higgs bosons at hadron colliders. We also find that the expected accuracies at the $Z$-pole and at a  Higgs factory are quite complementary in constraining mass splittings of heavy Higgs bosons.  The typical results are $|\cos(\beta-\alpha)| < 0.008,  |\Delta m_\Phi | < 200\ {\rm GeV}$, and $\tan\beta \sim 0.2 - 5$. The reaches from CEPC, FCC-ee and ILC are also compared, for both Higgs and $Z$-pole precision measurements.}


\begin{document}
\maketitle
\flushbottom


\section{Introduction}
\label{sec:intro}

With the milestone discovery of the Higgs boson ($h$) at the CERN Large Hadron Collider (LHC)~\cite{Aad:2012tfa,Chatrchyan:2012xdj}, particle physics has entered a new era. All the indications from the current measurements seem to confirm the validity of the Standard Model (SM) up to the electroweak (EW) scale of a few hundred GeV, and the observed Higgs boson is SM-like. Yet, there are compelling arguments, both from theoretical and observational points of view, in favor of the existence of new physics beyond the Standard Model (BSM) \cite{Giudice:2008bi}.
As such, searching for new Higgs bosons would be of high priority since they are present in many extensions of BSM theories. One of the most straightforward, but well-motivated extensions is the two Higgs doublet model (2HDM)~\cite{Branco:2011iw}, in which there are five massive spin-zero states in the spectrum ($h, H,A,H^\pm$) after the electroweak symmetry breaking (EWSB).   Extensive searches for BSM Higgs bosons have been actively carried out, especially in the LHC experiments~\cite{Aaboud:2017sjh,CMS-PAS-HIG-17-020,Aaboud:2017gsl,Aaboud:2017rel,Sirunyan:2018qlb,Aaboud:2017yyg,Aaboud:2017cxo,Aaboud:2018eoy,Khachatryan:2016are,Aaboud:2018ftw,Sirunyan:2018iwt,ATLAS-CONF-2016-089,Aaboud:2018gjj,CMS-PAS-HIG-16-031}. Unfortunately, no signal observation has been reported thus far. This would imply either the non-SM Higgs bosons are much heavier and essentially decoupled from the SM, or their interactions are accidentally aligned with the SM configuration~\cite{Carena:2013ooa,Dev:2014yca}. In either situation, it would be challenging to observe those states in experiments.

Complementary to the direct searches,  precision measurements of SM parameters,  in particular, the Higgs boson properties
could lead to relevant insights into new physics. There have been proposals to build a Higgs factory in the pursuit of precision Higgs measurements,  including the Circular Electron Positron Collider (CEPC) in China~\cite{CEPC-SPPCStudyGroup:2015csa, CEPCStudyGroup:2018ghi}, the electron-positron stage of the Future Circular Collider (FCC-ee) at CERN (previously known as TLEP~\cite{Gomez-Ceballos:2013zzn,fccpara,fccplan}), and the International Linear Collider (ILC) in Japan~\cite{Baer:2013cma}.  With about $10^6$ Higgs bosons produced at the Higgs factory, one would expect to reach sub-percentage precision determination of the Higgs properties, and thus to be sensitive to new physics associated with the Higgs boson. As an integrated part of the program, one would like to return to the $Z$-pole. With about $10^{10} - 10^{12}\ Z$ bosons, the achievable precisions on the SM parameters could be improved by a factor of $20-200$ over the Large Electron Positron (LEP) Collider results~\cite{ALEPH:2005ab}.  Such a high precision would hopefully shed light on new physics associated with the electroweak sector.

In this paper, we set out to examine the impacts from the precision measurements of the SM parameters at the proposed $Z$-factories and Higgs factories on the extended Higgs sector. 
There is a plethora of articles in the literature to study the effects of the heavy Higgs states on the SM observables~\cite{Branco:2011iw}. 
We illustrate this by focusing on the Type-II 2HDM\footnote{The implication of Higgs factory precision measurements on four typical types of 2HDM has been studied in Ref.~\cite{Gu:2017ckc}, focusing on the tree level constraints as well as loop contributions under alignment limit individually.  In particular, for Type-I 2HDM, the allowed range of $\cos(\beta-\alpha)$ based on tree level constraints is about a factor of 10 larger than that of Type-II 2HDM, which leads to characteristically different behaviour once combined tree level effects and loop corrections are taken into account. Therefore, we focus on Type-II 2HDM in the current paper and leave the detailed analyses of Type-I 2HDM for a future work~\cite{typeI2hdm}.}.
In our analyses, we include the tree-level corrections to the SM-like Higgs couplings and one-loop level contributions from the heavy Higgs bosons. A global fit is performed in the full model-parameter space. In particular, we study the extent to which the parametric deviations from the alignment and degenerate mass limits can be probed by the precision measurements. We find that the expected accuracies at the $Z$-pole and at a Higgs factory are quite complementary in constraining mass splittings of heavy Higgs bosons. The reach in the heavy Higgs masses and couplings can be complementary to the direct searches of the heavy Higgs bosons at the LHC.

The rest of the paper is organized as follows. In~\autoref{sec:input}, we summarize the anticipated accuracies on determining the EW observables at the $Z$-pole and Higgs factories. Those expectations serve as the inputs for the following studies for BSM Higgs sector. We then present the Type-II 2HDM and the one-loop corrections, as well as the existing constraints to the model parameters in~\autoref{sec:2hdm}. \autoref{sec:results} shows our main results from the global fit, for the cases of mass degeneracy and non-degeneracy of heavy Higgs bosons.   We summarize our results and draw conclusions in~\autoref{sec:conclu}.

\section{The EW and Higgs Precision Measurements at Future Lepton Colliders}
\label{sec:input}

The EW precision measurements are not only important in understanding the SM physics, but also can impose strong constraints on new physics models~\cite{Gori:2015nqa,Su:2016ghg}. 
The benchmark scenarios of several proposed future $e^+e^-$ machines and the projected precisions on $Z$-pole and Higgs measurements are summarized below. These expected results serve as the inputs for the later studies in constraining the BSM Higgs sector.

\subsection{The electroweak precision measurements}

The current best precision measurements for $Z$-pole physics came mostly from the LEP-I, and partially from the Tevatron and the LHC~\cite{Baak:2014ora,Haller:2018nnx}. These measurements could be significantly improved by a $Z$-pole run at future lepton colliders with a much larger data sample~\cite{CEPC-SPPCStudyGroup:2015csa,Gomez-Ceballos:2013zzn,Asner:2013psa,fccplan,fccpara}.  For example, the parameter $\sin^2 \theta_{eff}^\ell$ can be improved by more than one order of magnitude at the future $\ee$ collider; the $Z$-mass precision can be measured four times better in CEPC.  Precisions of other observables, including $m_W$,  $m_t$,  $m_h$,  $A_{FB}^{b,c,l}$,  $R_b$, \etc., can be improved  as well, depending on different machine parameter choices. Given the complexity of a full $Z$-pole precision fit, we study the implications of $Z$-pole precision measurements on the 2HDM adopting the Peskin-Takeuchi oblique parameters $S$, $T$ and $U$~\cite{Peskin:1991sw}.

\begin{table}[tbh]
\centering
\setlength{\tabcolsep}{.3em}
\begin{tabular}{|c|c|c|c|}
\hline
& CEPC  & ILC & FCC-ee \\
\hline
$\alpha_s(M_Z^2)$ &
$\pm 1.0 \times 10^{-4}$ &
$\pm 1.0 \times 10^{-4}$ &
$\pm 1.0 \times 10^{-4}$  
\\
$\Delta\alpha_{\rm had}^{(5)}(M_Z^2) $ &
$\pm 4.7 \times 10^{-5}$&
$\pm 4.7 \times 10^{-5}$ &
$\pm 4.7 \times 10^{-5}$  
\\
$m_Z$ [GeV] &
$\pm 0.0005 $& 
$\pm0.0021$& 
$\pm 0.0001_{\rm exp}$ 
\\
$m_t$ [GeV] (pole)&
$\pm 0.6_{\rm exp} \pm 0.25_{\rm th}$& 
$\pm 0.03_{\rm exp} \pm 0.1_{\rm th}$ & 
$\pm 0.6_{\rm exp} \pm 0.25_{\rm th}$  
\\
$m_h$ [GeV] &
$<\pm 0.1$& 
$<\pm 0.1$ &
$<\pm 0.1$ 
\\
$m_W$ [GeV] &
$\left(\pm 3_{\rm exp} \pm 1_{\rm th}\right) \times 10^{-3}$&
$\left(\pm 5_{\rm exp} \pm 1_{\rm th}\right) \times 10^{-3}$& 
$\left(\pm 8_{\rm exp} \pm 1_{\rm th}\right) \times 10^{-3}$ 
   \\
$\sin^2\theta^{\ell}_{\rm eff}$  &
$\left(\pm 4.6_{\rm exp} \pm 1.5_{\rm th}\right) \times 10^{-5}$&   
$\left(\pm 1.3_{\rm exp} \pm 1.5_{\rm th}\right) \times 10^{-5} $  &    
$\left(\pm 0.3_{\rm exp} \pm 1.5_{\rm th}\right) \times 10^{-5}$      
   \\
$\Gamma_{Z}$ [GeV] &
$\left(\pm 5_{\rm exp} \pm 0.8_{\rm th}\right) \times 10^{-4}$& 
$\pm 0.001$   &   
$\left(\pm 1_{\rm exp} \pm 0.8_{\rm th}\right) \times 10^{-4}$ 
   \\
\hline
\end{tabular}
\caption{Anticipated precisions of the EW observables at the future lepton colliders. The results are mainly from~\cite{Fan:2014vta,Lepage:2014fla,Baak:2013fwa,Baak:2014ora,LiangTalk}. 
}
\label{tab:stu-ee-input}
\end{table}

\begin{table}[tb]
\centering
\resizebox{\textwidth}{!}{
  \begin{tabular}{|l|c|r|r|r|c|r|r|r|c|r|r|r|c|r|r|r|c|r|r|r|}
   \hline
    & \multicolumn{4}{c|}{Current ($1.7 \times 10^{7}\ Z$'s)}& \multicolumn{4}{c|}{CEPC ($10^{10}Z$'s)}& \multicolumn{4}{c|}{FCC-ee ($7\times 10^{11}Z$'s)}&\multicolumn{4}{c|}{ILC ($10^{9}Z$'s)} \\
   \hline
   \multirow{2}{*}{}
   &\multirow{2}{*}{$\sigma$} &\multicolumn{3}{c|}{correlation}
   &{$\sigma$} &\multicolumn{3}{c|}{correlation}
   &{$\sigma$} &\multicolumn{3}{c|}{correlation}
   &{$\sigma$} &\multicolumn{3}{c|}{correlation} \\
   \cline{3-5}\cline{7-9}\cline{11-13}\cline{15-17}
   &&$S$&$T$&$U$&($10^{-2}$)&$S$&$T$&$U$&($10^{-2}$)&$S$&$T$&$U$&($10^{-2}$)&$S$&$T$&$U$\\
   \hline
   $S$& $0.04 \pm 0.11$& 1 & 0.92 & $-0.68$ & $2.46$  & 1     & 0.862       & $-0.373$ &   $0.67$    &  1     &   0.812    &    0.001   &   $3.53$    &   1    &    0.988   & $-0.879$ \\
\hline
   $T$&$0.09\pm 0.14$& $-$ & 1 & $-0.87$ & $2.55$  &  $-$   &  1      &  $-0.735$   &   $0.53$    &   $-$    &    1   &    $-0.097$   &    $4.89$   &   $-$    &   1    &   $-0.909$\\
\hline
   $U$& $-0.02 \pm 0.11$& $-$ & $-$ & 1 &$2.08$  &  $-$   &  $-$     &  1   &   $2.40$    &   $-$    &   $-$    &    1   &  $3.76$     &   $-$    &   $-$    & 1 \\

   \hline
  \end{tabular}
  }
  \caption{Estimated $S$, $T$, and $U$ ranges and correlation matrices $\rho_{ij}$  from $Z$-pole precision measurements  of the current results, mostly  from LEP-I~\cite{ALEPH:2005ab},  and at future lepton colliders CEPC~\cite{CEPC-SPPCStudyGroup:2015csa},     FCC-ee~\cite{Gomez-Ceballos:2013zzn}  and  ILC ~\cite{Asner:2013psa}. {\tt Gfitter} package~\cite{Baak:2014ora} is used in obtaining those constraints.  }
\label{tab:STU}
\end{table}

The anticipated precisions on the measurements of $\alpha_s$, $\Delta \alpha_{\rm had}^{(5)} (M_Z^2)$, $m_Z$, $m_t$, $m_h$, $m_W$, $\sin^2\theta_{\rm eff}^\ell$ and $\Gamma_Z$ are summarized in~\autoref{tab:stu-ee-input}~\cite{Fan:2014vta,Lepage:2014fla,Baak:2013fwa,Baak:2014ora,LiangTalk} for various benchmark scenarios of future $Z$-factories with the indicated $Z$ data samples. The corresponding constrained $\stu$ ranges and the error correlation matrices are listed in~\autoref{tab:STU}.
The results listed as ``current'' are obtained directly from the {\tt Gfitter} results which use the current $Z$-pole precision measurements~\cite{Baak:2014ora,Haller:2018nnx}, with reference values of the SM Higgs boson mass of $m_{h\,,{\rm ref}}=125$ GeV and $m_{t\,,{\rm ref}}=172.5$ GeV~\cite{Haller:2018nnx}. The predictions for future colliders are obtained by using the {\tt Gfitter} package~\cite{Baak:2014ora} with corresponding precisions for different machines, using the best-fit SM point with the current precision measurements as the central value.   For  the $Z$-pole observables with estimated precisions not yet available at future colliders, the current precisions are used instead.   As seen from the table, CEPC could reach the sensitivities of
\begin{equation}
\Delta S=\pm 0.0246\,, \ \ \  \Delta T=\pm 0.0255\,,\ \ \ \Delta U=\pm 0.0208
\end{equation}
 at $1\sigma$ level. FCC-ee would further improve the accuracy.
  In our analyses  as detailed in a later section, the 95\% C.L. $\stu$ contours are adopted to constrain the 2HDM parameter spaces,  using the $\chi^2$-fit with error-correlation matrices .

\subsection{Higgs precision measurements}

At a future $e^+e^-$  collider of the Higgs factory with the center-of-mass energy of 240$-$250\,GeV, the dominant channel to measure the Higgs boson properties is the Higgsstrahlung process of
\begin{equation}
\eehz\,.
\end{equation}
Due to the clean experimental environment and well-determined kinematics at the lepton colliders, both the inclusive cross section $\sigma(hZ)$ independent of the Higgs decays, and the exclusive ones of different Higgs decays in terms of $\sigma(hZ)\times {\rm BR}$, can be measured to remarkable precisions. The invisible decay width of the Higgs boson can also be very well constrained. In addition, the cross sections of $WW,ZZ$ fusion processes for the Higgs boson production grow with the center-of-mass energy logarithmically. While their rates are still rather small and are not very useful at 240$-$250 GeV, at higher energies in particular for a linear collider,  such fusion processes become significantly more important and can provide crucial complementary information. For $\sqrt{s}>500$ GeV, $t \bar t h$ production can also be used as well.

\begin{table}[tb]
 \begin{center}
  \begin{tabular}{|l|r|r|r|r|r|r|r|r|r|r|}
   \hline
   collider& \multicolumn{1}{c|}{CEPC}& \multicolumn{1}{c|}{FCC-ee}&\multicolumn{6}{c|}{ILC} \\
   \hline
   $\sqrt{s}$     &  $\text{240\,GeV} $ &  $\text{240\,GeV}$  &  \text{250\,GeV}  &
   \multicolumn{2}{c|}{\text{350\,GeV}}  & \multicolumn{3}{c|}{\text{500\,GeV}} \\
   $\int{\mathcal{L}}dt $     &  $\text{5 ab}^{-1} $ &  $\text{5 ab}^{-1}$   &  $\text{2 ab}^{-1} $  &
   \multicolumn{2}{c|}{$\text{200 fb}^{-1}$}  & \multicolumn{3}{c|}{$\text{4 ab}^{-1}$} \\
   \hline
    \hline
production& $Zh$  & $Zh$   & $Zh$      & $Zh$     & $\nu\bar{\nu}h$     & $Zh$     & $\nu\bar{\nu}h$ & $t\bar{t}h$ \\
   \hline
  $\Delta \sigma / \sigma$ & 0.51\%  & 0.57\% & 0.71\% & 2.1\% & $-$ & 1.06 & $-$ & $-$ \\ \hline \hline
   decay & \multicolumn{8}{c|}{$\Delta (\sigma \cdot BR) / (\sigma \cdot BR)$}  \\
  \hline
   $h \to b\bar{b}$              &  0.28\%               & 0.28\%                      &  0.42\%    & 1.67\%         & 1.67\%                  & 0.64\%    & 0.25\%           & 9.9\%        \\

   $h \to c\bar{c}$              & 2.2\%                   & 1.7\%                   & 2.9\%          & 12.7\%    & 16.7\%                   & 4.5\%     & 2.2\%           & $-$           \\

   $h \to gg$                    & 1.6\%                   & 1.98\%                  & 2.5\%           & 9.4\%    & 11.0\%                  & 3.9\%     & 1.5\%           & $-$           \\

   $h \to WW^*$                  & 1.5\%                   & 1.27\%                   & 1.1\%          & 8.7\%    & 6.4\%                  & 3.3\%    & 0.85\%           & $-$           \\

   $h \to \tau^+\tau^-$         & 1.2\%                   & 0.99\%                   &2.3\%           &4.5\%          & 24.4\%                  & 1.9\%    & 3.2\%           & $-$           \\

   $h \to ZZ^*$                  & 4.3\%                   & 4.4\%                   & 6.7\%        & 28.3\%     & 21.8\%                   & 8.8\%     & 2.9\%           & $-$           \\

   $h \to \gamma\gamma$          & 9.0\%                   & 4.2\%                   & 12.0\%     & 43.7\%     &50.1\%                   & 12.0\%   &6.7\% & $-$ \\

   $h \to \mu^+\mu^-$           & 17\%                   & 18.4\%                  & 25.5\%        & 97.6\%     & 179.8\%                  & 31.1\%     & 25.5\%            & $-$           \\
   \hline
    $(\nu\bar\nu)h \to b\bar{b}$  & 2.8\%       &   3.1\%       &       3.7\% & $-$  & $-$  & $-$  & $-$  & $-$  \\
    \hline
  \end{tabular}
  \caption{Estimated statistical precisions for Higgs boson measurements obtained at  the proposed CEPC program with 5 ab$^{-1}$ integrated luminosity~\cite{CEPC-SPPCStudyGroup:2015csa}, FCC-ee program with 5 ab$^{-1}$ integrated luminosity~\cite{Gomez-Ceballos:2013zzn},  and  ILC with various center-of-mass energies~\cite{Barklow:2015tja}.
   }
\label{tab:mu_precision}
  \end{center}
\end{table}

To set up the baseline of our study, we hereby list the running scenarios of various machines in terms of their center-of-mass energies and the corresponding integrated luminosities, as well as the estimated precisions of relevant Higgs boson measurements that are used in our global analyses in~\autoref{tab:mu_precision}.   The anticipated accuracies for CEPC and FCC-ee are comparable for most channels, except for $h\rightarrow \gamma\gamma$. 
There are several factors that contribute to the difference for this channel, which 
include  the superior resolution of the CMS-like electromagnetic calorimeter that was used in FCC-ee analyses, and the absence of background from beamstrahlung photons \cite{Gomez-Ceballos:2013zzn}.
In our global fit to the Higgs boson measurements, we only include the rate information for the Higgsstrahlung $Zh$ and  the $WW$ fusion process. Some other measurements, such as the angular distributions, the diboson process $\eeww$, can provide important information in addition to the rate measurements alone~\cite{Beneke:2014sba,Craig:2015wwr,Durieux:2017rsg}.

%

\section{Type-II Two Higgs Doublet Model}
\label{sec:2hdm}

\subsection{Model Setup}

Two ${\rm SU}(2)_L$ scalar doublets $\Phi_i\ (i=1,2)$ with a hyper-charge assignment $Y=+1/2$ are introduced in 2HDM,
\begin{equation}
\Phi_{i}=\begin{pmatrix}
  \phi_i^{+}    \\
  (v_i+\phi^{0}_i+iG_i)/\sqrt{2}
\end{pmatrix}\,.
\end{equation}
Each obtains a vacuum expectation value (vev)  $v_i\ (i=1,2)$ after EWSB with $v_1^2+v_2^2 = v^2 = (246\ {\rm GeV})^2$, and  $v_2/v_1=\tan\beta$.

The 2HDM Lagrangian for the Higgs sector can be written as
\begin{equation}\label{equ:Lall}
\mathcal{L}=\sum_i |D_{\mu} \Phi_i|^2 - V(\Phi_1, \Phi_2) + \mathcal{L}_{\rm Yuk}\,,
\end{equation}
with the Higgs potential of
\begin{eqnarray}
 V(\Phi_1, \Phi_2)=&& m_{11}^2\Phi_1^\dag \Phi_1 + m_{22}^2\Phi_2^\dag \Phi_2 -m_{12}^2(\Phi_1^\dag \Phi_2+ h.c.) + \frac{\lambda_1}{2}(\Phi_1^\dag \Phi_1)^2 + \frac{\lambda_2}{2}(\Phi_2^\dag \Phi_2)^2  \notag \\
 & &+ \lambda_3(\Phi_1^\dag \Phi_1)(\Phi_2^\dag \Phi_2)+\lambda_4(\Phi_1^\dag \Phi_2)(\Phi_2^\dag \Phi_1)+\frac{1}{2} \lambda_5 \Big[ (\Phi_1^\dag \Phi_2)^2 + h.c.\Big]\,,
\end{eqnarray}
by assuming CP-conserving and a soft $\mathbb{Z}_2$ symmetry breaking term $m_{12}^2$.

After EWSB, one of the four neutral components and two of the four charged components are eaten by the SM gauge bosons $Z$, $W^\pm$, providing their masses.  The remaining physical mass eigenstates are two CP-even neutral Higgs bosons $h$ and $H$, with $m_h<m_H$, one CP-odd neutral Higgs boson $A$, as well as a pair of charged ones $H^\pm$.  Instead of the eight parameters appearing in the Higgs potential $m_{11}^2, m_{22}^2, m_{12}^2, \lambda_{1,2,3,4,5}$, a more convenient choice of the parameters is $v, \tan\beta, \alpha, m_h, m_H, m_A, m_{H^\pm}, m_{12}^2$, where $\alpha$ is the rotation angle diagonalizing the CP-even Higgs mass matrix\footnote{$\beta$ can also be viewed as the mixing angle  of the CP-odd scalars (the basis has been chosen when we write down the Yukawa couplings). In Ref.~\cite{Haber:2006ue}, the authors presented a basis-independent method for 2HDM and discussed the significance of $\tan\beta$. In a general 2HDM model, $\tan\beta$ is basis-dependent and it cannot be a physical parameter as we can always choose the Higgs basis, in which only one Higgs doublet acquires vev and the other does not. However, once we choose a preferred basis when we specify the Yukawa couplings, $\tan\beta$ can be a meaningful parameter.}.

The Type-II 2HDM is characterized by the choice of the Yukawa couplings to the SM fermions and is  given in the form of
 \begin{equation}\label{eq:Lyuk}
  -\mathcal{L}_{\rm Yuk}=Y_{d}{\overline Q}_L\Phi_1d_R^{}+Y_{e}{\overline L}_L\Phi_1 e_R^{}+ Y_{u}{\overline Q}_Li\sigma_2\Phi^*_2u_R^{}+\text{h.c.}\,.
\end{equation}
 After EWSB, the effective Lagrangian for the light CP-even Higgs couplings to the SM particles can be parameterized as
 \begin{eqnarray}\label{eq:Leff}
&&\mathcal{L}= \kappa_Z \frac{m_Z^2}{v}Z_{\mu}Z^{\mu}h+\kappa_W \frac{2m_W^2}{v}W_{\mu}^+ W^{\mu-}h + \kappa_g \frac{\alpha_s}{12 \pi v} G^a_{\mu\nu}G^{a\mu\nu}h + \kappa_{\gamma} \frac{\alpha}{2\pi v} A_{\mu\nu}A^{\mu\nu} h \nonumber \\
&& + \kappa_{Z\gamma}\frac{\alpha}{\pi v}A_{\mu\nu}Z^{\mu\nu}h -\Big( \kappa_u \sum_{f=u,c,t} \frac{m_f}{v}f \bar f + \kappa_d \sum_{f=d,s,b} \frac{m_f}{v}f \bar f + \kappa_{e} \sum_{f=e,\mu,\tau} \frac{m_f}{v}f \bar f \Big)h\,,
\end{eqnarray}
where
\begin{equation}
\kappa_i = \frac{g_{hii}^{\rm BSM}}{g_{hii}^{\rm SM}}\,,
\end{equation}
for $i$ indicates individual Higgs coupling.
Their values at the tree level are
\begin{equation}
\kappa_Z=\kappa_W=\sin(\beta-\alpha)\,,\quad \kappa_u=\frac{\cos\alpha}{\sin\beta}\,,\quad
\kappa_{d,e}=-\frac{\sin\alpha}{\cos\beta}\,.
\end{equation}
Our sign convention is  $\beta \in (0, \frac{\pi}{2})$, $\beta -\alpha \in [0, \pi]$, so that $\sin (\beta-\alpha) \geq 0$.

The  CP-even Higgs couplings to the SM gauge bosons are $g_{hVV} \propto \sin(\beta-\alpha)$, and $g_{HVV} \propto \cos(\beta-\alpha)$.
The current measurements of the Higgs boson properties from the LHC are consistent with the SM Higgs boson interpretation. There are two well-known limits in 2HDM that would lead to a SM-like Higgs sector. The first situation is the alignment limit~\cite{Carena:2013ooa, Bernon:2015qea} of $\cos(\beta-\alpha)=0$, in which the light CP-even Higgs boson couplings are identical to the SM ones, regardless of the other scalar masses, potentially leading to rich BSM physics. For $\sin(\beta-\alpha)=0$, the opposite situation occurs with the heavy $H$ being identified as the SM Higgs boson. While it is still a viable option for the heavy Higgs boson being the observed 125 GeV SM-like Higgs boson~\cite{Coleppa:2014cca, Bernon:2015wef}, the allowed parameter space is being squeezed with the tight direct and indirect experimental constraints.  Therefore, in our analyses below, we identify the light CP-even Higgs $h$ as the SM-like Higgs with $m_h$ fixed to be 125 GeV. The other well-known case is the ``decoupling limit'', in which the heavy mass scales are all large $m_{A, H, H^\pm}\gg 2m_Z$~\cite{Haber:1994mt}, so that they decouple from the low energy spectrum.  For masses of heavy Higgs bosons much larger than $\lambda_i v^2$, $\cos(\beta-\alpha)\sim \mathcal{O}(m_Z^2 / m_A^2)$  under perturbativity and unitarity requirement. Therefore, the light CP-even Higgs boson $h$ is again SM-like. Although it is easier and natural to achieve the decoupling limit by sending all the other mass scales to be heavy, there would be little BSM observable effects given the nearly inaccessible heavy mass scales. We will thus primarily focus on the alignment limit.

Note that while $\kappa_g$, $\kappa_\gamma$ and $\kappa_{Z\gamma}$ are zero at the tree-level for both the SM and 2HDM, they are generated at the loop-level. In the SM, $\kappa_g$, $\kappa_\gamma$ and $\kappa_{Z\gamma}$ all receive contributions from fermions (mostly top quark) running in the loop, while $\kappa_\gamma$ and $\kappa_{Z\gamma}$  receive contribution from $W$-loop in addition~\cite{Henning:2014wua}.  In 2HDM, the corresponding $hff$ and $hWW$ couplings that enter  the loop corrections need to be modified to the corresponding 2HDM values. Expressions for the dependence of $\kappa_g$, $\kappa_\gamma$ and $\kappa_{Z\gamma}$ on $\kappa_V$ and $\kappa_f$ can be found in Ref.~\cite{Heinemeyer:2013tqa}. There are, in addition, loop corrections to $\kappa_g$, $\kappa_\gamma$  and $\kappa_{Z\gamma}$ from extra Higgs bosons in 2HDM.

It is of particular importance to include a discussion for  the triple couplings among Higgs bosons themselves.
At the alignment limit,
\begin{equation}
\lambda_{h \Phi\Phi}=-\frac{C_\Phi  }{2v} (m_h^2+2m_\Phi^2-\frac{2 m_{12}^2}{\sin\beta \cos\beta })\,,
\end{equation}
with $C_\Phi = 2(1)$ for $\Phi = H^\pm (H,A)$. In 2HDM with degenerate masses of $m_\Phi \equiv m_H = m_A = m_{H^\pm}$,  we can introduce a new parameter $\lambda$ defined as
\begin{equation}
\label{eq:triple}
\lambda v^2 \equiv m_\Phi^2 - \frac{m_{12}^2}{\sin \beta \cos \beta}\,,
\end{equation}
which is the parameter that enters the Higgs self-couplings and relevant for the loop corrections to the SM-like Higgs boson couplings. This parameter could be used interchangeably with $m_{12}^2$ as we will do for convenience.  For the rest of our analysis, we fix $v=246$ GeV and $m_h=125$ GeV.   The remaining free parameters are
\begin{equation}\label{eq:para}
\tan\beta\,,\  \cos(\beta-\alpha)\,,\ m_H\,,\ m_A\,,\ m_{H^\pm}\ {\rm and}\ \lambda\,.
\end{equation}
Note that while these six parameters are independent of each other, their allowed ranges under perturbativity, unitarity, and stability consideration are correlated.

For simplicity with important consequences, one often starts from the degenerate case where all heavy Higgs boson masses are set the same. We will explore both the degenerate and non-degenerate cases specified as
\begin{eqnarray}\label{eq:degen}
{\rm Degenerate\ Case:\ }&&
m_\Phi \equiv m_H = m_A = m_{H^\pm} \\
{\rm Non\  Degenerate\ Case:\ }&&
\Delta m_{A,C} \equiv m_{A, H^\pm} -m_{H}\,.
\end{eqnarray}

Given the current LHC Higgs boson measurements~\cite{Coleppa:2013dya, Craig:2013hca,Barger:2013ofa,Belanger:2013xza}, deviations of the Higgs boson couplings from the decoupling and alignment limits are still allowed  at about $10\%$ level.    All the tree-level deviations from the SM Higgs boson couplings are parametrized by only two parameters:  $\tan\beta$ and $\cos(\beta-\alpha)$. Once additional loop corrections are included, dependences on the heavy Higgs boson masses as well as $\lambda v^2$ also enter. In our analyses below, we study the combined contributions to the couplings of the SM-like Higgs boson with both tree-level and loop corrections.

 Before concluding this section, a special remark is in order. The model parameters introduced in this section and henceforth are all at the electroweak scale, identified as on-shell parameters to directly compare with experimental measurements. We do not consider the running effects due to other new physics at a higher scale such as in Supersymmetry or Grand Unified theories. This would become relevant if one asks whether the alignment behavior could be a natural result due to some symmetry or other principles~\cite{Dev:2014yca}. In such scenarios, the alignment may take place at a higher scale but could be modified at the electroweak scale. Our results here, on the other hand, could be viewed as the acceptable deviations from the exact alignment conditions in a more fundamental theory.

 \subsection{Loop corrections to  the SM-like  Higgs couplings}
 \label{sec:hloop}

We define the normalized  SM-like Higgs boson couplings including loop effects as
\begin{eqnarray}
\kappa^{\rm 2HDM}_{\rm loop}  &\equiv & \frac{g_{\rm  tree}^{\rm 2HDM}+g_{\rm loop}^{\rm 2HDM}}{g_{\rm tree}^{\rm SM}+g_{\rm loop}^{\rm SM}} \nonumber\\
 &=&  \kappa_{\rm tree}+\frac{g_{\rm loop}^{\rm 2HDM}(\Phi)}{g_{\rm tree}^{\rm SM}}
\frac{1}{1+\frac{ g_{\rm loop}^{\rm SM}}{g_{\rm tree}^{\rm SM}}}
 + \Big[\frac{g_{\rm loop}^{\rm 2HDM}({\rm SM})}{g_{\rm tree}^{\rm SM}} -
         \kappa_{\rm tree} \frac{ g_{\rm loop}^{\rm SM}}{g_{\rm tree}^{\rm SM}} \Big]
         \frac{1}{1+\frac{ g_{\rm loop}^{\rm SM}}{g_{\rm tree}^{\rm SM}}}\,,
         \label{eq:loop}
\end{eqnarray}
where $\kappa_{\rm tree}\equiv g_{\rm tree}^{\rm 2HDM}/g_{\rm tree}^{\rm SM}$. $g_{\rm loop}^{\rm 2HDM}({\rm \Phi})$ and $g_{\rm loop}^{\rm 2HDM}({\rm SM})$ are the 2HDM Higgs boson couplings including loop corrections with heavy Higgs bosons or with SM particles only, respectively.

To the leading order in 1-loop corrections, \autoref{eq:loop} simplifies to
\begin{equation}
\kappa^{\rm 2HDM}_{\rm 1-loop}  = \kappa_{\rm tree} + \Delta \kappa^{\rm 2HDM}_{\rm 1-loop}+ \Big[\frac{g_{\rm 1-loop}^{\rm 2HDM}({\rm SM})}{g_{\rm tree}^{\rm SM}} -
         \kappa_{\rm tree} \frac{ g_{\rm 1-loop}^{\rm SM}}{g_{\rm tree}^{\rm SM}} \Big]\,,
 \end{equation}
with $\Delta \kappa^{\rm 2HDM}_{\rm 1-loop}\equiv g_{\rm 1-loop}^{\rm 2HDM}(\Phi)/g_{\rm tree}^{\rm SM}$.  In the alignment limit of $\kappa_{\rm tree}=1$, the term in the bracket is exactly zero, and $\kappa^{\rm 2HDM}_{\rm 1-loop} |_{\rm alignment} = 1 + \Delta \kappa^{\rm 2HDM}_{\rm 1-loop}$.

In our calculations, we adopt the on-shell renormalization scheme~\cite{FeynArts-SM}. The conventions for the renormalization constants and the renormalization conditions are mostly following Refs.~\cite{FeynArts-SM,Kanemura:2004mg}. All related counter terms, renormalization constants and renormalization conditions are implemented according to the on-shell scheme and incorporated into model files of 
{\tt FeynArts} \cite{Hahn:2000kx}\footnote{Note that in this scheme, there will be gauge-dependence in the calculation of the counter term of $\beta$~\cite{Freitas:2002um}. For convenience, we will adopt this convention and the Feynman-'t-Hooft gauge is used throughout the calculations. For more sophisticated gauge-independent renormalization scheme to deal with $\alpha$ and $\beta$, see~\cite{Krause:2016oke,Denner:2016etu,Altenkamp:2017ldc,Kanemura:2017wtm}. Corresponding implementations have been uploaded to \href{https://github.com/ycwu1030/THDMNLO_FA}{https://github.com/ycwu1030/THDMNLO\_FA}.}.
One-loop corrections are generated using {\tt FeynArts} and {\tt FormCalc}~\cite{Hahn:2016ebn} including all possible one-loop diagrams. {\tt FeynCalc}~\cite{Shtabovenko:2016sxi,Mertig:1990an} is also used to simplify the analytical expressions.   {\tt LoopTool}~\cite{Hahn:1998yk} is used to evaluate the numerical value of all the loop-induced amplitude. The numerical results have been cross-checked with another numerical program {\tt H-COUP}~\cite{Kanemura:2017gbi} in some cases.

For the couplings of the SM-like Higgs boson to a pair of gauge bosons and fermions,   the general renormalized $hff$ and $hVV$ vertices take the following forms
 \begin{eqnarray}
\hat\Gamma^R_{hff}(p_1^2,p_2^2,q^2) &=& \hat\Gamma^S_{hff} + \hat\Gamma^P_{hff}\gamma^5 + \hat\Gamma^{V_{p_1}}_{hff}\slashed{p}_1 + \hat\Gamma^{V_{p_2}}_{hff}\slashed{p}_2 \nonumber \\
&& + \hat\Gamma_{hff}^{A_{p_1}}\slashed{p}_1\gamma^5 + \hat\Gamma_{hff}^{A_{p_2}}\slashed{p}_2\gamma^5 + \hat\Gamma_{hff}^{T}\slashed{p}_1\slashed{p}_2 + \hat\Gamma_{hff}^{PT}\slashed{p}_1\slashed{p}_2\gamma^5\,, \\
\hat\Gamma^{R,\mu \nu}_{hVV}(p_1^2,p_2^2,q^2) &=& \hat\Gamma_{hVV}^1g^{\mu\nu} + \hat\Gamma_{hVV}^2\frac{p_1^\mu p_2^\nu}{m_V^2}+i\hat\Gamma_{hVV}^3\epsilon^{\mu\nu\rho\sigma}\frac{p_{1\rho}p_{2\sigma}}{m_V^2}\,,
\end{eqnarray}
where $q^\mu$, $p_1^\mu$, and $p_2^\mu$ are the momenta of the Higgs boson and two other particles, respectively, and $q^2$ is the typical momentum transfer of the order $m_h^2$.
 $\kappa_i$ for each vertex is given by $\hat\Gamma_{hff}^S$ and $\hat\Gamma_{hVV}^1$ for $hff$ and $hVV$, which includes both the tree-level and one-loop corrections:
\begin{equation}
 {\kappa}_V^{} = \frac{\hat{\Gamma}_{hVV}^1(m_V^2,m_h^2,q^2)_{\text{2HDM}}}{\hat{\Gamma}_{hVV}^1(m_V^2,m_h^2,q^2)_{\text{SM}}}\,,\ \ \
{\kappa}_f^{} = \frac{\hat{\Gamma}^S_{hff}(m_f^2,m_f^2,q^2)_{\text{2HDM}}}{\hat{\Gamma}^S_{hff}(m_f^2,m_f^2,q^2)_{\text{SM}}}\,.
  \label{BottomUp:kappa_hat}
\end{equation}

\subsection{Loop corrections to $Z$-pole precision observables}
\label{sec:Zloop}

The 2HDM contributions to the Peskin-Takeuchi oblique parameters~\cite{Peskin:1991sw} are given by~\cite{He:2001tp}\footnote{Here, we fix a typo in~\cite{He:2001tp} in the expression for $\Delta\,U$.}
\begin{eqnarray}
\label{eqs:2HDM_EWPD}
\Delta\,S&=&\frac{1}{\pi\, m_Z^2} \Big\{  \Big[{\cal B}_{22}( m_Z^2\,; m_H^2\,, m_A^2 ) - {\cal B}_{22}( m_Z^2\,; m_{H^\pm}^2\,, m_{H^\pm}^2)\Big] \nonumber \\
&&+ \Big[ {\cal B}_{22}( m_Z^2\,; m_h^2\,, m_A^2 )- {\cal B}_{22}( m_Z^2\,; m_H^2\,, m_A^2 ) + {\cal B}_{22}( m_Z^2\,; m_Z^2\,, m_H^2 ) - {\cal B}_{22}( m_Z^2\,; m_Z^2\,, m_h^2 ) \nonumber \\
&& - m_Z^2 {\cal B}_0 ( m_Z\,; m_Z\,, m_H^2 ) + m_Z^2 {\cal B}_0 ( m_Z\,; m_Z\,, m_h^2 )    \Big] \cos^2(\beta-\alpha)  \Big\}\,,\\
\Delta T&=& \frac{1}{ 16\pi\, m_W^2\, s_W^2 } \Big\{   \Big[ F( m_{H^\pm}^2 \,, m_A^2 ) +  F( m_{H^\pm}^2\,, m_H^2) - F(m_A^2\,, m_H^2)  \Big]\nonumber \\
&&+   \Big[  F( m_{H^\pm}^2 \,, m_h^2 ) -  F( m_{H^\pm}^2\,, m_H^2) - F( m_A^2\,, m_h^2 )+ F(m_A^2\,, m_H^2)   \nonumber \\
&& + F( m_W^2\,, m_H^2) - F(m_W^2\,, m_h^2 ) - F(m_Z^2 \,, m_H^2 ) + F(m_Z^2\,, m_h^2) \nonumber \\
&& + 4 m_Z^2 \overline B_0 (m_Z^2\,, m_H^2\,, m_h^2)   - 4 m_W^2 \overline B_0 ( m_W^2\,, m_H^2\,, m_h^2 )  \Big] \cos^2(\beta-\alpha) \Big\}\,, \\
\Delta\,U&=& -\Delta\,S + \frac{1}{\pi m_W^2}\Big\{ \Big[{\cal B}_{22}(m_W^2,m_A^2,m_{H^\pm}^2)-2{\cal B}_{22}(m_W^2,m_{H^\pm}^2,m_{H^\pm}^2) + {\cal B}_{22}(m_W^2,m_H^2,m_{H^\pm}^2)\Big] \nonumber \\
&& + \Big[{\cal B}_{22}(m_W^2,m_h^2,m_{H^\pm}^2) - {\cal B}_{22}(m_W^2,m_H^2,m_{H^\pm}^2) + {\cal B}_{22}(m_W^2,m_W^2,m_H^2) - {\cal B}_{22}(m_W^2,m_W^2,m_h^2) \nonumber \\
&& - m_W^2{\cal B}_0(m_W^2,m_W^2,m_H^2) + m_W^2{\cal B}_0(m_W^2,m_W^2,m_h^2)\Big]\cos^2(\beta-\alpha) \Big\}\,,
\end{eqnarray}
where we explicitly split these expressions into terms independent of or dependent on the alignment parameter of $\cos(\beta-\alpha)$. The expression for various ${\cal B}$ and $F$-functions can be found in Ref.~\cite{He:2001tp}.
The mass splittings among heavy Higgs bosons of $(m_H\,, m_A\,, m_{H^\pm})$ violate the SU(2) custodial symmetry and thus will lead to contributions to the $T$ and $U$ parameters.

\begin{figure}[tb]
\centering
\includegraphics[width=0.45\textwidth]{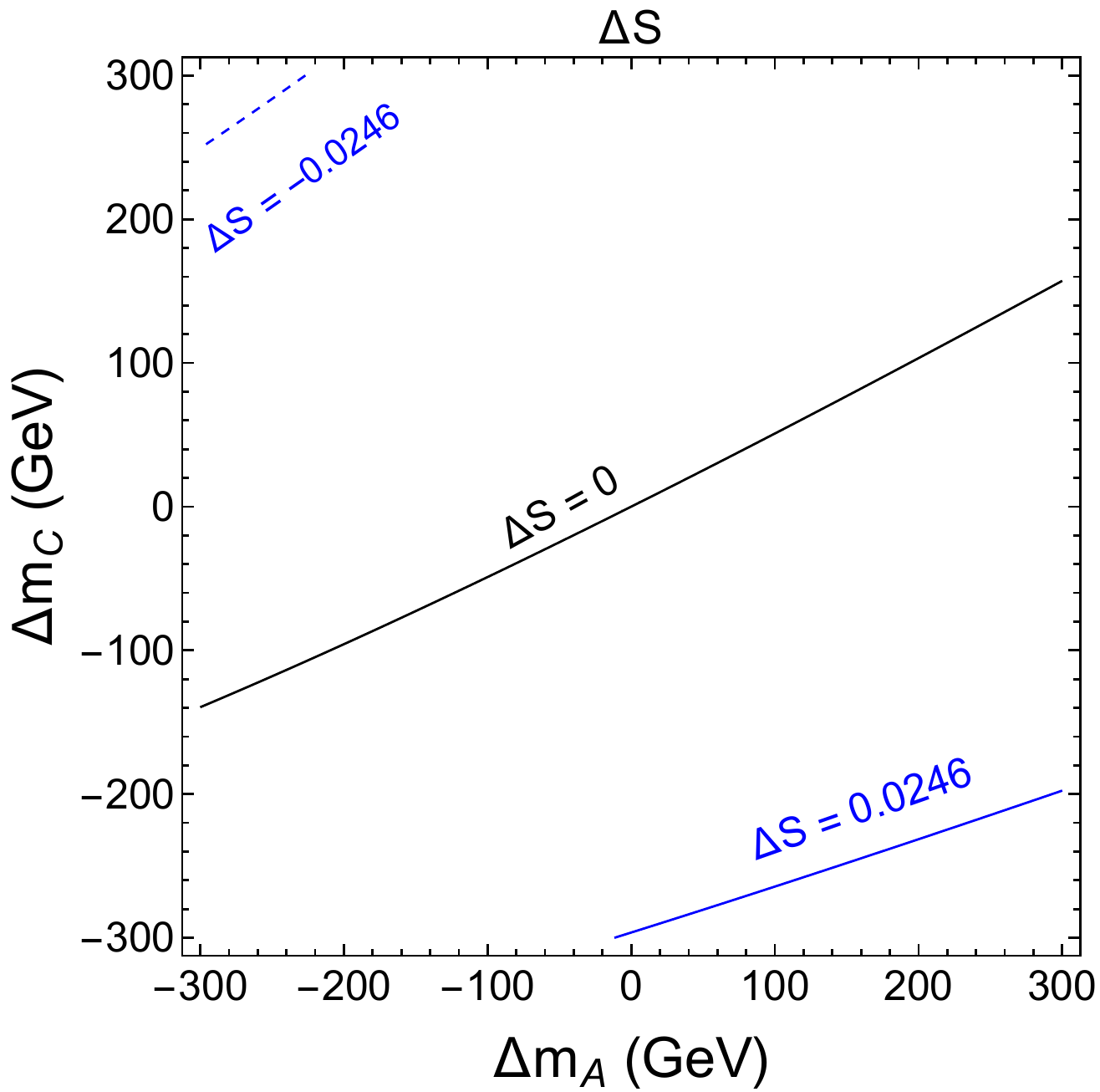}
\includegraphics[width=0.45\textwidth]{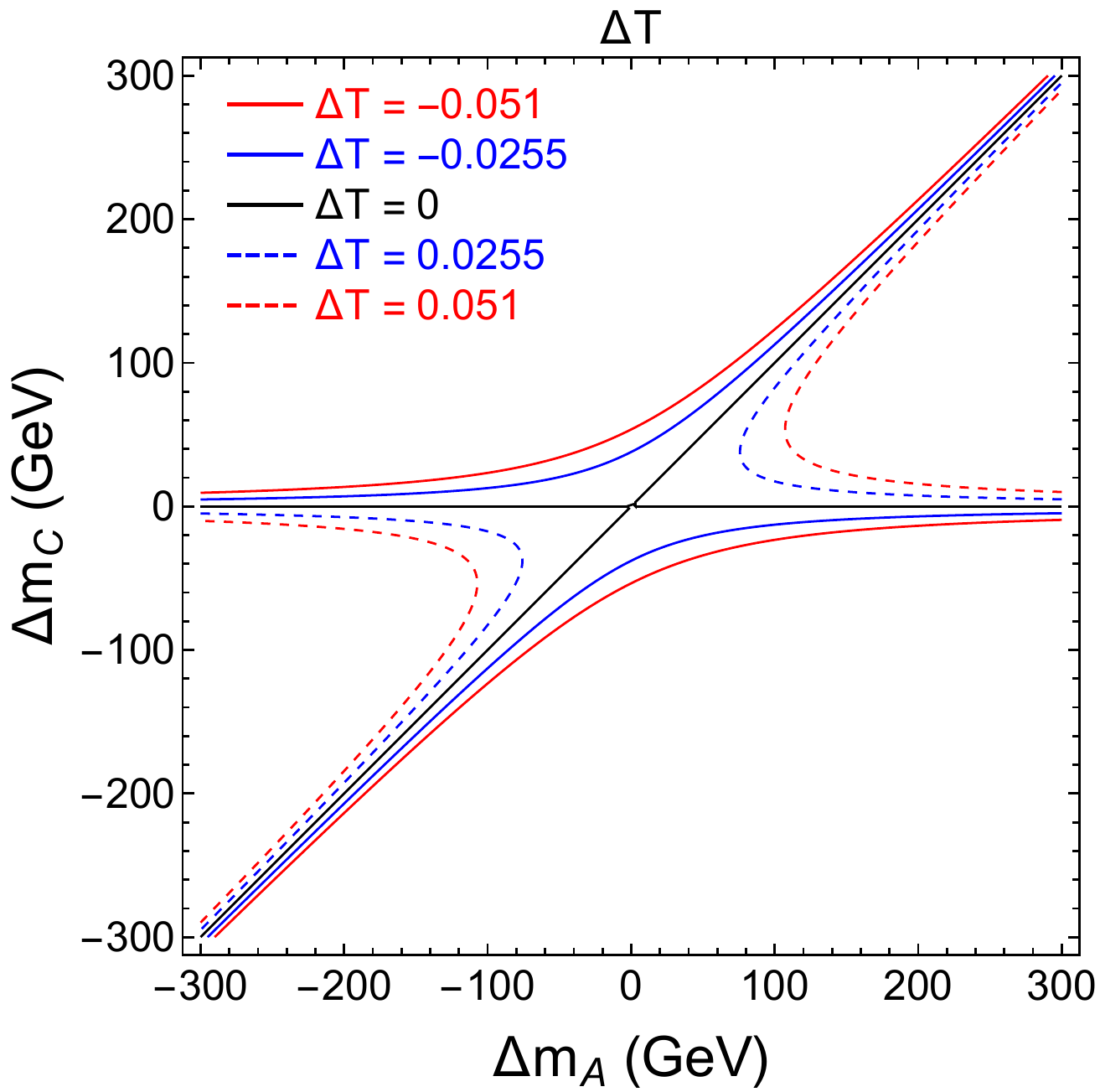}\\
 \caption{2HDM contributions to $\Delta S$ (left panel) and $\Delta T$ (right panel) in the $\Delta m_A$-$\Delta m_C$ plane. We fix $m_H=800$ GeV under the alignment limit. The blue and red lines represent the $1\sigma$ and $2\sigma$ CEPC precisions of $(\Delta S\,, \Delta T)$ respectively.  
}
\label{fig:STU_THDM_mamc}
\end{figure}

In~\autoref{fig:STU_THDM_mamc},  we show the contributions to $\Delta S$ (left panel) and $\Delta T$ (right panel) in 2HDM varying $\Delta m_A \equiv m_A - m_H$ and $\Delta m_C \equiv m_{H^\pm} - m_H$ between $\pm$ 300 GeV, for $\cos({\beta-\alpha})=0$.  While the contribution to $\Delta S$ is typically small $|\Delta S| \lesssim 0.03$, the contribution to $\Delta T$ quickly increases when $m_{H^\pm}$ is non-degenerate with either $m_A$ or $m_H$.
Therefore, an improved determination of $\Delta T$ from $Z$-pole precision measurement would severely constrain the mass splitting between the charged Higgs and its neutral partners. Furthermore, non-alignment case also breaks the symmetric pattern between $\Delta m_A$ and $\Delta m_C$ for $\Delta T$ contribution, preferring a slightly negative value of mass splittings.

  \subsection{Theoretical constraints and current experimental bounds}
 \label{sec:bounds}

Heavy Higgs loop corrections would involve the Higgs boson masses and self-couplings, which are constrained by various theoretical considerations and experimental measurements, such as vacuum stability,  perturbativity and unitarity, as well as electroweak precision measurements, flavor physics constraints,  and LHC direct searches.  We briefly summarize below the theoretical considerations and experimental constraints.

\begin{itemize}
\item \textbf{Vacuum stability}

In order to have a stable vacuum, the following conditions on the quartic couplings need to be satisfied~\cite{Gunion:2002zf}:
\begin{align}
\lambda_1>0, \quad \lambda_2>0,\quad \lambda_3 > -\sqrt{\lambda_1\lambda_2},\quad \lambda_3+\lambda_4-|\lambda_5|> -\sqrt{\lambda_1\lambda_2}\,.
\end{align}

\item \textbf{Perturbativity and unitarity}

We adopt a general perturbativity condition of $|\lambda_i| \leq 4\pi$ and the tree-level unitarity of the scattering matrix in the 2HDM scalar sector~\cite{Ginzburg:2005dt}.

\end{itemize}

\begin{figure}[tb]
\begin{center}
\includegraphics[width=7cm]{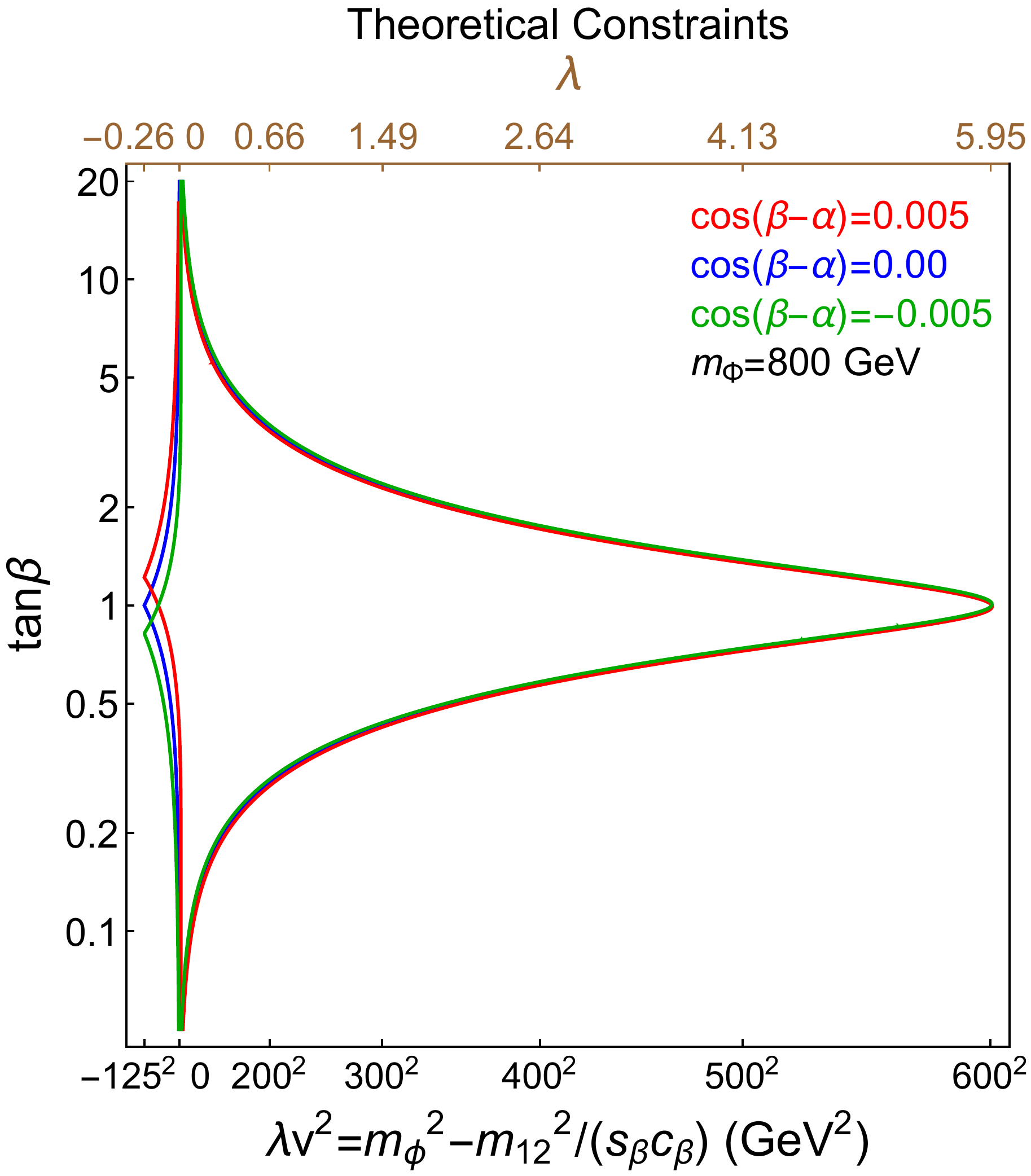}
 \includegraphics[width=7cm]{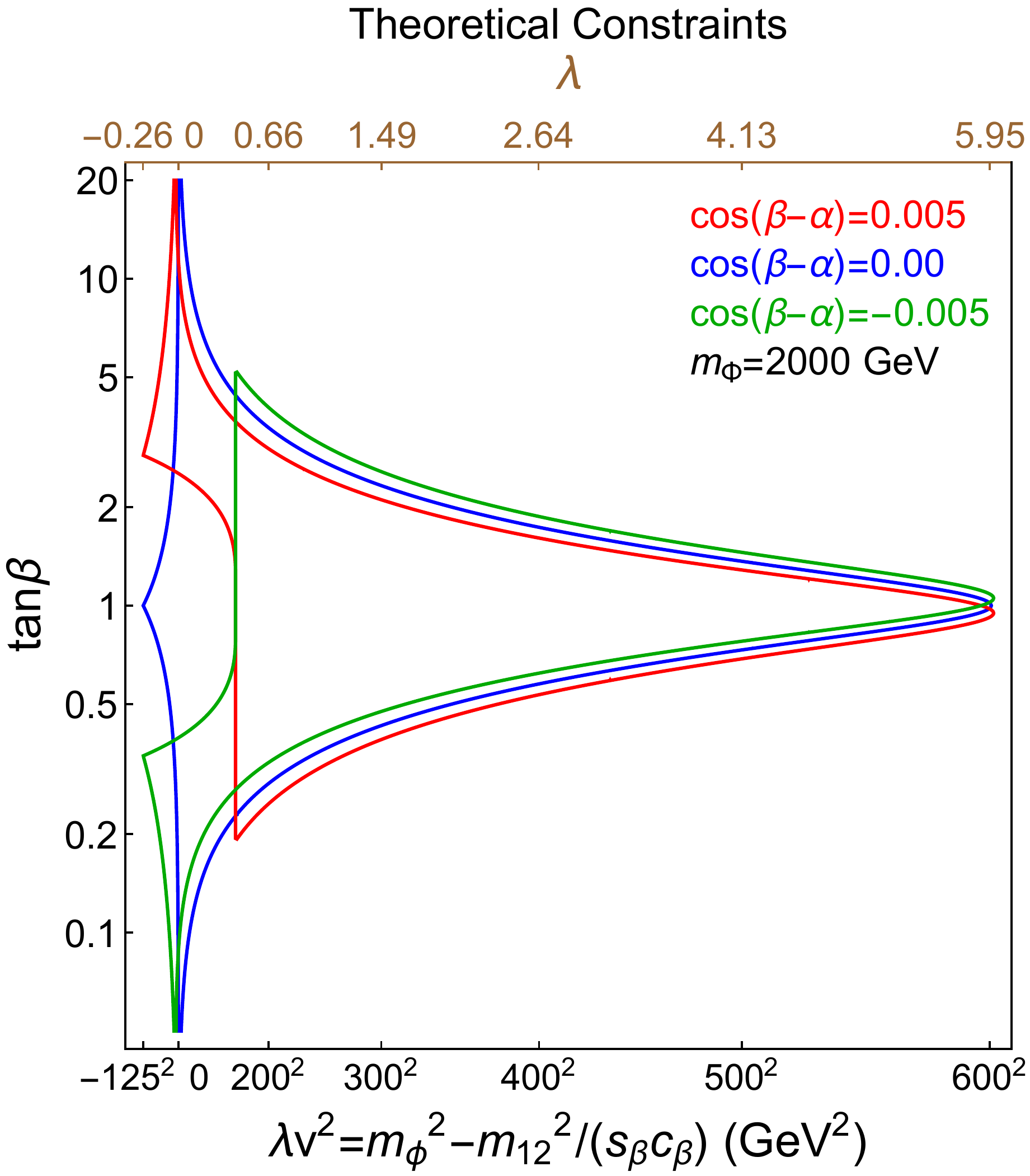}\\
 \vspace{5 mm}
 \includegraphics[width=7cm]{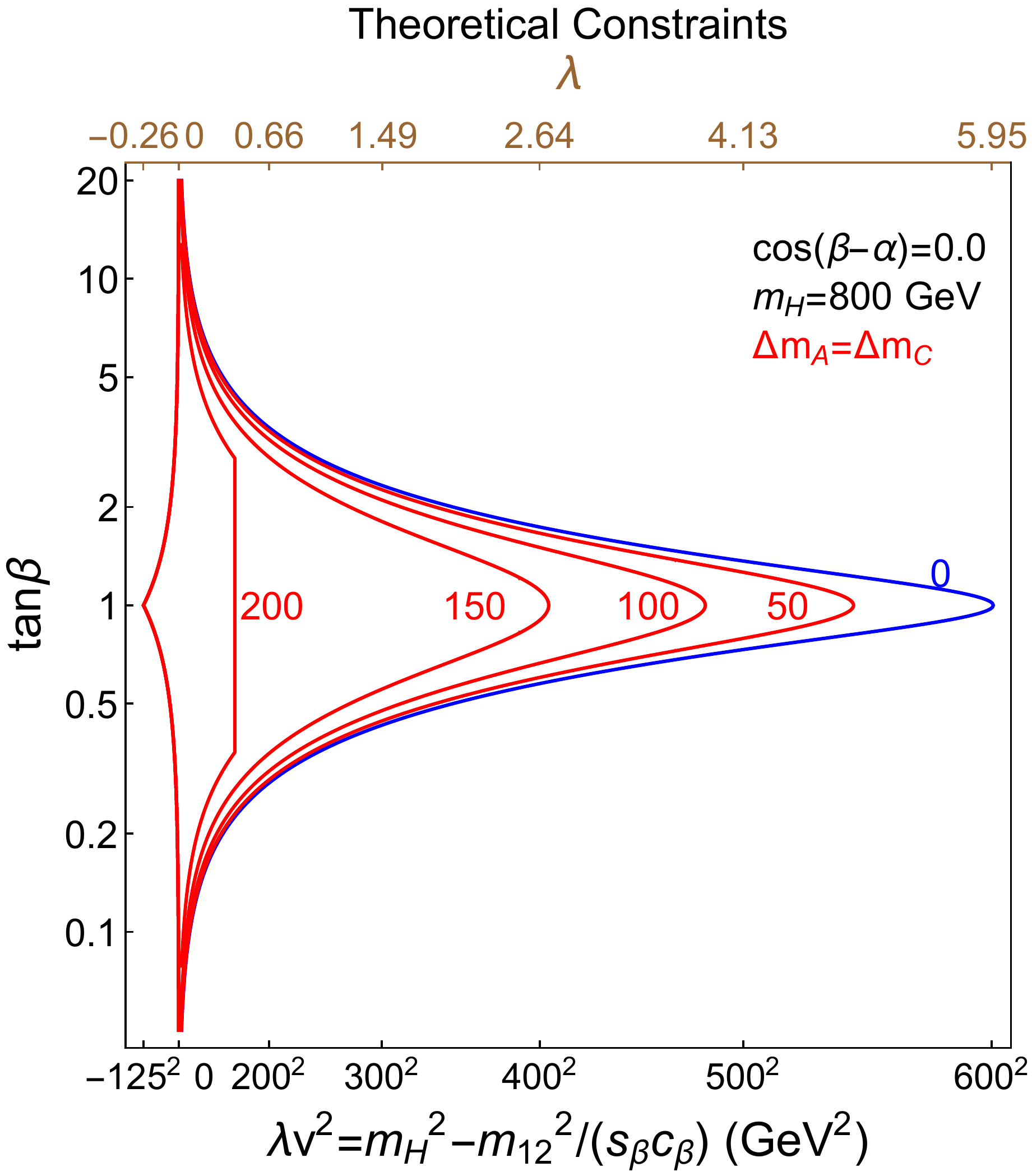}
  \includegraphics[width=7cm]{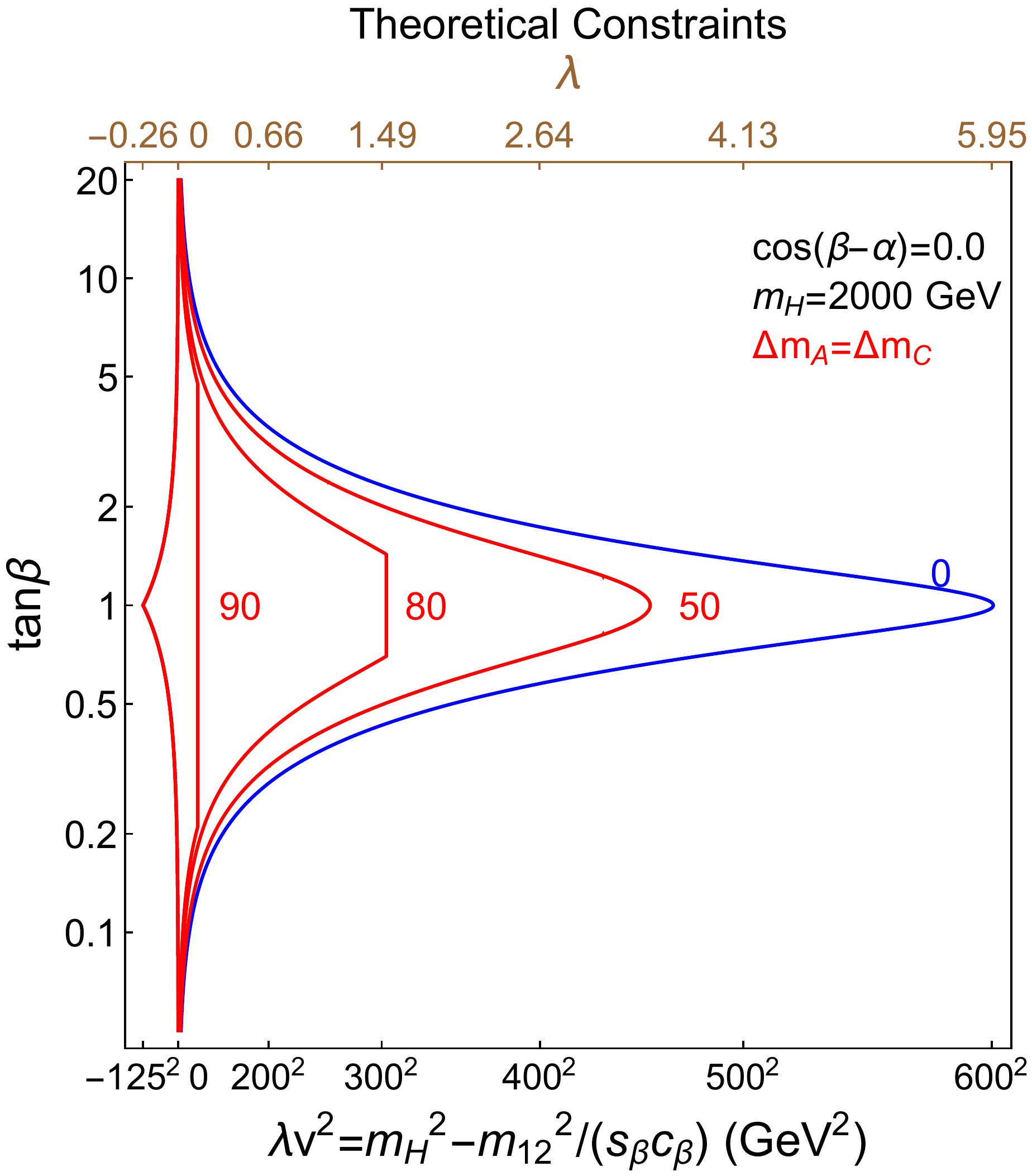}
\caption{Constraints in the $\lambda v^2$-$\tan\beta$ plane with all theoretical considerations taken into account.  The left panel is for $m_\Phi=800$ GeV and the right panel is for $m_\Phi=2000$ GeV.  The upper panels show $\cosba$ effects with $\cos(\beta-\alpha)=$0.005 (red curves), 0 (alignment limit, blue curves), and $-0.005$ (green curves) under degenerate heavy Higgs masses $m_{H^{\pm}}=m_{H}=m_{A}\equiv m_{\Phi}$ assumption.  The lower panels show the mass splitting effects with varying $\Delta m_A=\Delta m_C = m_{A/H^\pm}-m_H$.
}
\label{fig:theo_const_lambdav2TB}
\end{center}
\end{figure}

In~\autoref{fig:theo_const_lambdav2TB}, we show the constraints in the $\lambda v^2$-$\tan\beta$ plane once all the theoretical considerations are taken into account. For the upper panels, we work under the assumption with degenerate heavy Higgs boson masses $m_{H^{\pm}}=m_{H}=m_{A}\equiv m_{\Phi}$. The left panel is for $m_\Phi=800$ GeV and the right one is for $m_\Phi=2000$ GeV, with $\cos(\beta-\alpha)=$0.005 (red curves), 0 (alignment limit, blue curves), and $-0.005$ (green curves). Regions enclosed by the curves are theoretically preferred.
For a lower mass $m_\Phi=$ 800 GeV, the constraints vary very little with the values of $\cos(\beta-\alpha)$.
The largest range on $\lambda v^2 \equiv m_\Phi^2 -  m_{12}^2/\sin\beta \cos\beta  $ occurs at $\tan \beta =1$~\cite{Gu:2017ckc}:
\begin{equation}
  -m_h^2 < \lambda v^2 < (600\  \text{GeV})^2\,,
     \label{eq:lam_cons}
\end{equation}
which gives $ -0.29 < \lambda =- \lambda_4=- \lambda_5  < 5.95$ and $0 <  \lambda_3  < 6.21$.  For a large value of  $m_\Phi=2000$ GeV, a slight shift of $\cos(\beta-\alpha)$ leads to notable change in constraints on $\lambda v^2$,  as shown by the red and green curves in the top right panel of~\autoref{fig:theo_const_lambdav2TB}.

The theoretically preferred region also depends on the individual heavy Higgs boson masses, as well as the deviation from the degenerate condition. In the lower panels of~\autoref{fig:theo_const_lambdav2TB}, we show the constrained region for difference choices of $\Delta m_{A,C}$ with $m_H =800$ GeV (left) and 2000 GeV (right).  The degenerate case provides the weakest constraints, as shown by the blue line. Larger mass splittings lead to tighter constraints. For larger $m_H$, only smaller mass splittings between heavy Higgs bosons can be accommodated.  This is because  at large $m_H$, $ \Delta m \propto \lambda_i v^2/m_H$, with $\lambda_i$ being bounded by perturbativity and unitarity considerations.

\begin{itemize}
\item \textbf{LHC search bounds}
\end{itemize}
LHC Run-I at $7,\ 8$ TeV and Run-II at $13$ TeV have searched the heavy Higgs bosons in 2HDM via various channels.
The direct searches for neutral heavy Higgs bosons include the decay channels $\tau^+ \tau^-$~\cite{Aaboud:2017sjh,CMS-PAS-HIG-17-020}, $t\bar{t}$~\cite{Aaboud:2017hnm}, $WW/ZZ$~\cite{Aaboud:2017gsl,Aaboud:2017rel,Sirunyan:2018qlb}, $\gamma\gamma$~\cite{Aaboud:2017yyg},   $A\to hZ$~\cite{Aaboud:2017cxo},  $A/H\rightarrow HZ/AZ$~\cite{Aaboud:2018eoy,Khachatryan:2016are} and $H\rightarrow hh$~\cite{Aaboud:2018ftw,Sirunyan:2018iwt}.
The strongest bounds at large $\tan\beta$ come from $A/H\rightarrow \tau^+\tau^-$ mode, which excludes $m_{A/H}\sim 300-500 $ GeV for $\tan\beta\sim 10$, and about 1500 GeV for $\tan\beta\sim 50$. The strongest bounds at small $\tan\beta\lesssim 1$ come from $A/H\rightarrow  t\bar{t}$ mode.  The latest ATLAS search on such channel utilized the lineshape of $t\bar{t}$ invariant mass distribution, which exhibits a peak-dip structure due to the interference between the signal and the SM $t\bar{t}$ background~\cite{Dicus:1994bm,Carena:2016npr}.   A strong 95\% C.L. bound of $m_{A/H}$ around 600 GeV can be reached for $\tan\beta=1$ for degenerate mass of $m_A=m_H$ under the alignment limit. 
The direct searches for heavy charged Higgs bosons have been conducted with the $H^\pm \to (\tau\nu\,, tb)$ channels~\cite{ATLAS-CONF-2016-089,Aaboud:2018gjj,CMS-PAS-HIG-16-031}, and the bounds  are relatively weak given the rather small leading production cross section for $bg \to tH^\pm$,  the large SM backgrounds for the dominant $H^\pm\rightarrow tb$ channel and the relatively small branching fraction of $H^\pm \rightarrow \tau \nu$~\cite{Arbey:2017gmh}.

The search sensitivities at the high-luminosity LHC (HL-LHC) for the heavy Higgs bosons have been estimated in Ref.~\cite{Baglio:2015wcg}, with the rescaling of the LHC $7\oplus 8$ TeV search limits under  the alignment limit and mass-degenerate assumption.  The strongest constraints for the large $\tan\beta$ region come from the $A/H\to \tau^+ \tau^-$ searches: $m_{A/H}$ could be excluded to about 1000 GeV for $\tan\beta\sim 10 $, and even larger masses for larger $\tan\beta$. $H^\pm \rightarrow {tb}$ offers better exclusion at low $\tan\beta$, which excludes $m_{H^\pm}$ to about 600 GeV for $\tan\beta\sim 1$. Possible $A/H\rightarrow {t \bar t}$ mode might help to extend the exclusion reach to about 2000 GeV for $\tan\beta\sim 1$~\cite{Craig:2016ygr,Carena:2016npr}. At 100 TeV $pp$ collider with 3 ${\rm ab^{-1}}$ luminosity, $A/H\to \tau^+ \tau^-$ could extend the reach at large $\tan\beta$ to about 2000 GeV at $\tan\beta\sim 10$ and about $3$ TeV for $\tan\beta\sim 50$. The coverage at low $\tan\beta$ could also be extended to about $m_{H^\pm}\sim 1500$ GeV via $H^\pm \rightarrow {tb}$ and $m_A\sim 2500$ GeV via $A/H\rightarrow {t \bar t}$ for $\tan\beta\sim 1$~\cite{Baglio:2015wcg}.

Since the branching fractions of the conventional search channels could be highly suppressed once other exotic decay channels of the non-SM Higgs boson to light Higgs bosons and/or SM gauge bosons open up~\cite{Coleppa:2013xfa,Coleppa:2014hxa,Kling:2015uba}, it is important to note that the current exclusion limits could be relaxed. Current LHC limits on $m_{A,H}$ via searches of exotic decay modes $A/H\rightarrow HZ/AZ$ are up to about $700-800$ GeV, depending on the spectrum of non-SM Higgs bosons~\cite{Aaboud:2018eoy,Khachatryan:2016are}.   $m_{A,H}$ could be excluded to about 1500 GeV at HL-LHC and about 3000 GeV at 100 TeV $pp$ collider~\cite{Felix100TeV}.

\begin{figure}[tb]
\centering
\includegraphics[width=0.45\textwidth]{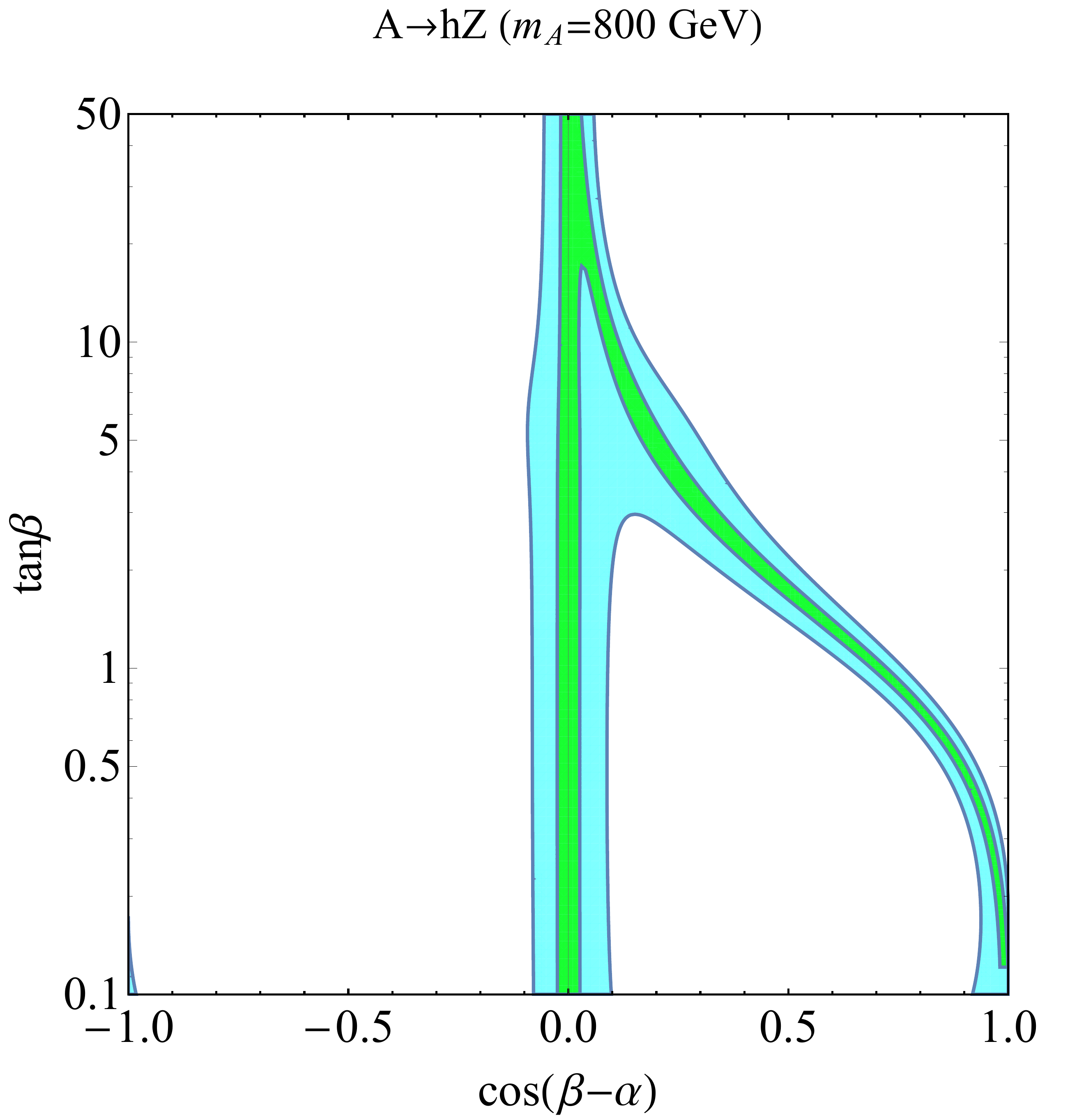}
\includegraphics[width=0.45\textwidth]{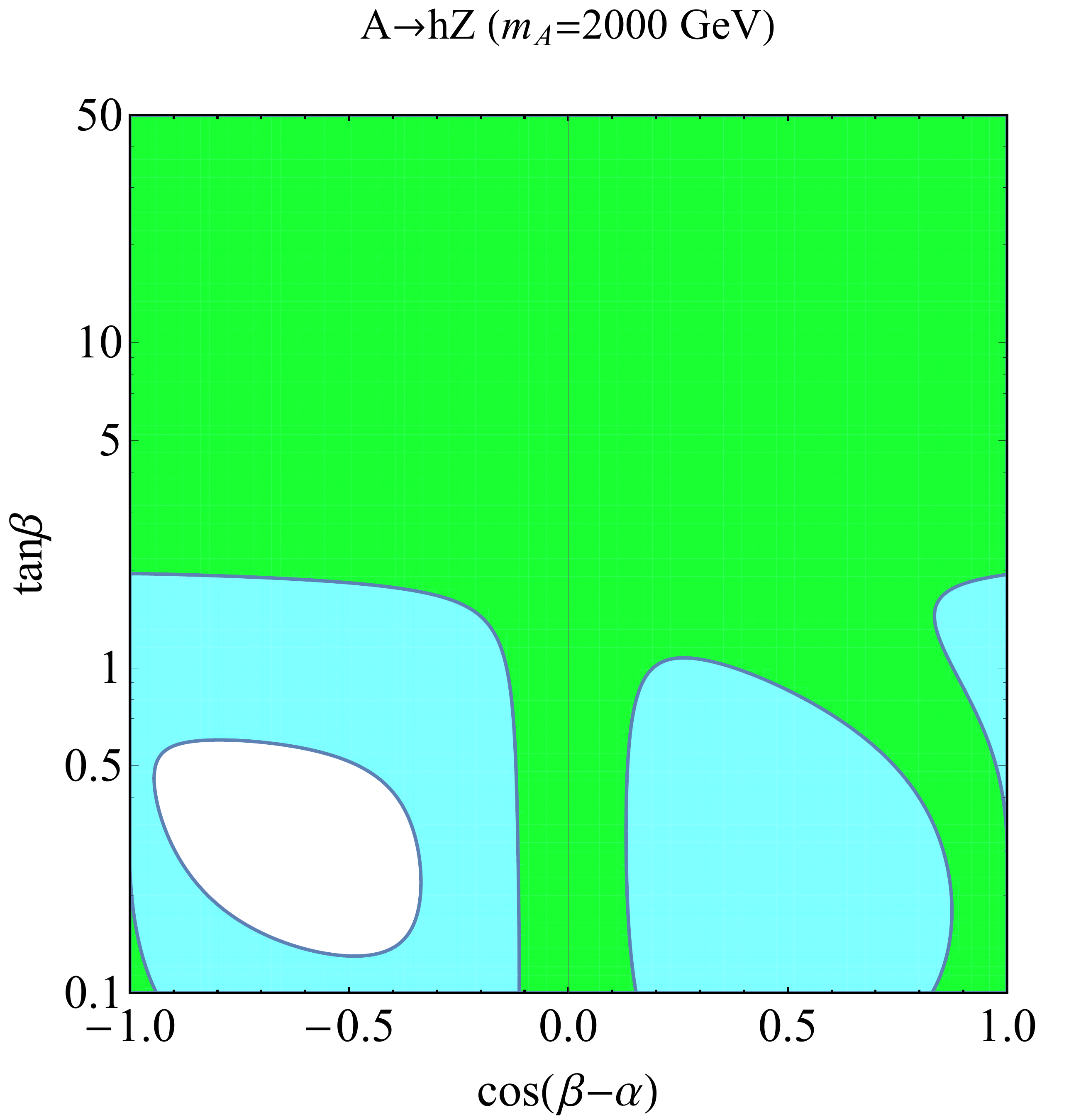}
\caption{Constraints in the $\cos(\beta-\alpha)$-$ \tan\beta$ plane with the LHC 13 TeV $36.1$ fb$^{-1}$ (cyan) and the projected HL-LHC 14 TeV $3000$ fb$^{-1}$ (green) $A\rightarrow hZ\rightarrow bb\ell\ell$ search limits~\cite{Aaboud:2017cxo,CMS-PAS-FTR-13-024}, for $m_A=800$ GeV (left) and $m_A=2000$ GeV (right).
The color-shaded regions are allowed. }
\label{fig:LHC_heavyH_limits}
\end{figure}

While the exotic Higgs decay channel of $A\rightarrow h(\to b \bar b, \tau^+\tau^-)Z$ is absent in the alignment limit,   this channel could be used to constrain $\cos(\beta-\alpha)$ and $\tan\beta$ when the deviation from the alignment limit is allowed. The projected $A\rightarrow hZ$ search  results in the $\cos(\beta-\alpha) $-$ \tan\beta$ plane of LHC 13 TeV with an integrated luminosity of $36$ fb$^{-1}$  (cyan) \cite{Aaboud:2017cxo} and future HL-LHC 14 TeV with an integrated luminosity of 3 ab$^{-1}$  (green)~\cite{CMS-PAS-FTR-13-024} for $m_A=800$ GeV  (left panel) and $m_A=2000$ GeV (right panel) are shown in~\autoref{fig:LHC_heavyH_limits} with the colored survival regions.
For the case of $m_A=800$ GeV, a narrow band within $|\cos(\beta-\alpha)|\lesssim 0.1$ or $|\cos(\beta-\alpha)|\lesssim 0.02$ is still allowed by the current LHC or the future HL-LHC data, as expected.
Another branch from $\cos(\beta-\alpha)=0$ to $\cos(\beta-\alpha)=1.0$ with $\tan\beta$ decreasing from $5-10$ to $\sim 0.1$ is also allowed, which corresponds to the region with a suppressed $\textrm{BR}(h\to b \bar b)$.
The constraint for the $m_A=2000$ GeV case is far less stringent for the LHC 13 TeV case.
Only the lower left region is excluded, in which both the production cross section  $\sigma(gg\to A)$ and  decay branching fraction of $\textrm{BR}(A\to hZ)\times \textrm{BR}(h\to b \bar b)$ are enhanced.
For the HL-LHC case, the $\tan\beta\lesssim 1$ regions are largely excluded, leaving the narrow band with $|\cos(\beta-\alpha)|\lesssim 0.1$ or a branch stretching from $\cos(\beta-\alpha)=0$ to $\cos(\beta-\alpha)=1.0$ with $\tan\beta$ decreasing from $\sim 1$ to $\sim 0.1$ allowed by the future HL-LHC data.  This is complementary to the SM-like Higgs boson signal strength measurements, which  constrain the range of $\cos(\beta-\alpha)$ to be less than about 0.1 around $\tan\beta\sim1$ and even narrower regions for small and large $\tan\beta$ for Type-II 2HDM~\cite{Gu:2017ckc} with the current LHC measurements, except for a small wrong-sign Yukawa coupling region at $\tan\beta\gtrsim 2$.

Flavor physics consideration usually constrains the charged Higgs mass to be larger than about 600 GeV for the Type-II  2HDM~\cite{Arbey:2017gmh}.  However, given the uncertainties involved in those flavor measurements, and that they are in general less stringent than the direct collider limits,  we thus will not pursue the flavor bounds further.


 \section{Study Strategy and Results}
 \label{sec:results}

In an earlier work~\cite{Gu:2017ckc}, constraints from the tree-level effects on $\cos(\beta-\alpha)$ and $\tan\beta$, as well as from loop contributions in the degenerate mass case $m_H=m_A=m_{H^\pm}=m_\Phi$ under the alignment limit are analyzed. In this work,  we extend the studies to more general cases of the non-degenerate masses and non-alignment, as well as including both the tree-level and one-loop contributions. We also incorporate the $Z$-pole precision results to show the complementarity between the Higgs and $Z$-pole precision measurements.

\subsection{Global fit framework}
\label{sec:fit}

To transfer the  anticipated accuracy on the experimental measurements to the constraints on the model parameters, we perform a global fit by constructing the $\chi^2$ with the profile likelihood method
\begin{equation}
  \chi^2 = \sum_{i} \frac{(\mu_i^{\rm BSM}-\mu_i^{\rm obs})^2}{\sigma_{\mu_i}^2}\,.
  \label{eq:chi2Higgs}
\end{equation}
 Here, $\mu_i^{\rm BSM} = (\sigma\times \text{BR})_{\rm BSM} / (\sigma\times \text{BR})_{\rm SM}$ for various Higgs search channels.    We note that the correlations among different $\sigma\times {\rm BR}$ are usually not provided, and are thus assumed to be zero in the fits.   $\mu_i^{\rm BSM}$ is predicted in each specific model, depending on model parameters. In our analyses, for the future colliders, $\mu_i^{\rm obs}$ are set to be the SM value $\mu_i^{\rm obs}=1$, assuming no deviations from the SM observables. The corresponding $\sigma_{\mu_i}$ are the estimated errors for each process, as already shown in~\autoref{tab:mu_precision} for the CEPC, FCC-ee and ILC. For the ILC with three different center-of-mass energies, we sum the contributions from each individual channel.

We fit directly to the signal strength $\mu_i$, instead of the effective couplings $\kappa_i$.  The latter are usually presented in most experimental papers.  While using the $\kappa$-framework is easy to map to specific models, unlike $\mu_i$, various $\kappa_i$ are not independent experimental observables.  Ultimately, fitting to either  $\mu_i$ or $\kappa_i$  should give the same results, if the correlations between $\kappa_i$ are properly included. Those correlation matrices, however, are typically not provided  from experiments. Therefore, fitting to $\kappa_i$ only, assuming no correlations, usually leads to more relaxed constraints.  For a comparison of $\mu$-fit versus $\kappa$-fit results, see Ref.~\cite{Gu:2017ckc}.

For $Z$-pole precision measurements, we fit into the oblique parameters $S$, $T$ and $U$, including the correlations between those oblique parameters, as given in~\autoref{tab:STU}. We define the $\chi^2$ as
\bea
\chi^2 &\equiv& \sum_{ij} ( X_i - \hat X_i) (\sigma^2)_{ij}^{-1} ( X_j - \hat X_j )\,,
\label{eq:chi2Zpole}
\eea
with $ X_i=(\Delta S\,, \Delta T\,, \Delta U)_{\rm 2HDM}$ being the 2HDM predicted values, and  $\hat{X}_i=(\Delta S\,, \Delta T\,, \Delta U)$ being the  current best-fit central value for current measurements, and 0 for future measurements.  The $\sigma_{ij}$ are the error matrix, $\sigma_{ij}^2\equiv \sigma_i \rho_{ij} \sigma_j$ with $\sigma_i$ and correlation matrix $\rho_{ij} $ given in~\autoref{tab:STU}.

For the comprehensive fit, including both Higgs boson and $Z$-pole measurements, $\chi^2$ in~\autoref{eq:chi2Higgs} and~\autoref{eq:chi2Zpole} are linearly combined.  For the one-, two- or three-parameter fit, the corresponding $\Delta \chi^2=\chi^2 -\chi^2_{\rm min}$ for 95\% C.L. is 3.84, 5.99 or 7.82,  respectively.

\subsection{Case with degenerate heavy Higgs boson masses}

\begin{figure}[tb]
\begin{center}
\includegraphics[width=10.5cm]{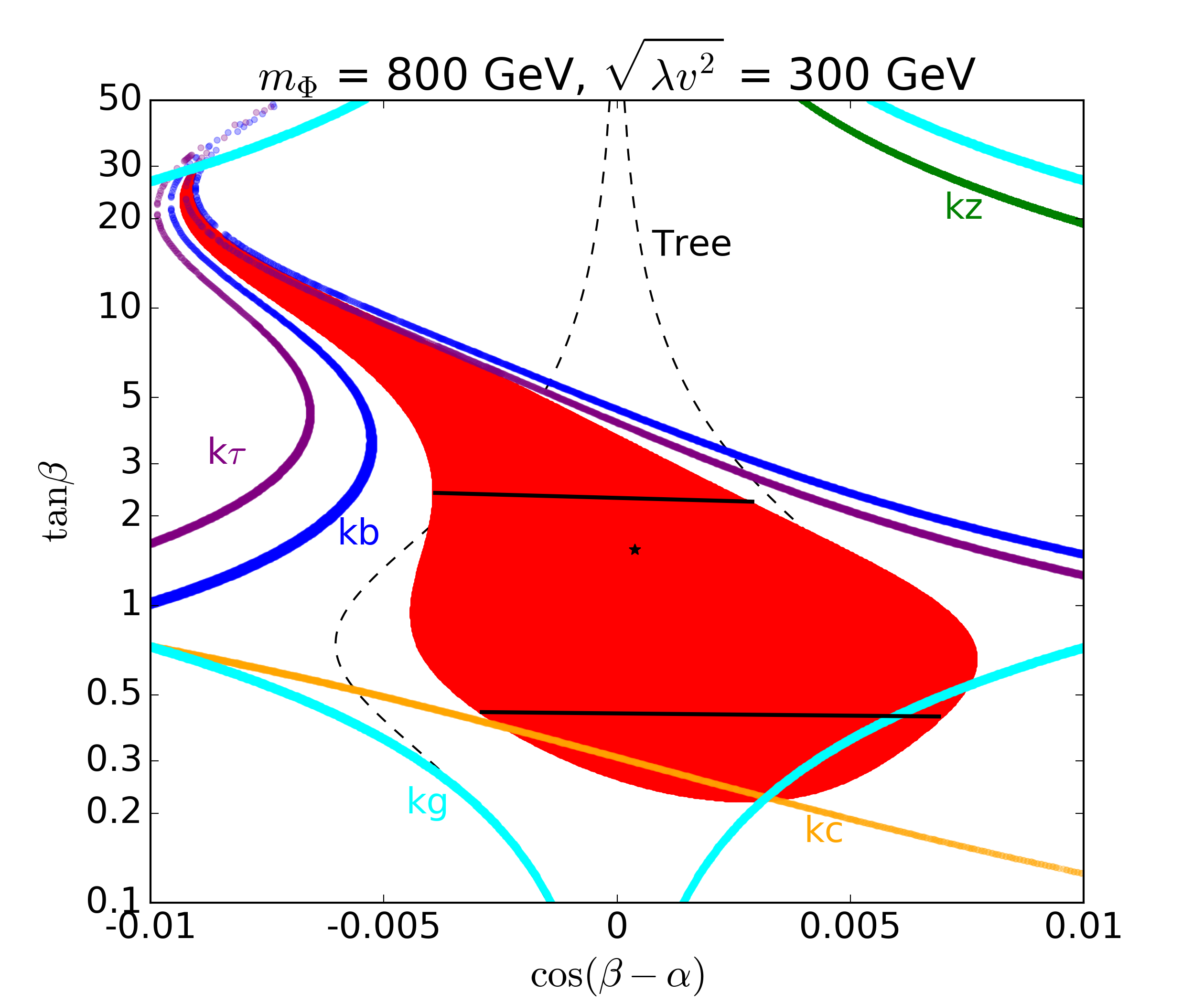}
\caption{95\% C.L. allowed region in the $\cosba$-$\tan \beta$  plane with CEPC Higgs precision measurements.   The central red region is the global fit result with the best-fit point indicated by the black star.  Benchmark point of $m_H=m_A=m_H^\pm\equiv m_\Phi = 800 \gev, \lambvs = 300 \gev $ is used here.
The constraints from individual couplings are given with the color codes: blue ($\kappa_b$), orange ($\kappa_c$), purple ($\kappa_\tau$), green ($\kappa_Z$), cyan ($\kappa_g$). The region enclosed by the dashed black lines  shows the  tree-level two-parameter global fit result for comparison. Two solid horizontal black lines  represent the upper and lower limit for parameter $\tanb$ from theoretical constraints.  }
 \label{fig:dmac_cepc_ana}
\end{center}
\end{figure}

We first consider the simple case of degenerate heavy Higgs boson masses $m_H=m_A=m_{H^\pm}\equiv m_\Phi$ such that the $Z$-pole precision are automatically satisfied.
As shown in Ref.~\cite{Gu:2017ckc}, in the Type-II 2HDM, the current LHC Higgs precision has already constrained $\cos(\beta-\alpha)$ to be less than about 0.1.
To explore the impact from the anticipated precision Higgs measurements at the CEPC, we perform a two-parameter global fit including the loop contributions. In~\autoref{fig:dmac_cepc_ana}, we show the 95\% C.L. allowed region in the two-parameter $\cosba$-$\tan \beta$ plane from the individual couplings by the colored curves: blue ($\kappa_b$), orange ($\kappa_c$), purple ($\kappa_\tau$), green ($\kappa_Z$), cyan ($\kappa_g$), for  a benchmark point of $m_\Phi = 800 \gev, \lambvs = 300 \gev$.   $\kappa_\gamma$ does not have a notable effect therefore not shown.  For large values of $\tan\beta$, regions below the colored curves are allowed,  while for small values of $\tan\beta$, regions above the colored curves are allowed. The central red region is the global fit result with the best-fit point indicated by the black star. The two solid horizontal black lines represent the upper and lower limit for parameter $\tanb$ from theoretical constraints, as shown in~\autoref{fig:theo_const_lambdav2TB} earlier.
The region enclosed by the dashed black lines shows the tree-level only result for comparison.   

For the Type-II 2HDM, the $\cos(\beta-\alpha)$ region gets smaller for larger and smaller values of $\tan\beta$.
At large $\tan\beta$, $\kappa_b$ and $\kappa_\tau$ provide the strongest constraint since they are enhanced by a universal $\tanb$ factor.
For small values of $\tan\beta$, $\kappa_g$ (or effectively, $\kappa_t$) rules out large values of $\cos(\beta-\alpha)$, followed by $\kappa_c$ for negative $\cos(\beta-\alpha)$. Combining all the channels, the 95\% C.L. region for the global fit  leads to
$0.2 \leq \tan\beta\leq 30$,  $ -0.01 \leq \cos(\beta-\alpha)\leq 0.008$,  for the benchmark point $m_\Phi = 800 \gev, \lambvs = 300 \gev$.
We note that the upper bound on $\tan\beta$ and the lower (negative) bound on $\cos(\beta-\alpha)$ coming from $\kappa_g$ is mainly due to the large contribution from $b$-quark loop with a enhanced $\kappa_b$.
The overall range is slightly smaller than that obtained from the tree-level only result, shown by region enclosed by the dashed lines.   The distorted shape of the global fit results, comparing to the tree-level only results is due to the interplay between both the tree-level contribution and loop corrections. Note that while $\kappa_Z$ can be measured with less than 0.2\% precision,  it is less constraining comparing to other couplings given the $1/\tan\beta$ ($\tan\beta$) enhanced sensitivities for $\kappa_{t,c}$ ($\kappa_{b,\tau}$) at small (large) $\tan\beta$ region.

\begin{figure}[tb]
\begin{center}
\includegraphics[width=7.5 cm]{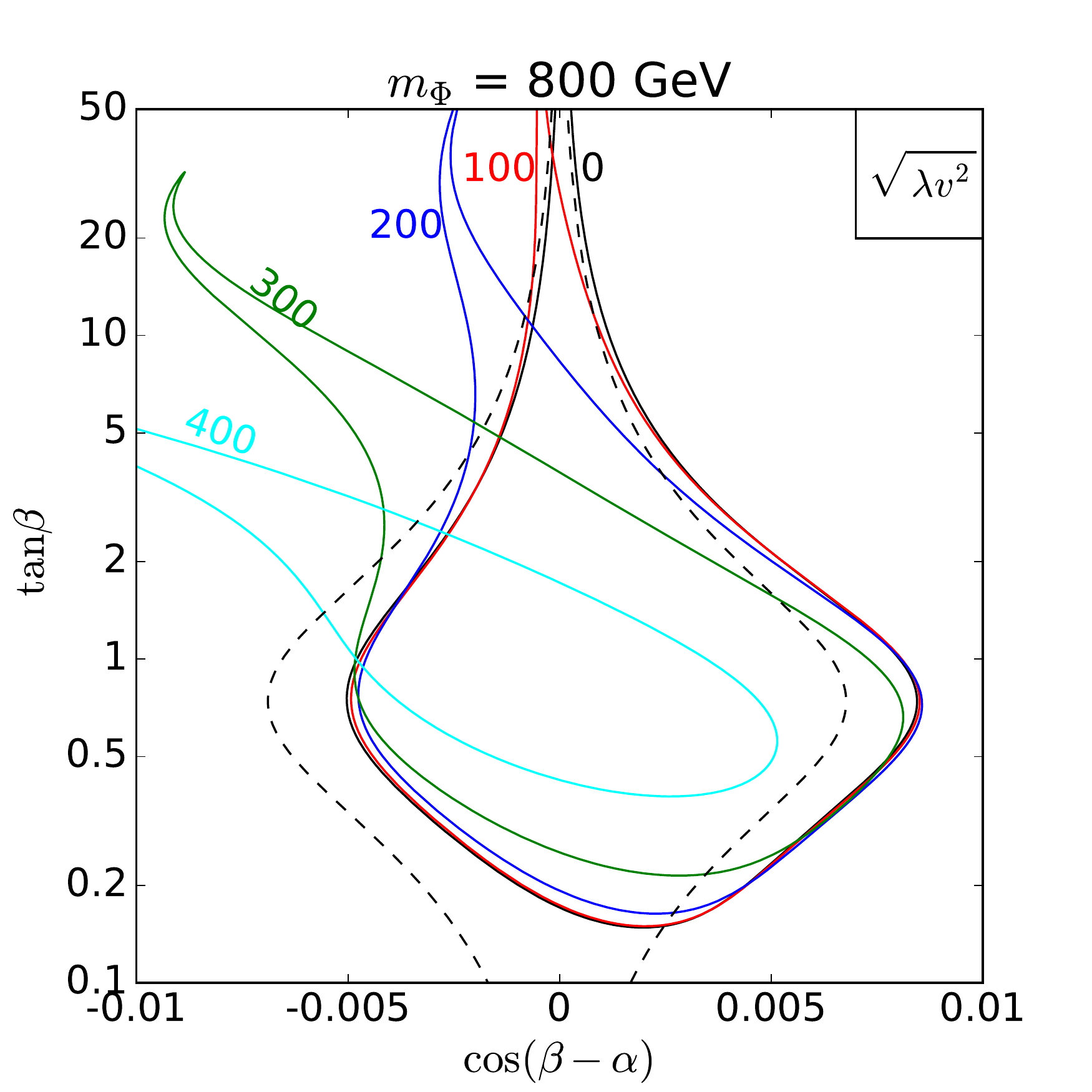}
 \includegraphics[width=7.5 cm]{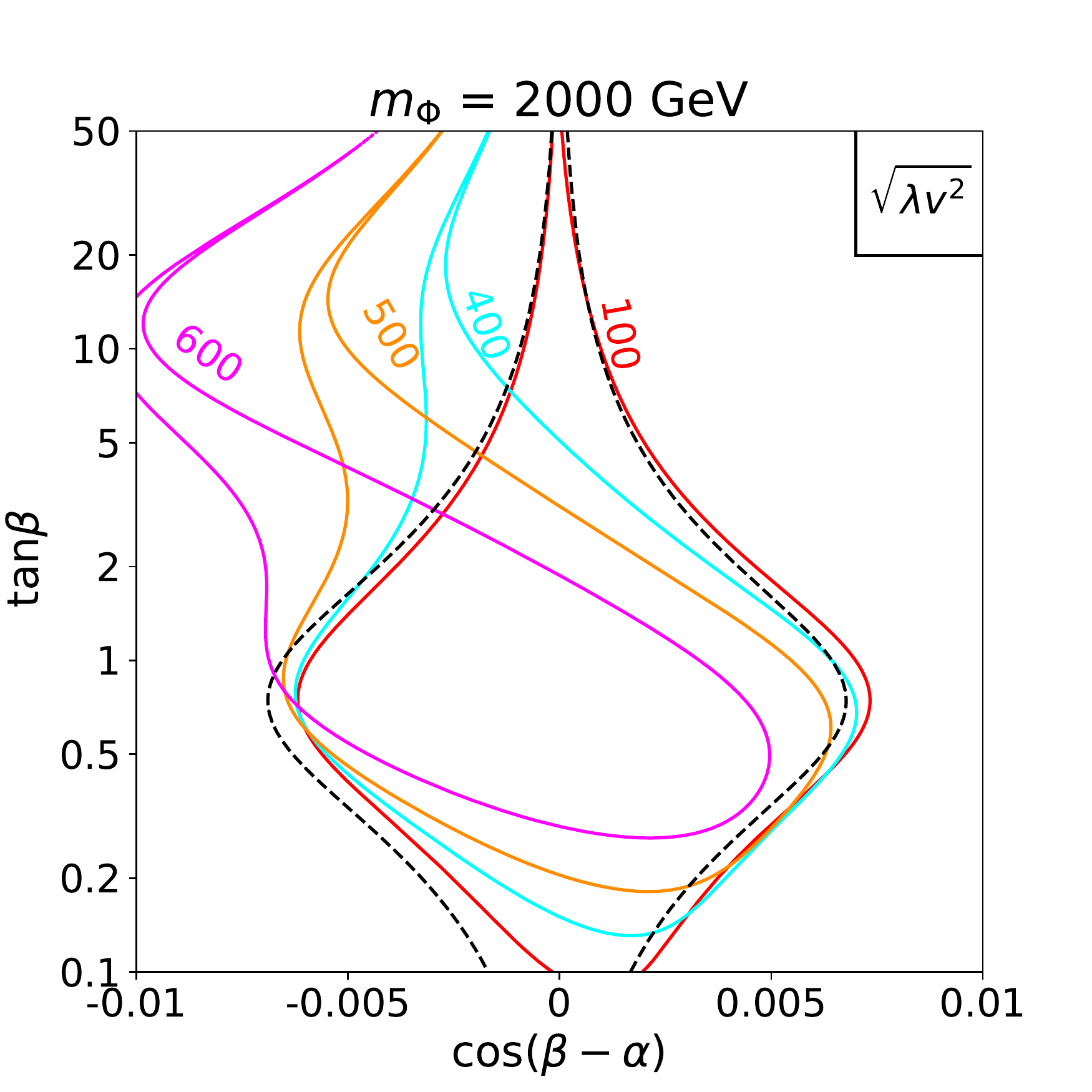}
\caption{Three-parameter fitting results at 95\% C.L. in the $\cos(\beta-\alpha)$-$\tan \beta$ plane for various  values of $\lambvs$ in GeV with CEPC precision.    $m_A=m_H=m_H^\pm=m_\Phi$ is set to be $800$ (left panel), $2000\gev$ (right panel).  As a comparison we also show the tree-level only global fit results, represented by the dashed black lines.}
\label{fig:dmac_cepc}
\end{center}
\end{figure}
To illustrate the dependence on $m_\Phi$ and $\lambda v^2$, which enter the loop corrections, in~\autoref{fig:dmac_cepc}, we show the 95\% C.L. allowed region in the $\cos(\beta-\alpha)$-$\tan\beta$ plane given CEPC Higgs precision, for $m_\Phi=800$ GeV, $\sqrt {\lambda v^2} = 0, 100, 200, 300, 400$ GeV (left panel) and  $m_\Phi=2000$ GeV,  $\sqrt {\lambda v^2} = 100,  400, 500, 600$ GeV (right panel), indicated by different colored lines. In general, including loop corrections shrinks the allowed parameter space, especially for extreme values of $\tan\beta$, and for small $m_\Phi$ and large $\lambda v^2$. The small (large) $\tan\beta$ regions are removed due to the excessive contributions from $c,t$ ($b, \tau$) contributions. For fixed $m_\Phi$, larger $\lambvs$ would lead to larger loop correction and thus larger shift from $\cos(\beta-\alpha)=0$ since $ \lambda v^2$ enters triple Higgs self-couplings. Comparing to the tree-level region which centers around the alignment limit of $\cos(\beta-\alpha)=0$, larger loop corrections distort the preferred $\cos(\beta-\alpha)$ region to more negative value.  For $m_\phi \lesssim1.5$ TeV, large $\lambvs$ values are excluded due to the deviation in  $\kappa_Z$.   As such, for $m_\Phi=800$ GeV,  no parameter space in the $\cos(\beta-\alpha)$-$\tan\beta$ plane survives at 95\% C.L. for $\sqrt{\lambda v^2}\gtrsim 450$ GeV. For large $m_\Phi$ about 2 $\tev$ (right panel), larger values of  $\sqrt{\lambda v^2}$ could be accommodated.  For $m_\Phi \gtrsim 3 \tev$,  the one-loop level effects almost decouple and the final allowed region is close to the tree-level results.
Comparing with the constraints on the $\cos(\beta-\alpha)$-$\tan\beta$ plane via LHC  searches  with $A \rightarrow hZ$ channel as shown in~\autoref{fig:LHC_heavyH_limits}, and the current  and HL-LHC Higgs coupling precision measurements~\cite{Gu:2017ckc}, the future Higgs factory can constrain the 2HDM parameter space at least an order of magnitude better in the allowed $\cos(\beta-\alpha)$ range.

\begin{figure}[tb]
\begin{center}
\includegraphics[width=7.5cm]{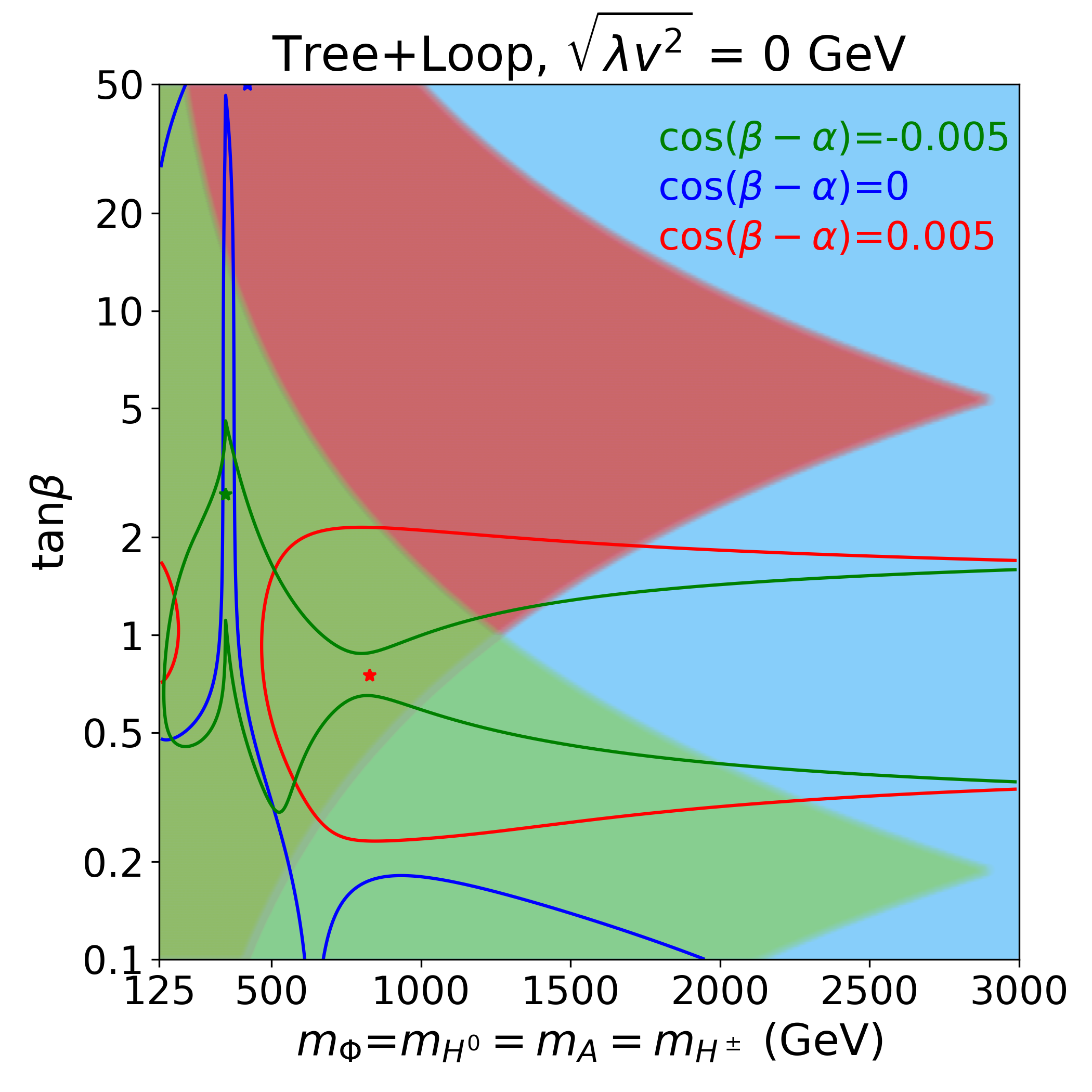}
\includegraphics[width=7.5cm]{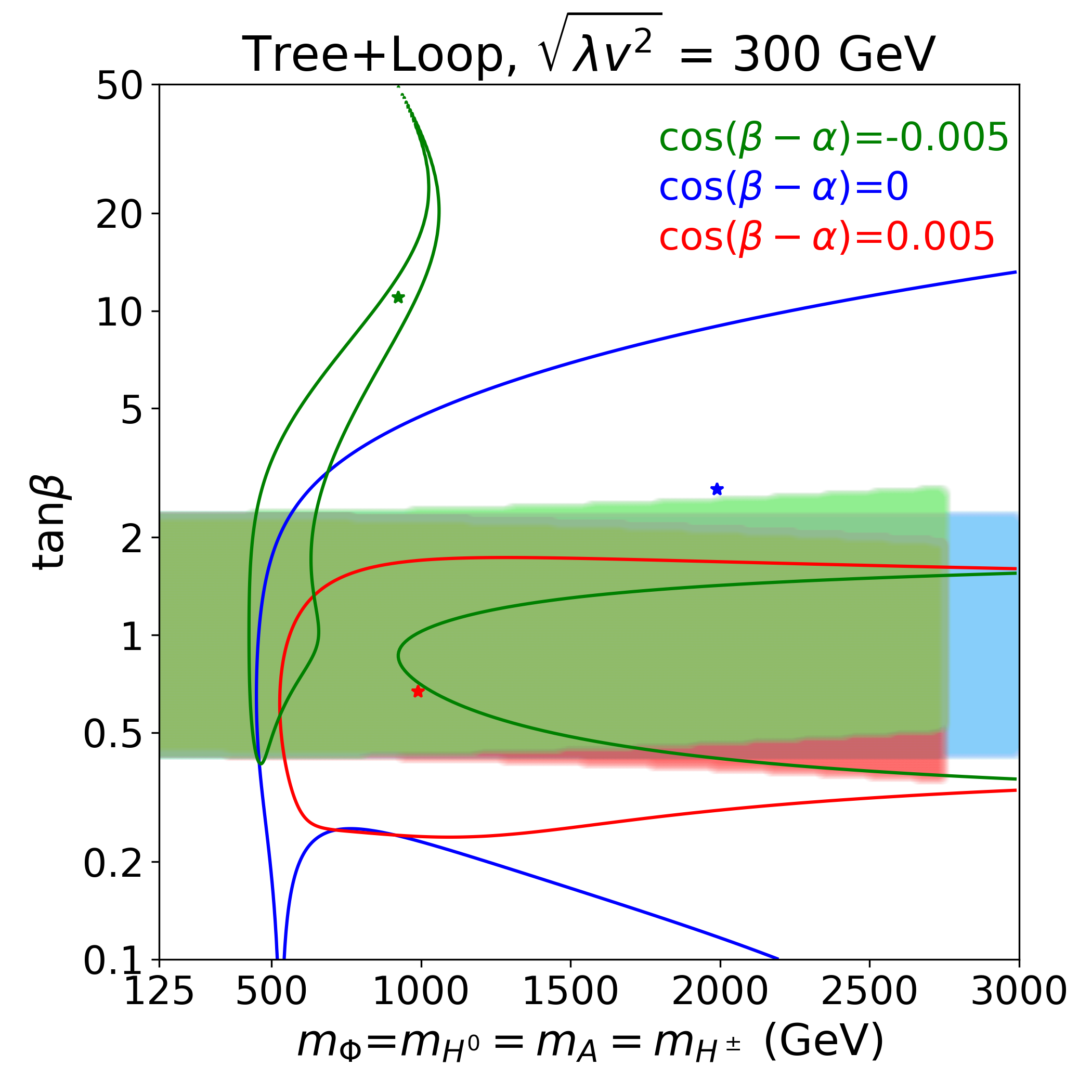}
  \caption{Three-parameter fitting results at 95\% C.L. in the $m_\Phi$-$\tan \beta$ plane with varying $\cosba$ with CEPC precision.   We set  $\sqrt{\lambda v^2}$ to be 0 (left panel) and $300 \gev$ (right panel).    Red, blue and green curves represent $\cosba = 0.005, 0, -0.005$ respectively. The colored stars show the corresponding best-fit point.    Also shown are the allowed regions under theoretical considerations under the same  color codes.    }
\label{fig:mphitanb_cepc_lambda}
\end{center}
\end{figure}

High precision on the Higgs coupling measurements can also be used to constrain the mass of the  heavy Higgs bosons running in the loop.
In~\autoref{fig:mphitanb_cepc_lambda}, we show the 95\% C.L. allowed region in the $m_{\Phi}$-$\tan\beta$ plane for $\sqrt{\lambda v^2}=0$ (left panel) and 300 GeV (right panel), for $\cos(\beta-\alpha)=-0.005$ (green lines), 0 (blue lines) and 0.005 (red lines). For $\sqrt{\lambda v^2}=0$ with minimal triple Higgs self-couplings, the most notable constraint takes place near $m_\Phi \approx 350$ GeV owing to the threshold contribution from the $t\bar t$ in the loop.
The alignment limit with loop corrections only (blue curve)   provides the most relaxed bounds for $m_\Phi\lesssim 350$ GeV and $\tan\beta\gtrsim 0.5$, as well as $m_\Phi\gtrsim 350$ GeV with a larger range of $\tan\beta$ surviving the CEPC Higgs precision. Once $\cos(\beta-\alpha)$ deviates from zero,  tree-level contributions become sizable. Even for a value of $\cos(\beta-\alpha)$ as small as 0.005, $\tan\beta$ region is shrunk to $0.2- 2$ with $m_\Phi \gtrsim 500$ GeV.  For negative $\cos(\beta-\alpha)=-0.005$, while $\tan\beta$ region further shrinks, the allowed $m_\Phi$ can be extended all the way down to about 130 GeV.

We also show  the allowed regions in the $m_\Phi$-$\tanb$ plane under theoretical considerations in~\autoref{fig:mphitanb_cepc_lambda} with the different colors for different choices of $\cos(\beta-\alpha)$. While all ranges of $m_\Phi$ and $\tan\beta$ are allowed in the alignment limit of $\cos(\beta-\alpha)=0$, once $\cos(\beta-\alpha)$ deviates away from 0, large $m_\Phi$ as well as small and large $\tan\beta$ regions are ruled out by theoretical considerations. Combining both the theoretical constraints and precision Higgs measurements, a constrained region in $m_\Phi$-$\tan\beta$ can be obtained for the non-alignment cases.

For $\sqrt{\lambda v^2}=300$ GeV,  larger loop corrections further modify the allowed region in $m_\Phi$ and $\tan\beta$. The $t\bar t$ threshold region $m_\Phi \approx 350$ GeV is inaccessible and the range of $\tanb$ is shrunk  to  0.3 $-$ 1.5 when $\cos(\beta-\alpha)$ varies from 0 to 0.005. For the negative $\cosba = -0.005$, the allowed region divides to two parts.  The part with $m_\Phi \leq 1000$ GeV has a wide range for parameter $\tanb$, while for $m_\Phi > 1000$ GeV, $0.4 < \tanb < 1.6$. Theoretical considerations further limit the range of $\tan\beta$ to be between 0.35 and 3, as shown by the shaded region.  For $\cos(\beta-\alpha)=\pm 0.005$, $m_\Phi$ has an upper limit of about  2750 GeV from theoretical considerations.

\begin{figure}[tb]
\begin{center}
 \includegraphics[width=7.5cm]{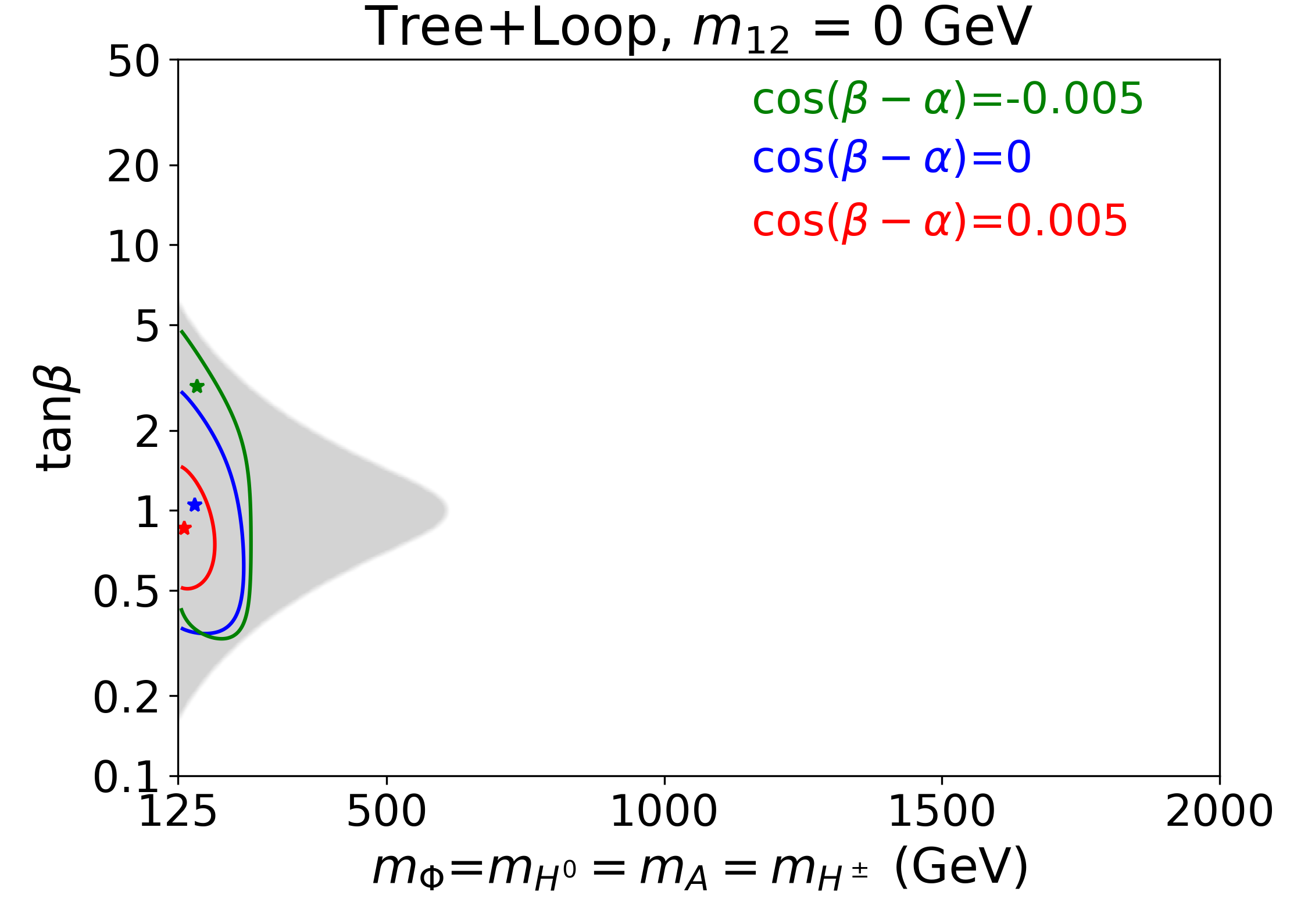}
\includegraphics[width=7.5cm]{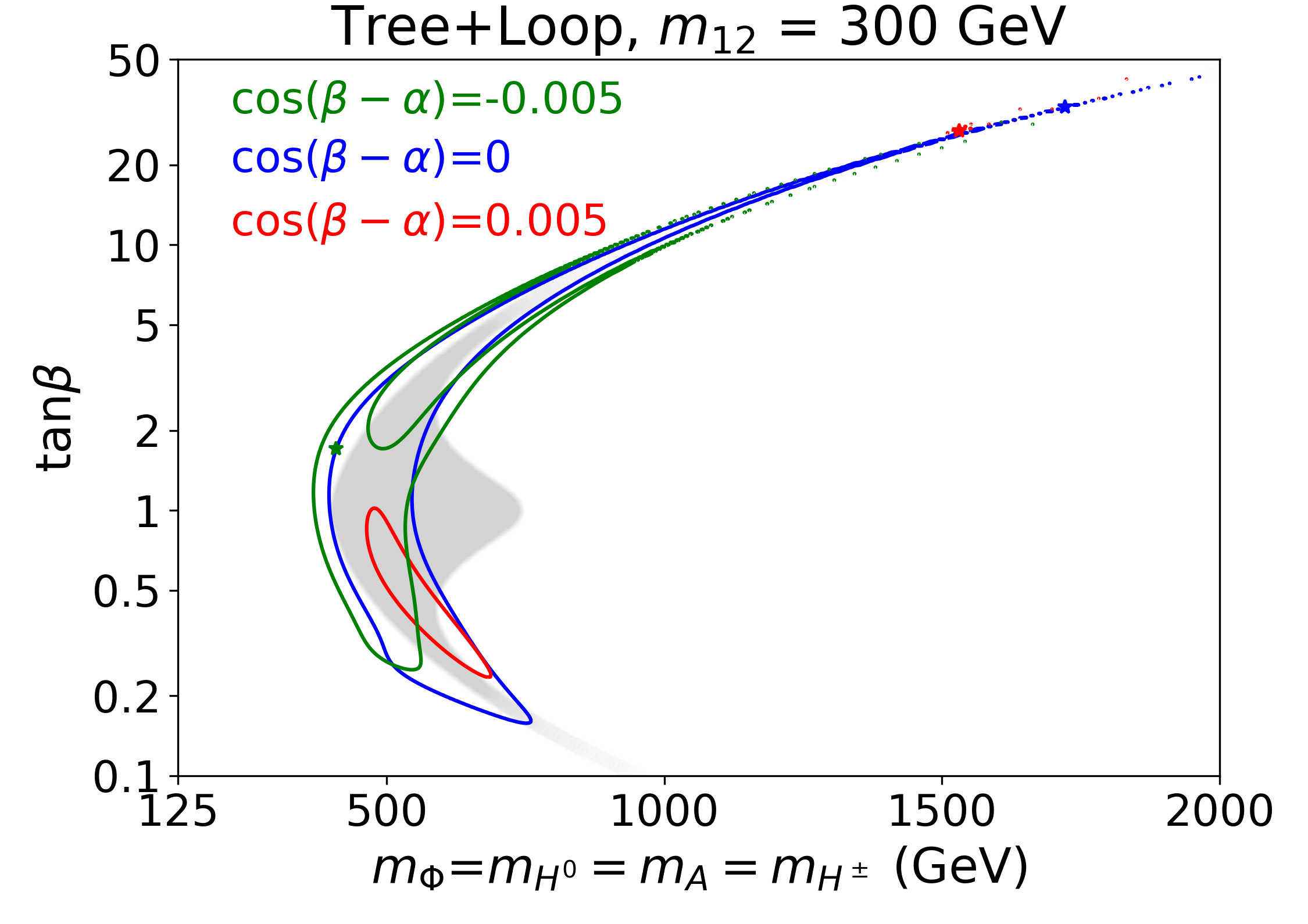}
\caption{Similar to~\autoref{fig:mphitanb_cepc_lambda}, except $m_{12}$ is fixed to be 0 (left panel) and 300 GeV (right panel) instead of fixing $\sqrt{\lambda v^2}$.  The colored stars show the corresponding best-fit point.  Gray shaded gray region shows the theoretical allowed region, which has little dependence on  $\cosba$.  }
\label{fig:mphitanb_cepc_m12}
\end{center}
\end{figure}

While $\lambda v^2 \equiv m_\Phi^2 - m_{12}^2/s_\beta c_\beta$ is a good parameter to use since it is directly linked to the triple Higgs self-couplings, sometime it is convenient to fix the soft $\mathbb{Z}_2$ breaking parameter $m_{12}^2$ instead.  The resulting 95\% C.L. allowed region in the $m_\Phi$-$\tan\beta$ plane is shown in~\autoref{fig:mphitanb_cepc_m12} for $m_{12}=0$ (left panel) and 300 GeV (right panel).  The theoretical constraints as discussed in the previous section are also indicated with the shaded gray regions. They have little dependence on the $\cosba$ value when $m_{12}^2$ is kept fixed.
For $m_{12}=0$, $m_\Phi = \sqrt{\lambda v^2}$ is constrained to be less than around 250 GeV. For larger values of  $m_{12}$, the rather narrow region in the plane as seen in the right panel indicates a strong correlation between $m_\Phi$ and $\tan\beta$ for large $\tan\beta$, approximately scaled as $\tan\beta \sim (m_\Phi / m_{12})^2$, which minimizes the corresponding $\lambda v^2$  value and thus its loop effects. The indirect probe in $m_\Phi$ via Higgs precision measurements complements the direct search limits at the LHC, especially in the intermediate $\tan\beta$ wedge region where the direct search limits are the most relaxed.

 \subsection{Case with non-degenerate heavy Higgs boson masses}

\begin{figure}[htb]
\begin{center}
\includegraphics[width=7.5cm]{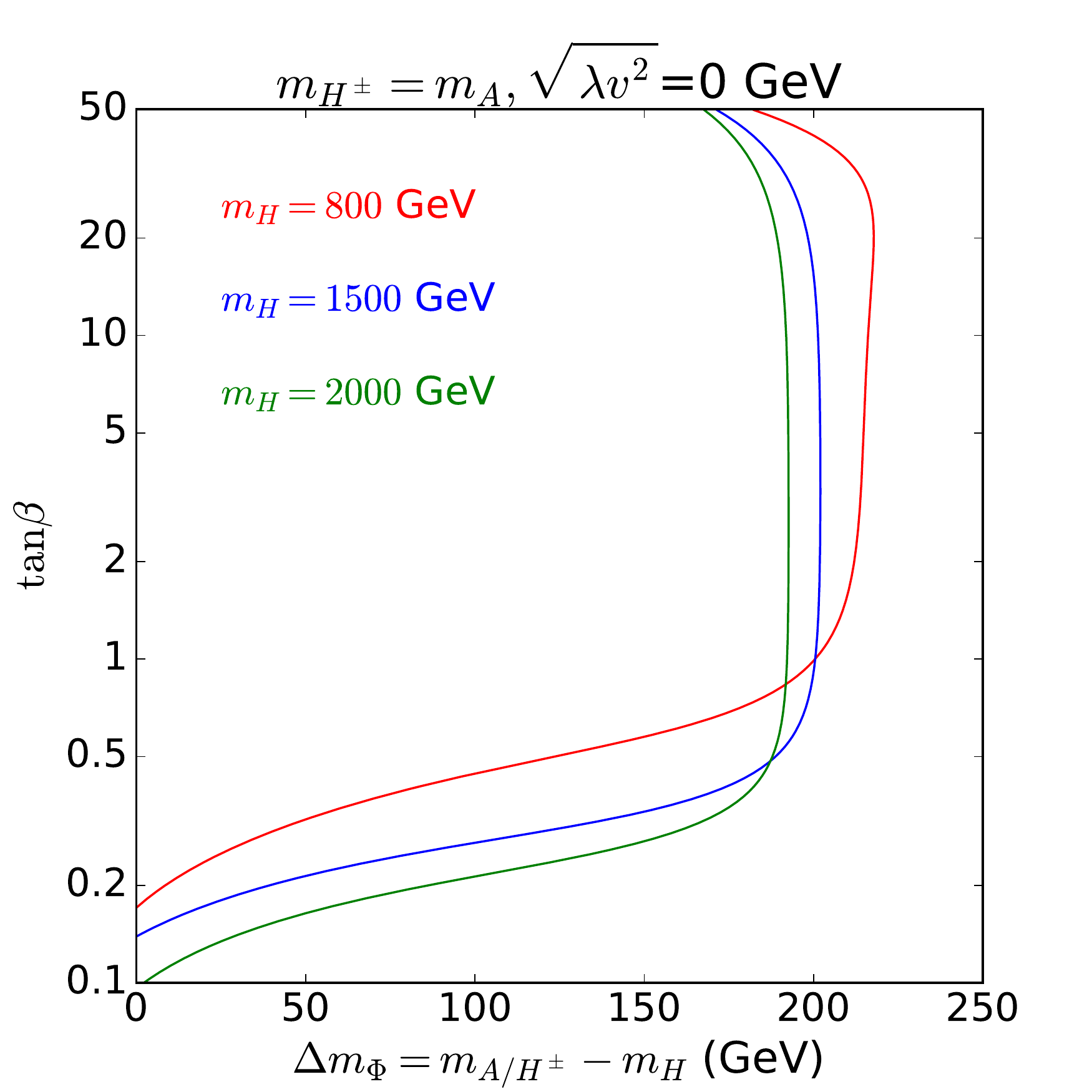}
\includegraphics[width=7.5 cm]{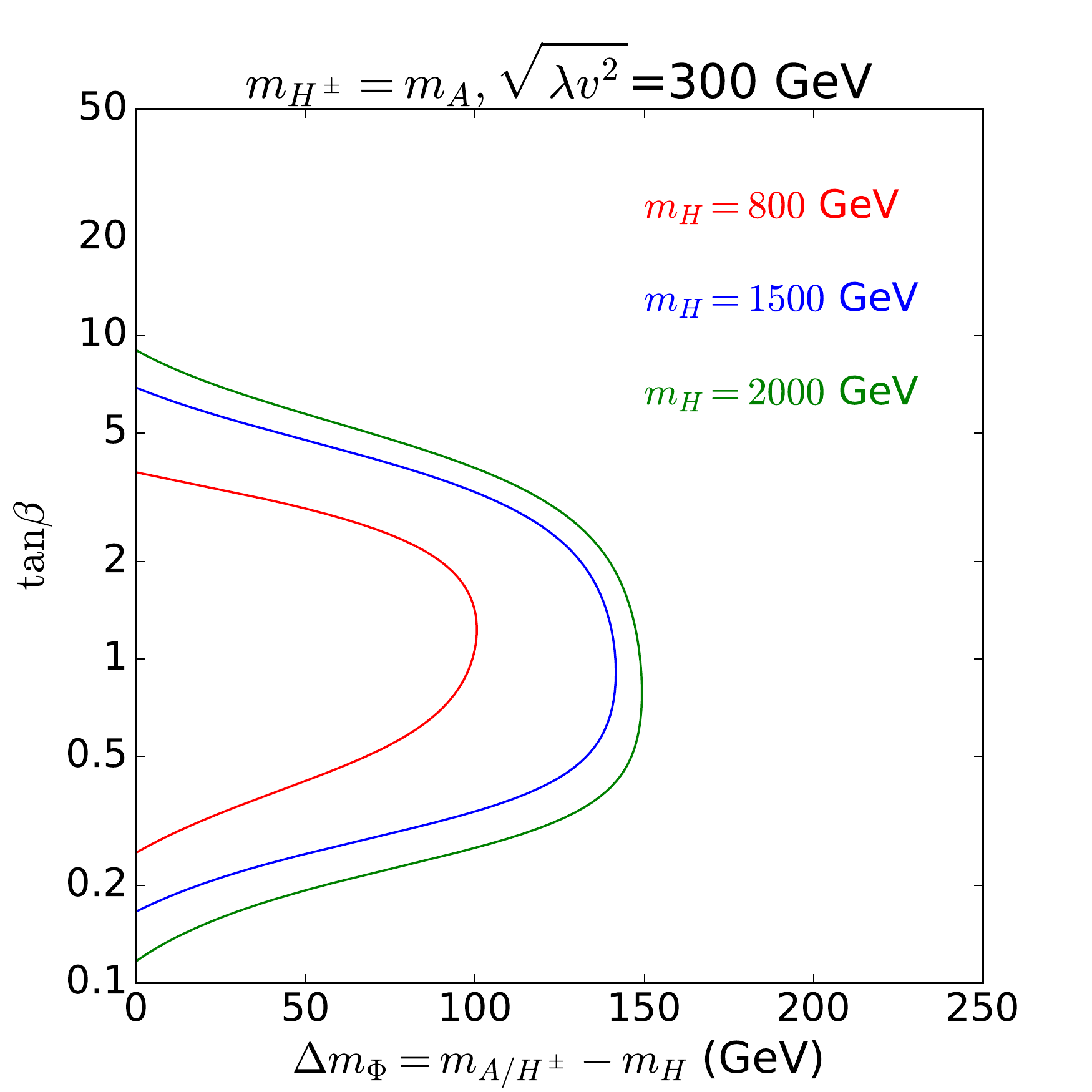}
 \includegraphics[width=7.5cm]{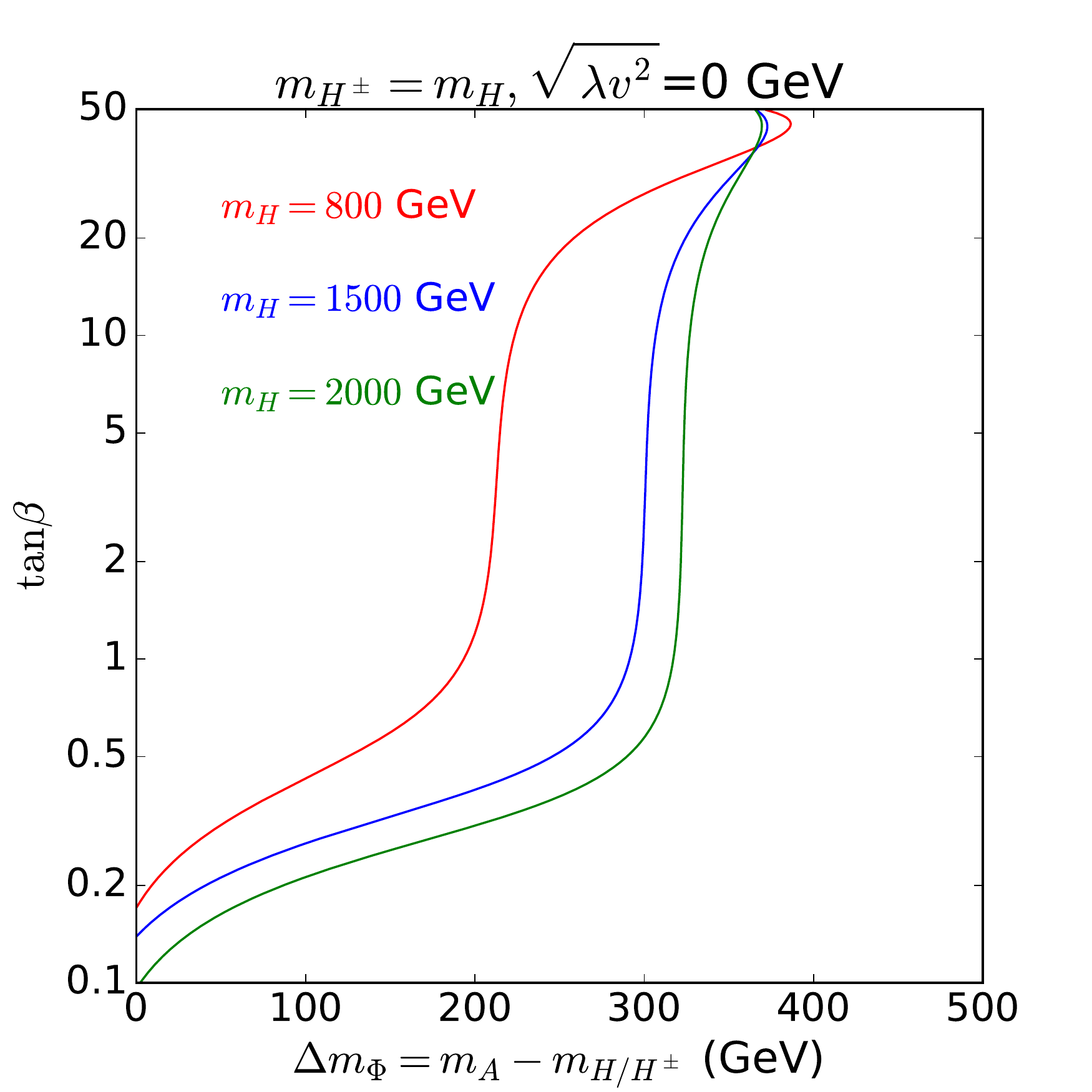}
\includegraphics[width=7.5 cm]{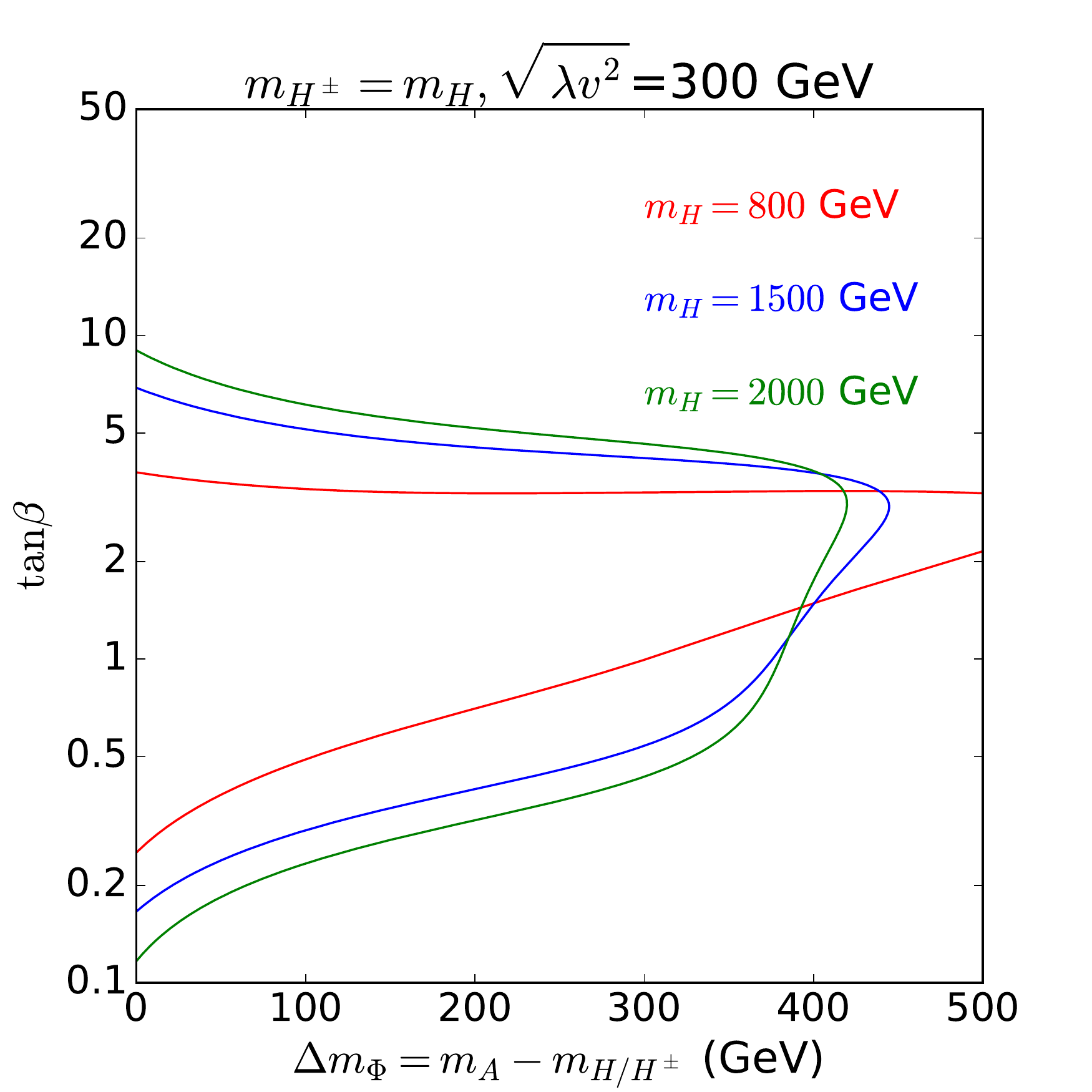}
\caption{Three-parameter fitting results at 95\% C.L. in the upper panels for $\Delta m_\Phi=m_{A/ H^\pm} - m_H$
 and lower panels for $\Delta m_\Phi=m_{A} - m_{H/ H^\pm}$,
with varying $m_H$ under the alignment limit   $\cos(\beta-\alpha)=0$ with CEPC Higgs precision. $\lambvs$ is taken to be  0   (left panels) and 300 GeV (right panels). $m_H=800, 1500, 2000$ GeV are shown in red, blue and green lines, respectively.  }
\label{fig:la_dm1}
\end{center}
\end{figure}

Going beyond the degenerate case, both the Higgs and $Z$-pole precision observables are sensitive to the mass splittings between the non-SM heavy Higgs bosons.
In~\autoref{fig:la_dm1}, we show the 95\% C.L. allowed region in the $\Delta m_\Phi$-$\tan\beta$ plane under the alignment limit for various values of $m_H$.
To satisfy the $Z$-pole precision constraints, we consider the heavy masses partially degenerate,  and take $m_A=m_{H^\pm}$ in the upper panels with $\Delta m_\Phi=m_{A/H^\pm} -m_H$, and $m_H=m_{H^\pm}$ in the lower panels with $\Delta m_\Phi=m_{A} - m_{H/H^\pm}$.
The left plots are for $\sqrt{\lambda v^2} = 0$ and the right plots are for $\sqrt{\lambda v^2} = 300$ GeV.

For the case of $m_A=m_{H^\pm}$ (upper panels),  $\Delta m_\Phi$ can be as large as 200 GeV for a wide range of $\tan\beta$ for $\sqrt{\lambda v^2} = 0$.   For $\sqrt{\lambda v^2} = 300$ GeV, the $\Delta m_\Phi$ region is more constrained: $\Delta m_\Phi \lesssim 150,\ 140,\ 90$ GeV, for $m_H=2000$, 1500 and 800 GeV, respectively.   The corresponding $\tan\beta$ range is also much more limited for larger values of $\lambda v^2$.

For the case of $m_H=m_{H^\pm}$ (lower panels),  the allowed range of $\Delta m_\Phi$ is larger, up to about 400 GeV for $\lambvs=0$, and up to about 500 GeV   for  $\lambvs=300$ GeV.
Note that the region for $\Delta m_\Phi=0$ corresponds to the situation of $\cos(\beta-\alpha)=0$ in~\autoref{fig:mphitanb_cepc_lambda},  which is much less restrictive than the non-degenerate case $\Delta m_\Phi \ne 0$.

\begin{figure}[htb]
\begin{center}
\includegraphics[width=5cm]{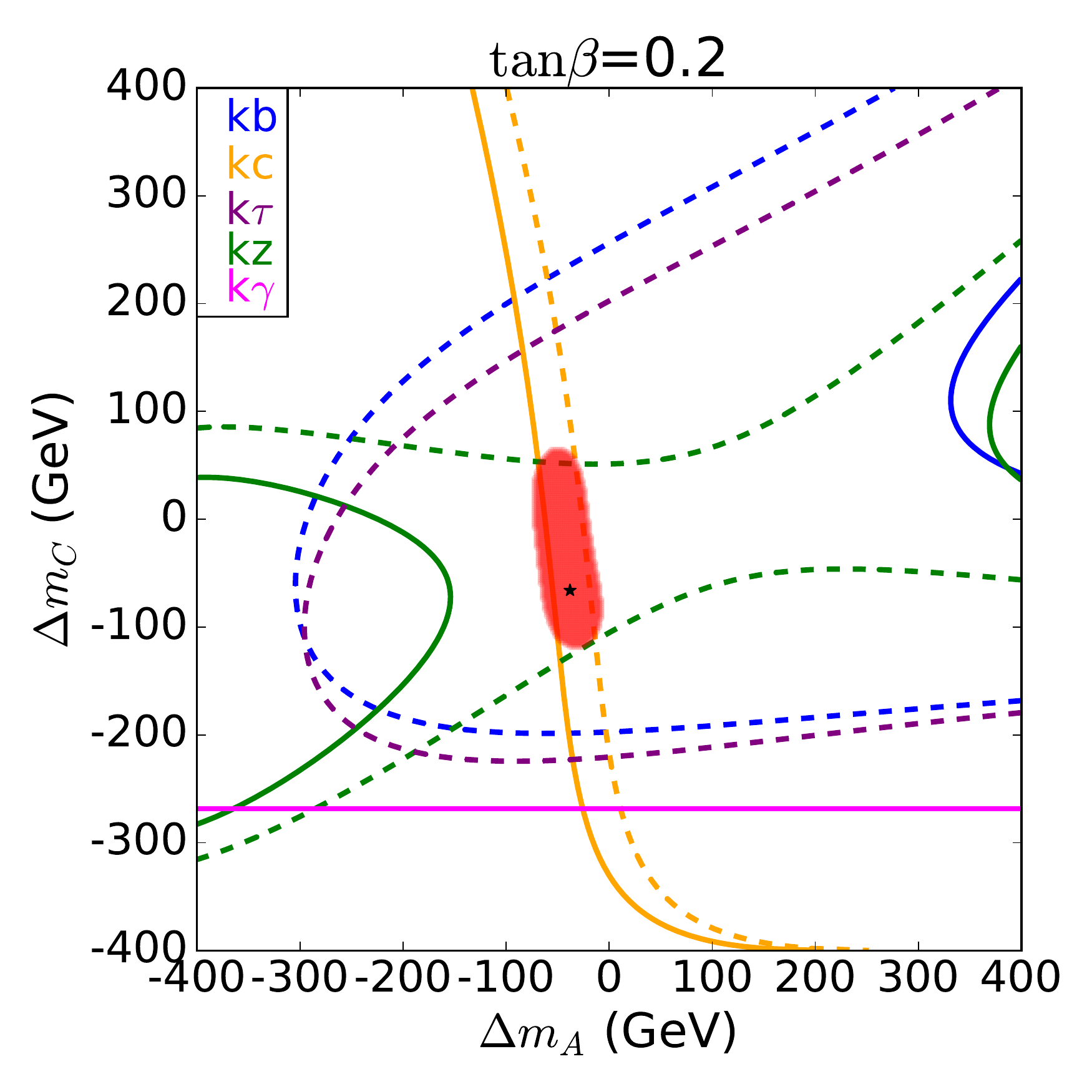}
\includegraphics[width=5cm]{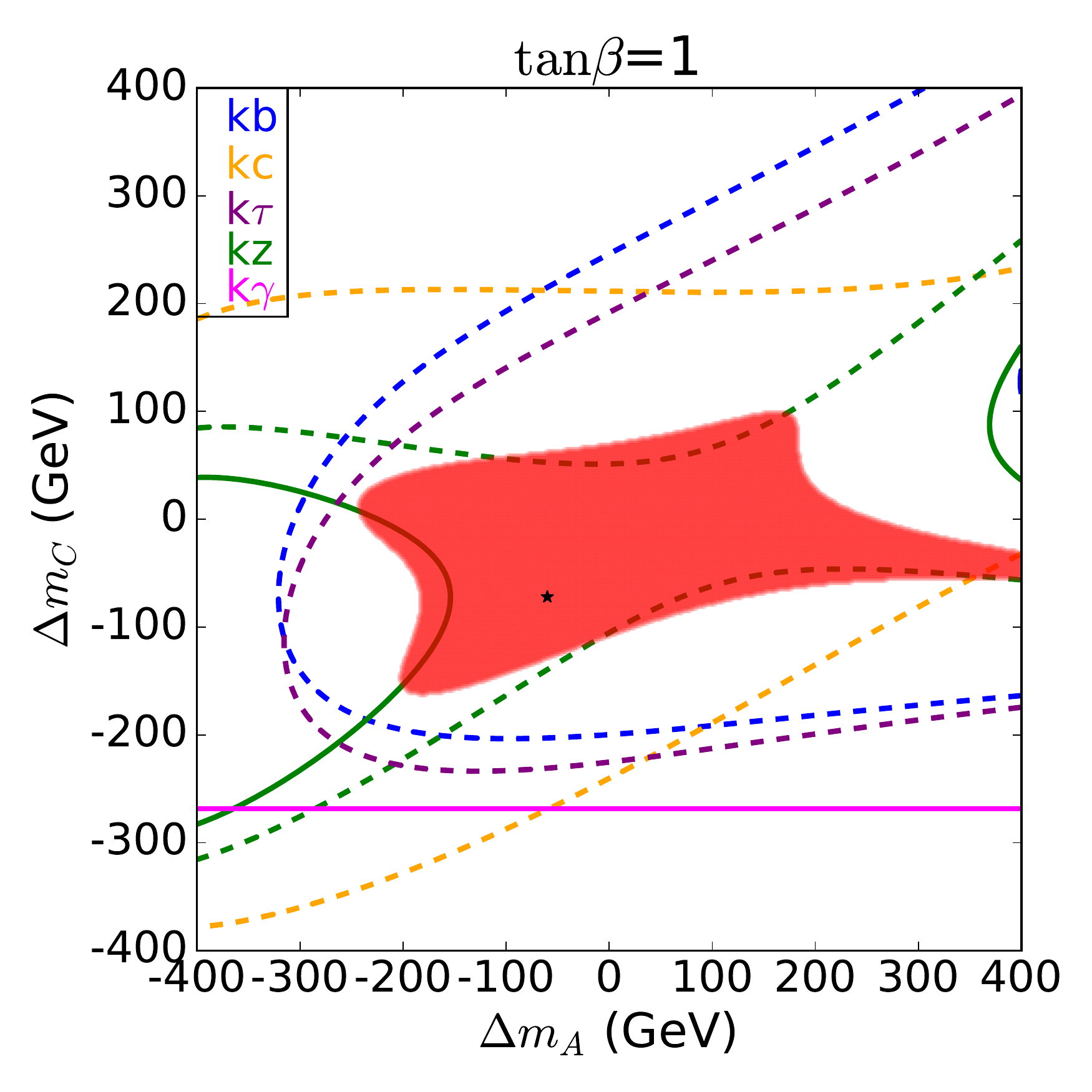}
\includegraphics[width=5cm]{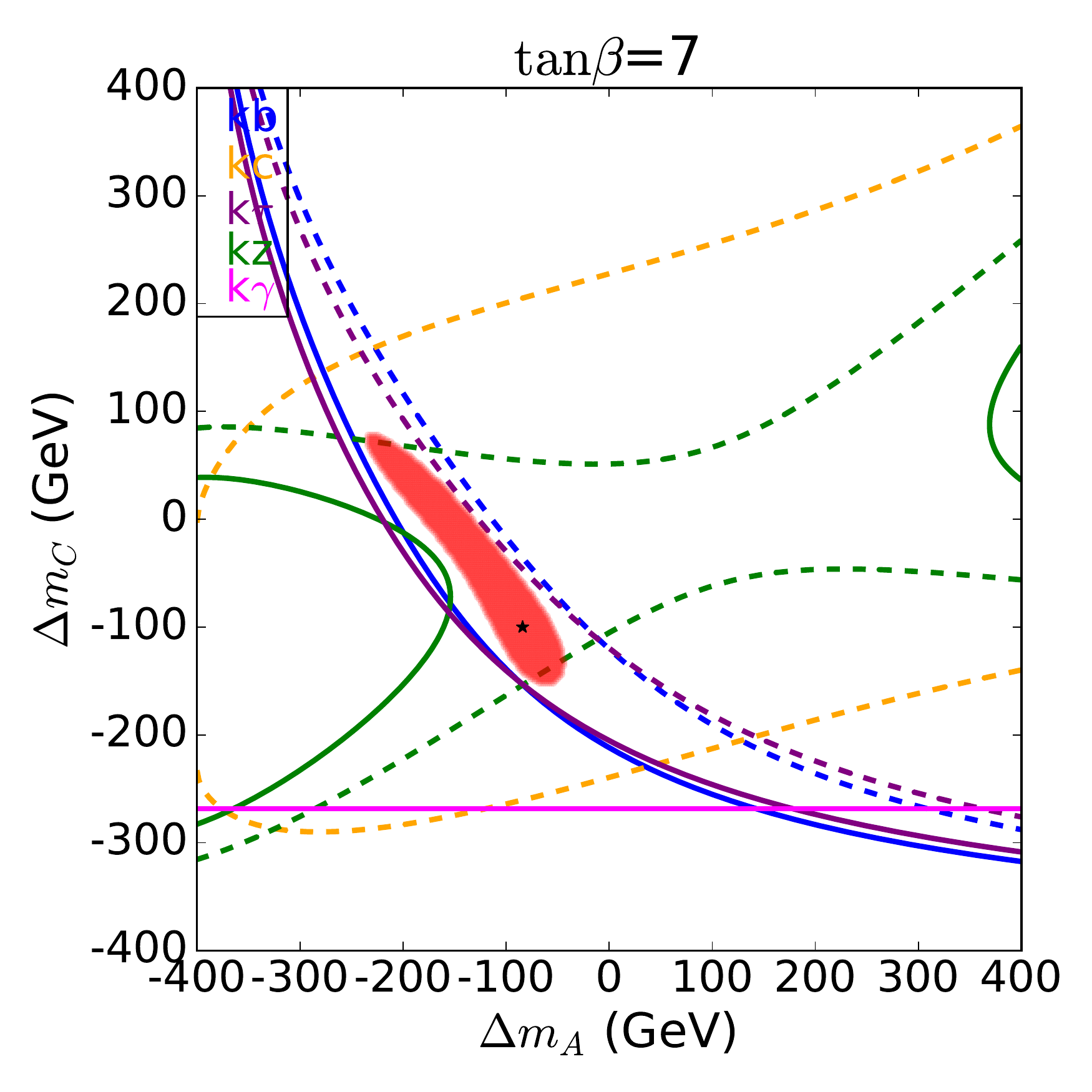}
\caption{Constraints on the $\Delta m_A$-$\Delta m_C$ plane from individual Higgs coupling measurement (color curves), and the 95\% C.L. global fit results (red shaded region), for $\tan\beta=0.2 (\text{left}), 1$ (middle), $\tan\beta=7$ (right) under alignment limit, with  $m_H=800$ GeV, $\sqrt{\lambda v^2}=300$ GeV.
For individual coupling constraint, the dashed line represents negative limit, while solid line represents the positive limit.  Regions between the solid and dashed curves are the allowed region. For $\kappa_\gamma$, region above the line is allowed.  }
\label{fig:da-dc-ana}
\end{center}
\end{figure}

In~\autoref{fig:da-dc-ana}, we show the constraints on the $\Delta m_A=m_A-m_H$ and $\Delta m_C=m_{H^\pm}-m_H$  plane from individual Higgs coupling measurements in color curves, and the 95\% C.L. global fit results in the red shaded region, for $\tan\beta=0.2$ (left panel), 1 (middle panel) and 7 (right panel) under alignment limit with  $m_H=800$ GeV, $\sqrt{\lambda v^2}=300$ GeV. For each individual coupling constraint  with a ``$\pm$'' error bar, the dashed line is for the negative limit, while the solid line is for the positive limit.  The range between the two lines is the survival region.
Under the alignment limit,  $\kappa_Z$ is independent of $\tanb$ as apparent in the figure.
For Type-II 2HDM, generally speaking, $\kappa_{b,\tau}$ are $\tanb$-enhanced, while $\kappa_{c}$ is $\cot \beta$-enhanced. Thus for small $\tanb$, the main constraint on the mass splitting comes from  $\kappa_c$ and leads to a small overlapping red region with $\kappa_Z$ as the global fit result of $\Delta m_A\sim -40$ GeV to 0 GeV (left panel).
For large $\tanb$, it is due to $\kappa_{b,\tau}$, resulting in $\Delta m_A\sim -50$ GeV to $-250$ GeV (right panel). For $\tan\beta\sim 1$,  constraints from both $\kappa_{b,\tau}$ and $\kappa_c$ are relatively relaxed, leading to a larger allowed region in the mass splittings $\Delta m_A\sim -250$ GeV to 400 GeV (middle panel) mostly due to $\kappa_Z$.  The range of $\Delta m_C$ is typically between $-$200 GeV to 100 GeV constrained from $\kappa_Z$.
$\kappa_\gamma$ mainly involves the charged Higgs loops and only constrains weakly.
Note that $\kappa_g$ does not constrain the mass splittings significantly  and therefore is not shown in the plots.

\begin{figure}[htb]
\begin{center}
\includegraphics[width=7.5cm]{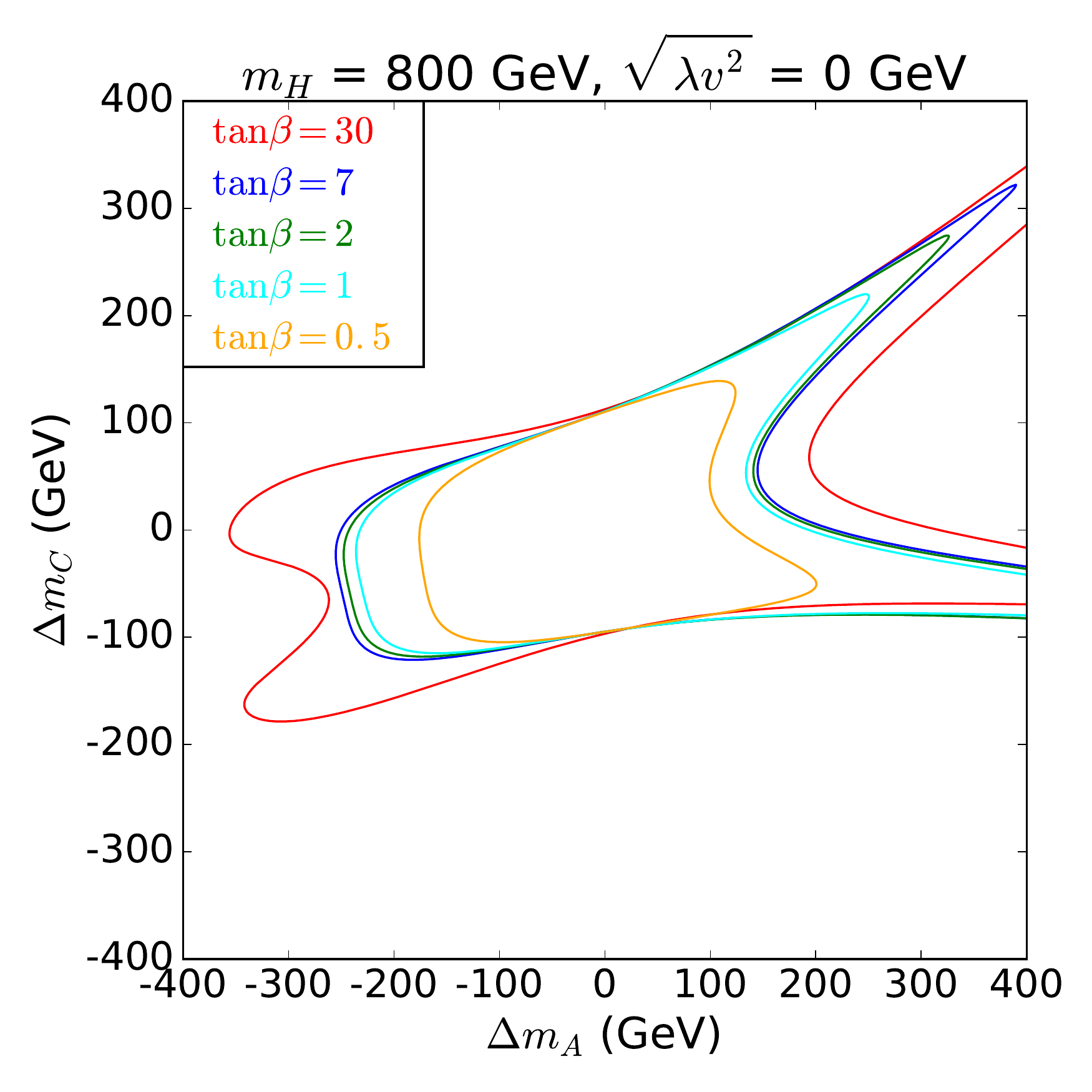}
 \includegraphics[width=7.5cm]{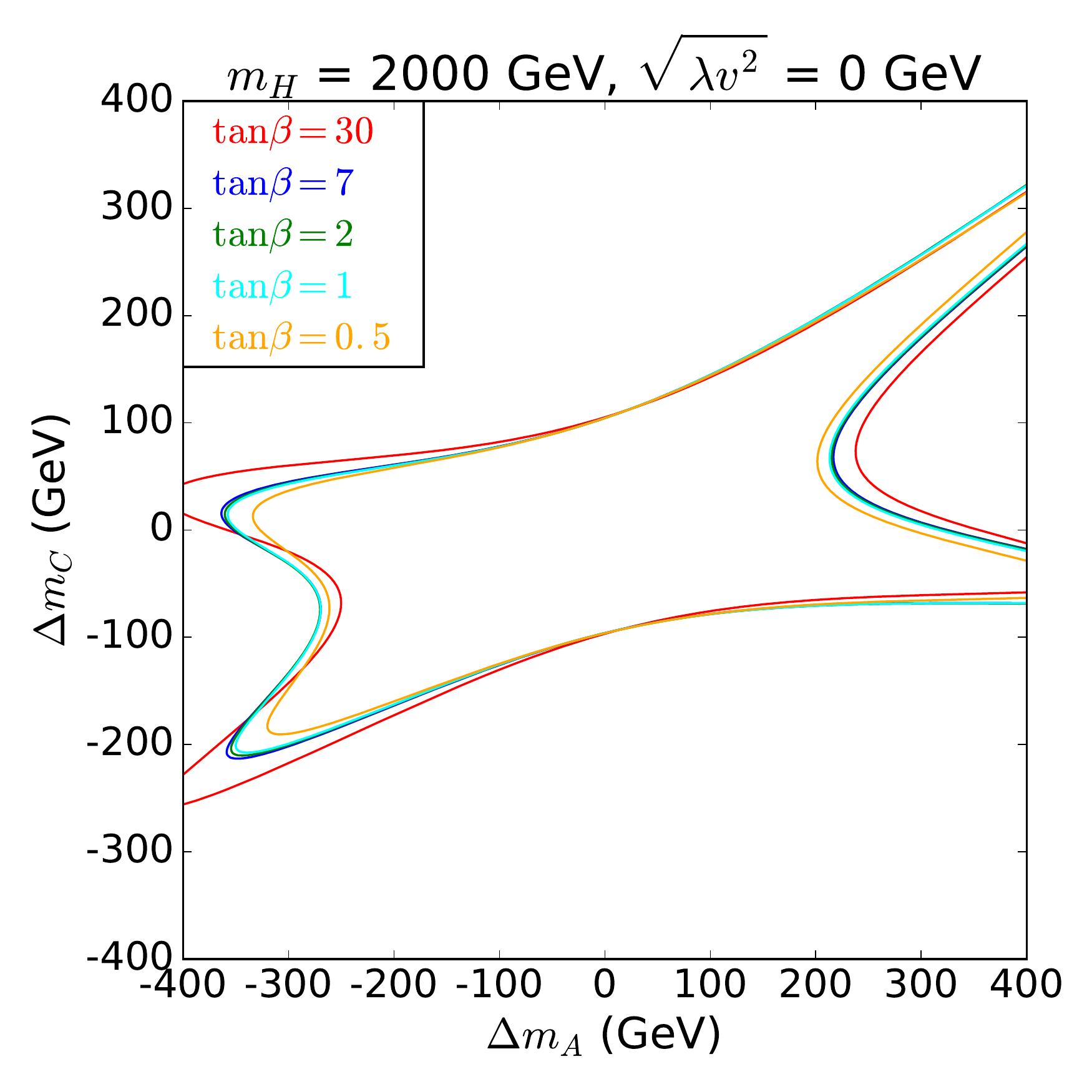}\\\vspace{1mm}
\includegraphics[width=7.5cm]{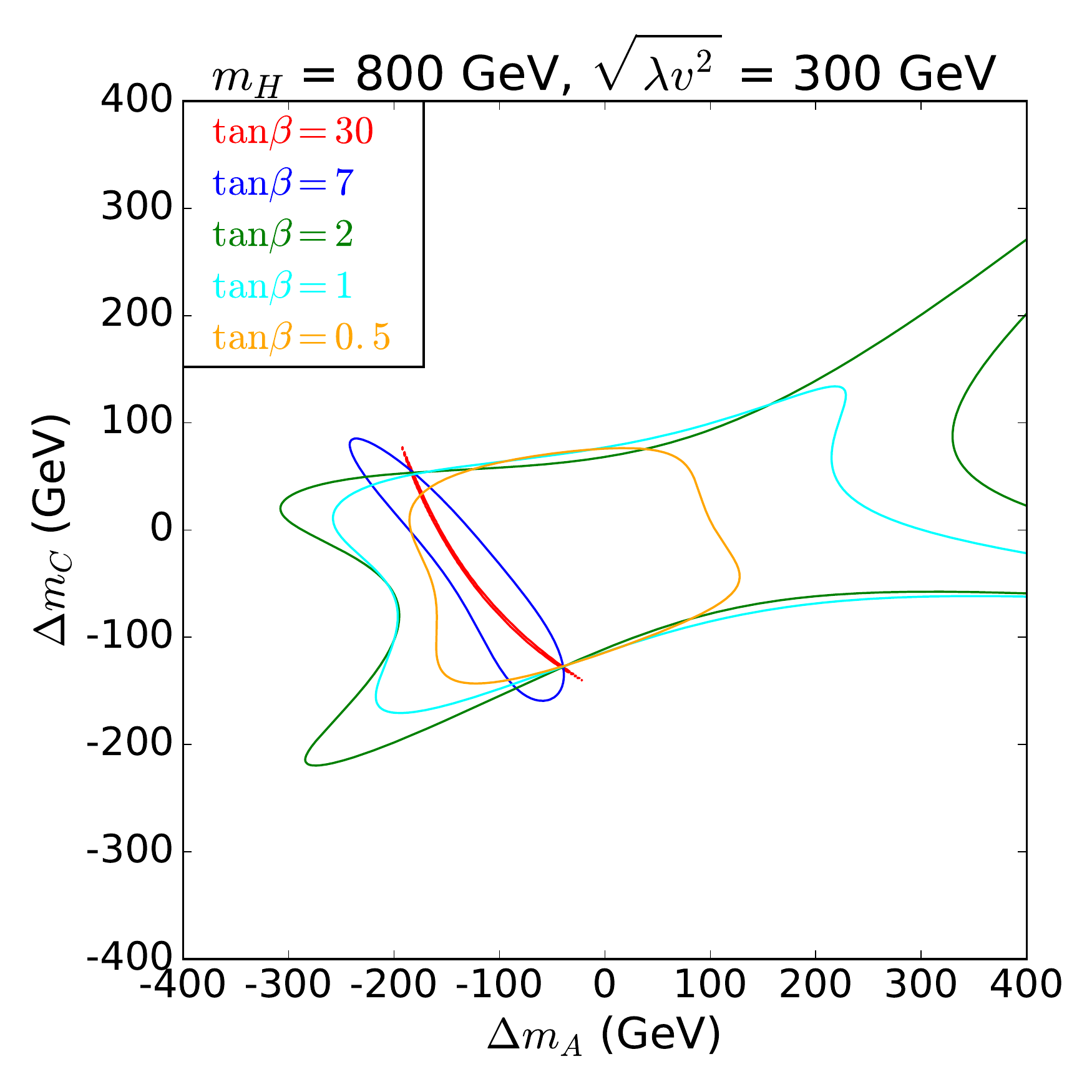}
 \includegraphics[width=7.5cm]{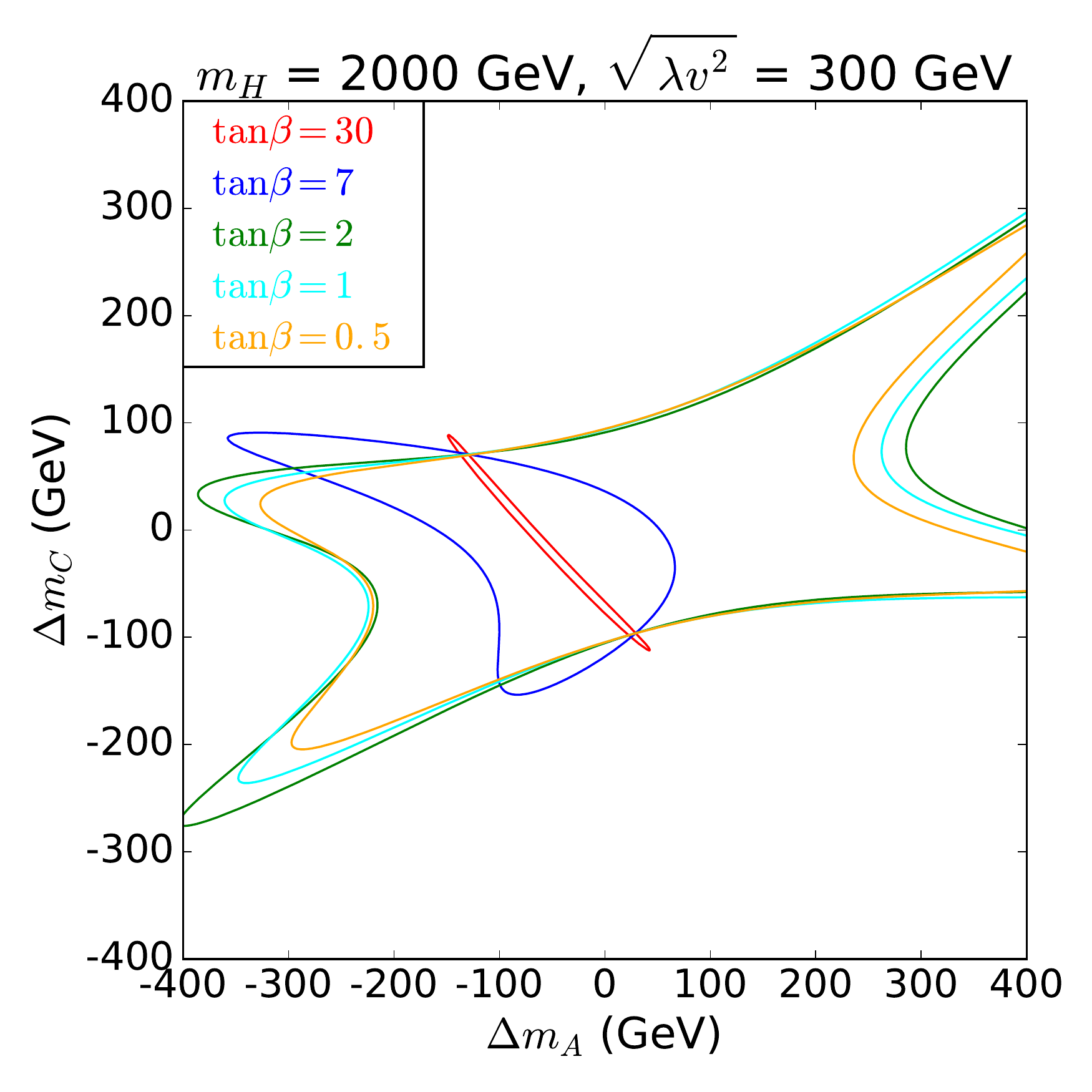}
\caption{Three-parameter fitting results at 95\% C.L. in the $\Delta m_A$-$\Delta m_C$ plane with varying $\tanb$ under the alignment limit condition $\cos(\beta-\alpha)=0$. The upper panels are for $\lambvs = 0$, while the lower panels are for $\lambvs = 300 \gev$. The masses are set $m_H  =$ 800 GeV (left panels), 2000 GeV (right panels). The colors represent different $\tanb =$ 30 (red), 7 (blue), 2 (green), 1 (cyan) and 0.5 (orange). }
\label{fig:dadc}
\end{center}
\end{figure}

In~\autoref{fig:dadc}, we present the 95\% C.L. allowed region in the $\Delta m_A$-$\Delta m_C$ plane, for $ m_H=$ 800 GeV (left panels) and 2000 GeV (right panels), again under the alignment limit. The upper panels are for $\sqrt{\lambda v^2}=0$ and lower panels are for $\sqrt{\lambda v^2}=300$ GeV, with various color codes for different values of $\tan\beta$.

For $\sqrt{\lambda v^2}=0$, large values of $\Delta m_{C}$ and $\Delta m_{A}$ around $\pm400$ GeV or larger could be accommodated, but strongly correlated with each other. For small $m_H$ with relatively large loop corrections, the ranges for $\Delta m_{C,A}$ shrink for smaller $\tan\beta$: with $\tan\beta=0.5$, only around 200 GeV mass difference could be accommodated. For larger values of $m_H$ around 2000 GeV, the allowed ranges of the mass difference are  {much more relaxed and are} almost independent of $\tan\beta$.
For $\sqrt{\lambda v^2}=300$ GeV, however, the largest ranges for $\Delta m_{C,A}$ could be achieved for $\tan\beta\sim 2$, for both  benchmark choices of $m_\Phi$, due to the constraints from individual couplings, as illustrated in~\autoref{fig:da-dc-ana}. For $m_H=2000$ GeV, the allowed ranges of the mass difference varies little with $0.5<\tan\beta<2$,  but shrink quickly for larger $\tan\beta$.

\begin{figure}[tbh]
\begin{center}
\includegraphics[width=11.5cm]{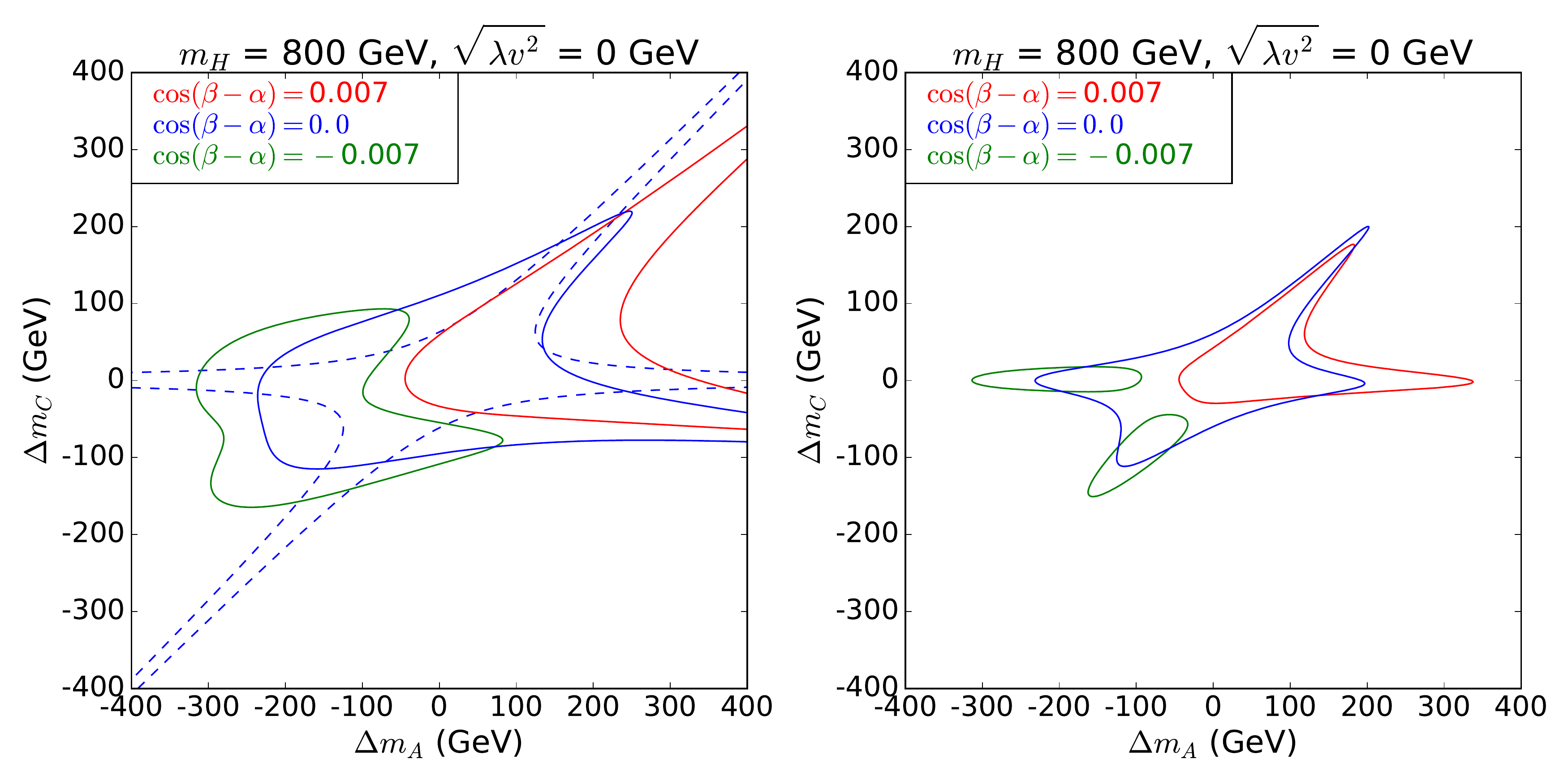}\\
\includegraphics[width=11.5cm]{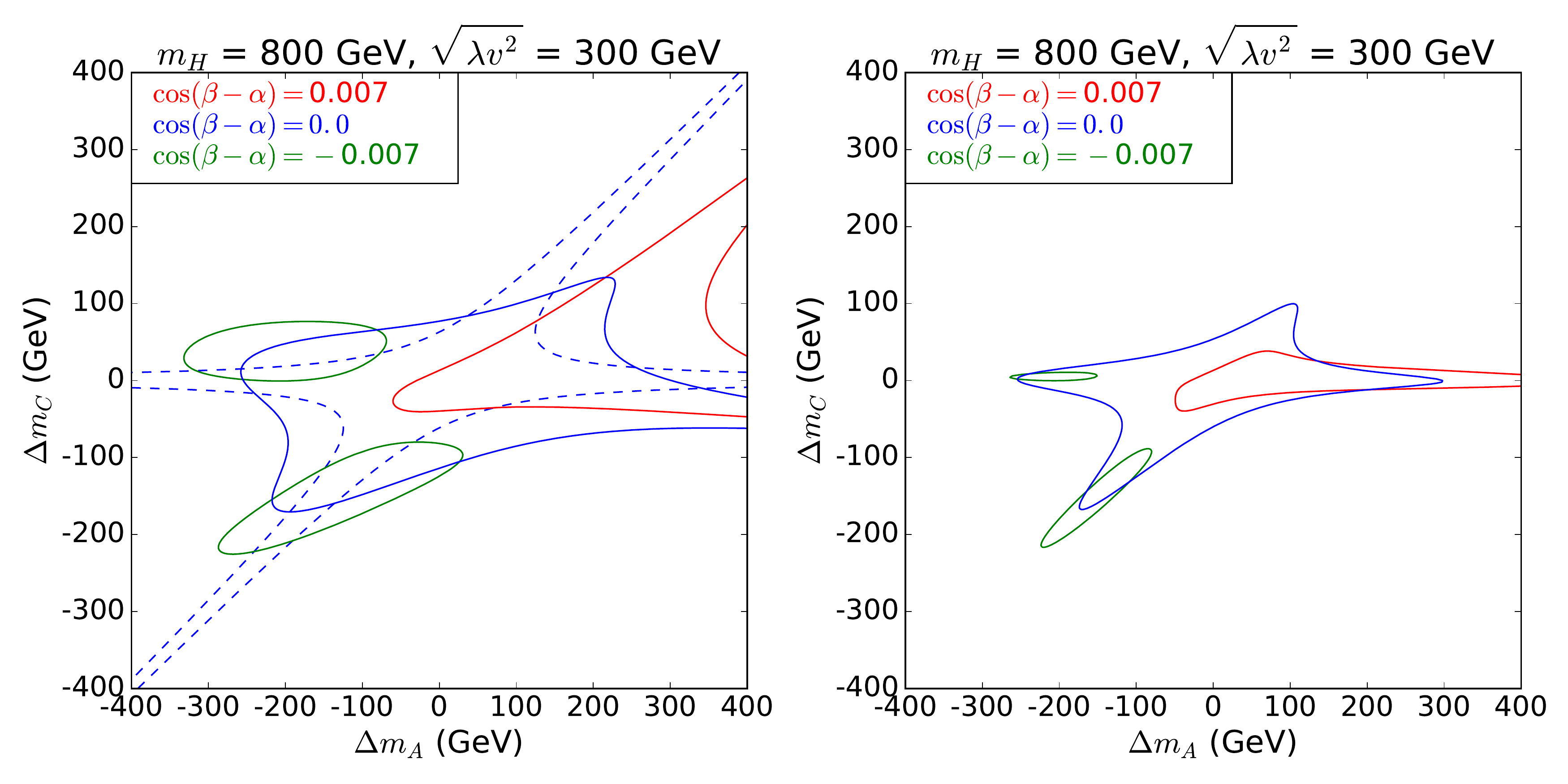}\\
\includegraphics[width=11.5cm]{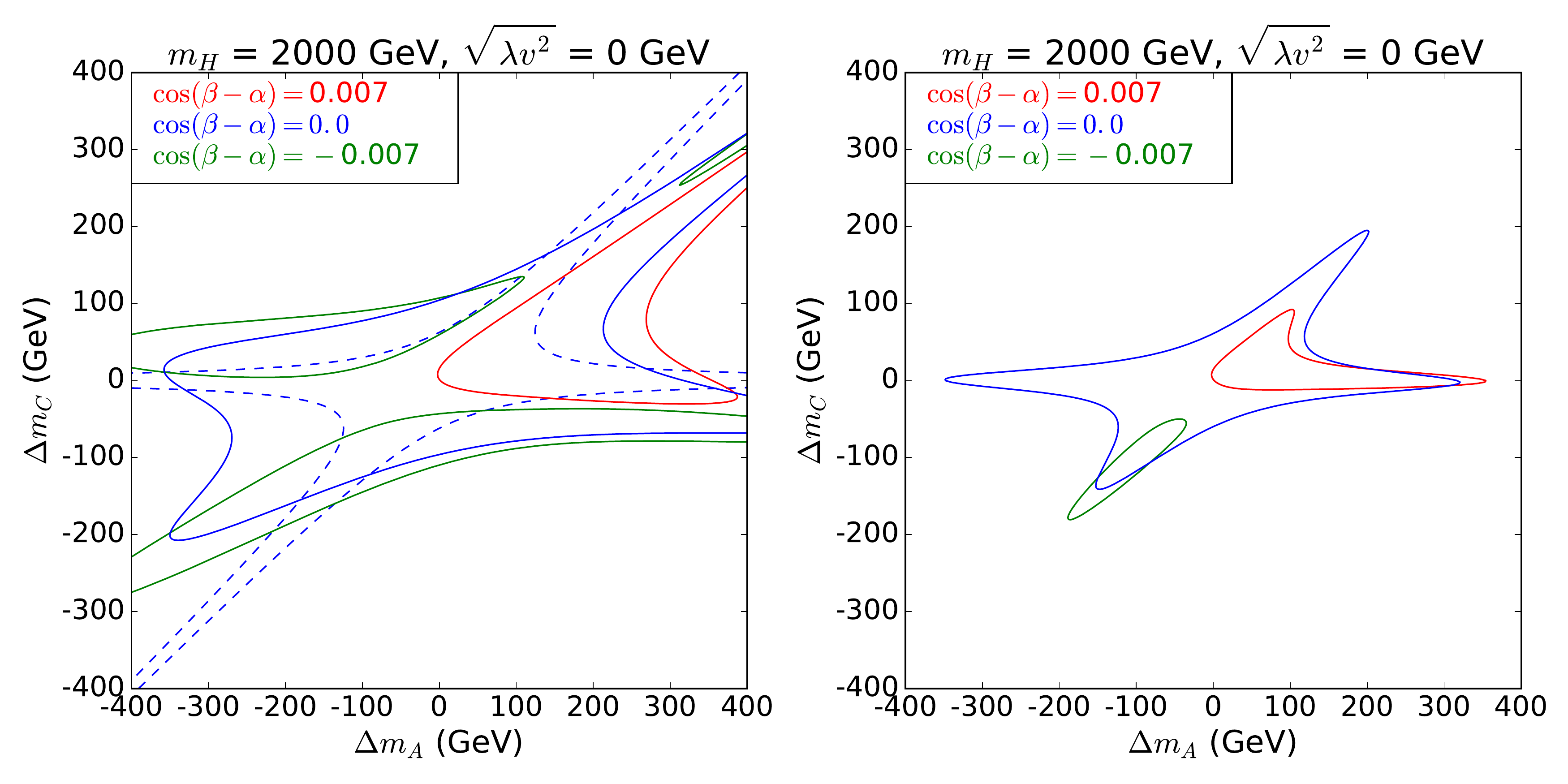}\\
  \caption{ Three-parameter fitting results at 95\% C.L. in the $\Delta m_A$-$\Delta m_C$ plane for various values of $\cos(\beta-\alpha)$, for the Higgs (solid curves) and $Z$-pole (dashed curves) constraints (left panels),  and combined constraints (right panels), with upper rows for $m_H=800$ GeV, $\sqrt{\lambda v^2}=0$, middle rows for $m_H=800$ GeV, $\sqrt{\lambda v^2}=300$ GeV, and bottom rows for $m_H=2000$ GeV, $\sqrt{\lambda v^2}=0$. $\tan\beta=1$ is assumed for all plots. }
\label{fig:da-dc-cos}
\end{center}
\end{figure}

In~\autoref{fig:da-dc-cos}, we show the 95\% C.L. contours in the $\Delta m_A$-$\Delta m_C$ plane, focusing on the $\cos(\beta-\alpha)$ dependence given by different color codes, for Higgs (solid curves) and $Z$-pole precision (dashed curves) constraints individually (left panels),  and combined (right panels), with upper rows for  $m_H=800$ GeV, $\sqrt{\lambda v^2}=0$, middle rows for $m_H=800$ GeV, $\sqrt{\lambda v^2}=300$ GeV, and bottom rows for $m_H=2000$ GeV, $\sqrt{\lambda v^2}=0$.  $\tan\beta=1$ is assumed for the plots.

For the Higgs precision fit, the alignment limit $\cos(\beta-\alpha)= 0$ (blue curve) typically gives the largest allowed ranges.  Even for small deviation away from the alignment limit, $\cos(\beta-\alpha)=\pm 0.007$, $\Delta m_A$ is constrained to be positive for $\cos(\beta-\alpha)= 0.007$, and it splits into two branches for
$\cos(\beta-\alpha)= -0.007$.
The $Z$-pole precision measurements force the mass splittings to either $\Delta m_C\sim 0$ or $\Delta m_C \sim \Delta m_A$, equivalent to  $m_{H^\pm} \sim m_{H, A}$.
The dependence on $\cos(\beta-\alpha)$ for $Z$-pole constraints is almost non-noticeable given the small range of $\cos(\beta-\alpha)$ allowed under the current LHC Higgs precision measurements.

Combining both the Higgs and $Z$-pole precisions (right panels), the range of $\Delta m_{C,A}$ are further constrained to be less than about 200 GeV in the alignment limit for  $m_H=800$ GeV, $\sqrt{\lambda v^2}=0$, with positive (negative) values for the mass splittings preferred for positive (negative) $\cos(\beta-\alpha)$.
For $\sqrt{\lambda v^2}=300$ GeV, loop corrections play a more important role. For $\cos(\beta-\alpha)= 0.007$, only thin strip of $\Delta m_C\sim 0$ and $0\lesssim \Delta m_A \lesssim 500$ GeV is allowed.
 For $\cos(\beta-\alpha)= -0.007$, $-250\ {\rm GeV} \lesssim \Delta m_C \sim \Delta m_A  \lesssim -100$ GeV  as well as  thin slice of $\Delta m_C\sim 0$ for negative $\Delta m_A$ could be accommodated.
For larger $m_H=2000$ GeV,  while the ranges for mass splittings are typically larger under the alignment limit, deviation from the alignment limit leads to tighter constraints due to the suppressed loop contributions.

The Higgs and $Z$-pole precision measurements at future lepton colliders provide complementary information.  While the $Z$-pole precision is more sensitive to the mass splittings between the charged Higgs boson and the neutral ones (either $m_H$ or $m_A$), the Higgs precision measurements  in addition could impose an upper bound on the mass splitting between the neutral ones. Furthermore, the Higgs precision measurements are more sensitive to the parameters $\cos(\beta-\alpha)$, $\tan\beta$, $\sqrt{\lambda v^2}$ and the masses of heavy Higgs bosons.

 \subsection{Comparison between different lepton colliders}
\begin{figure}[tb]
\begin{center}
\includegraphics[width=7.5 cm]{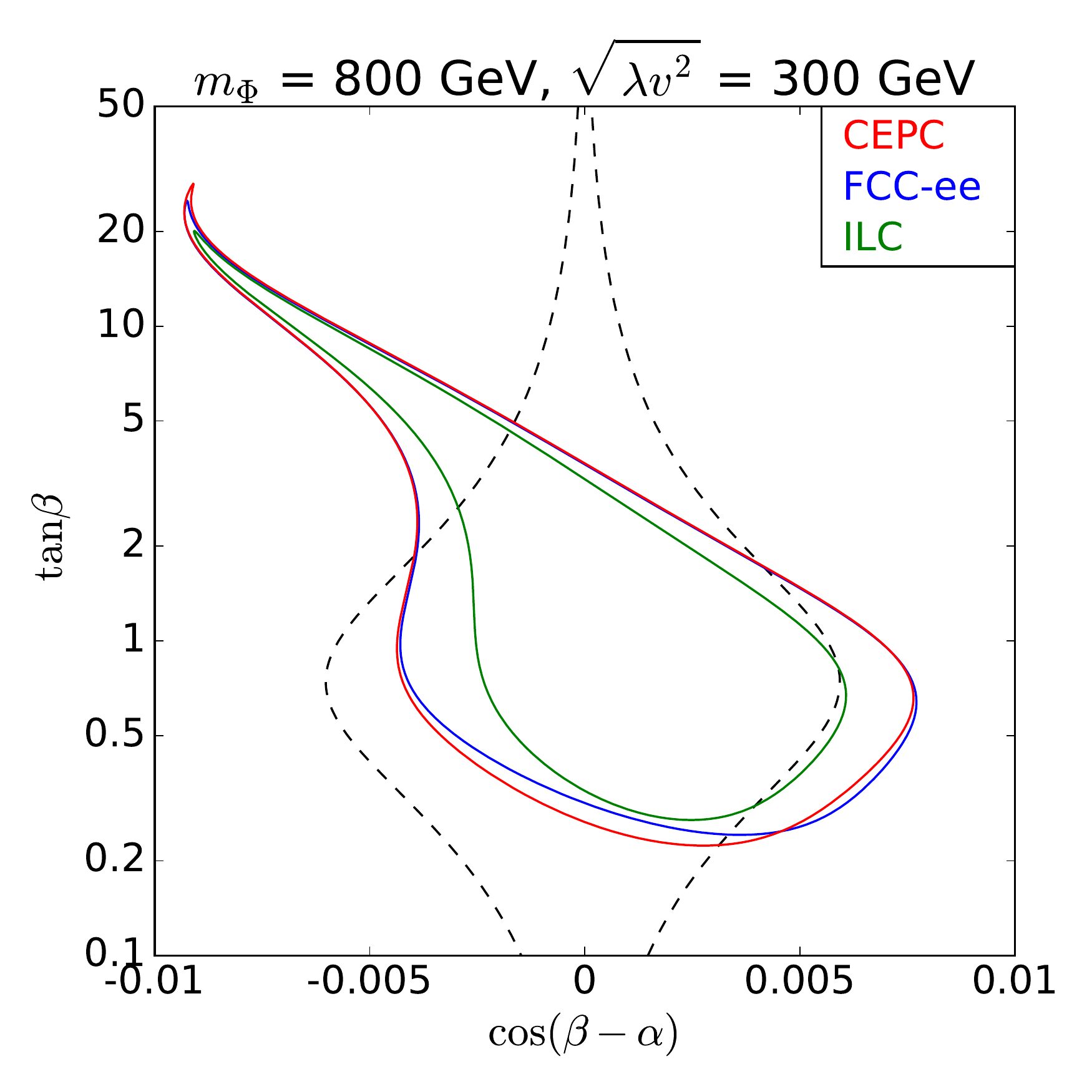}
\includegraphics[width=7.5 cm]{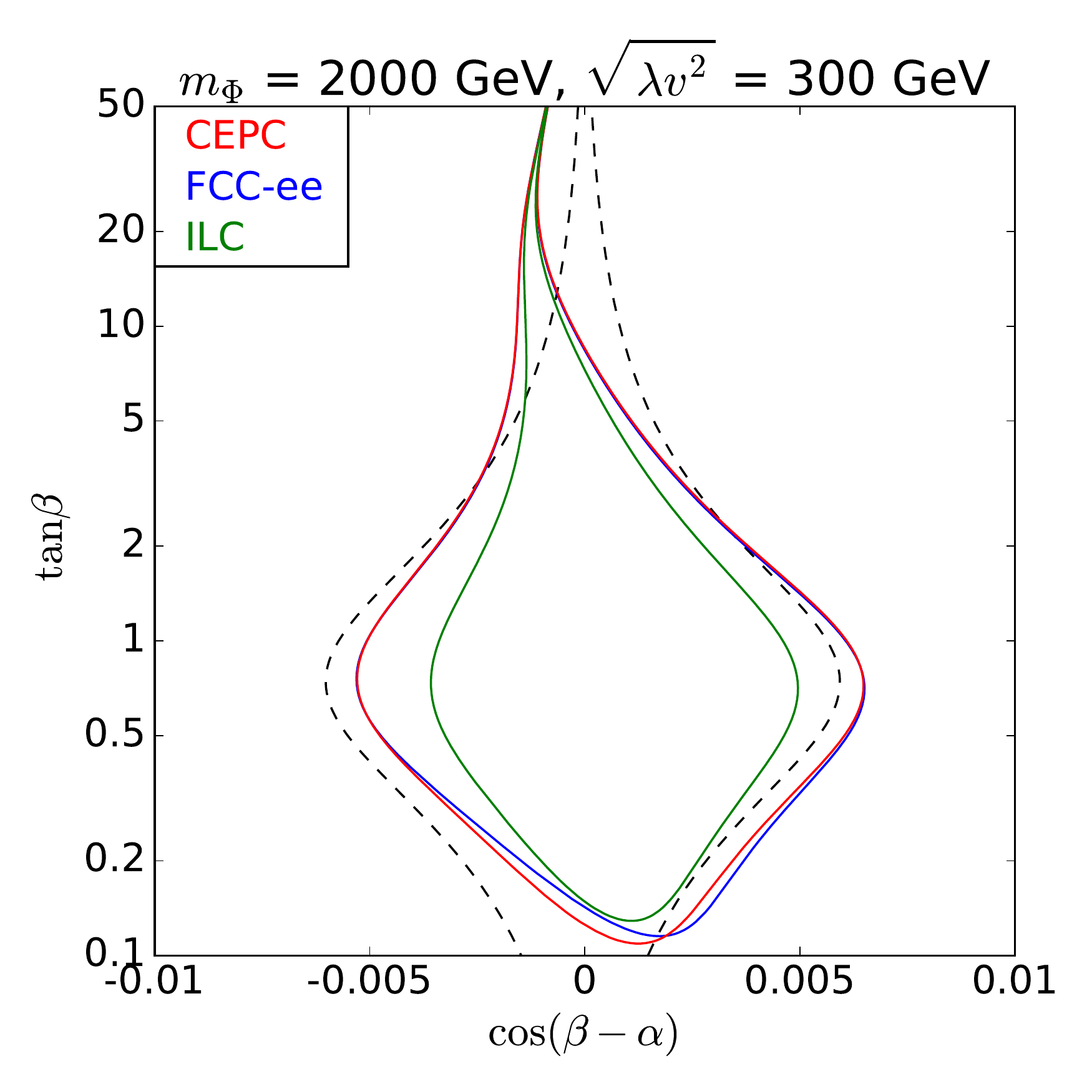}
 \caption{ Two-parameter fitting results at 95\% C.L. in the $\cos(\beta-\alpha)$-$\tan \beta$ plane  with CEPC (red), FCC-ee (blue) and ILC (green) precisions.  The black dashed line indicates the CEPC tree-level only results as a comparison. For the left panel, $m_\Phi =800 \gev, \lambvs =300$ GeV, and the right panel $m_\Phi =2000 \gev, \lambvs =300$ GeV.  
}
\label{fig:ee-costanb}
\end{center}
\end{figure}

\begin{figure}[tb]
\begin{center}
 \includegraphics[width=15.cm]{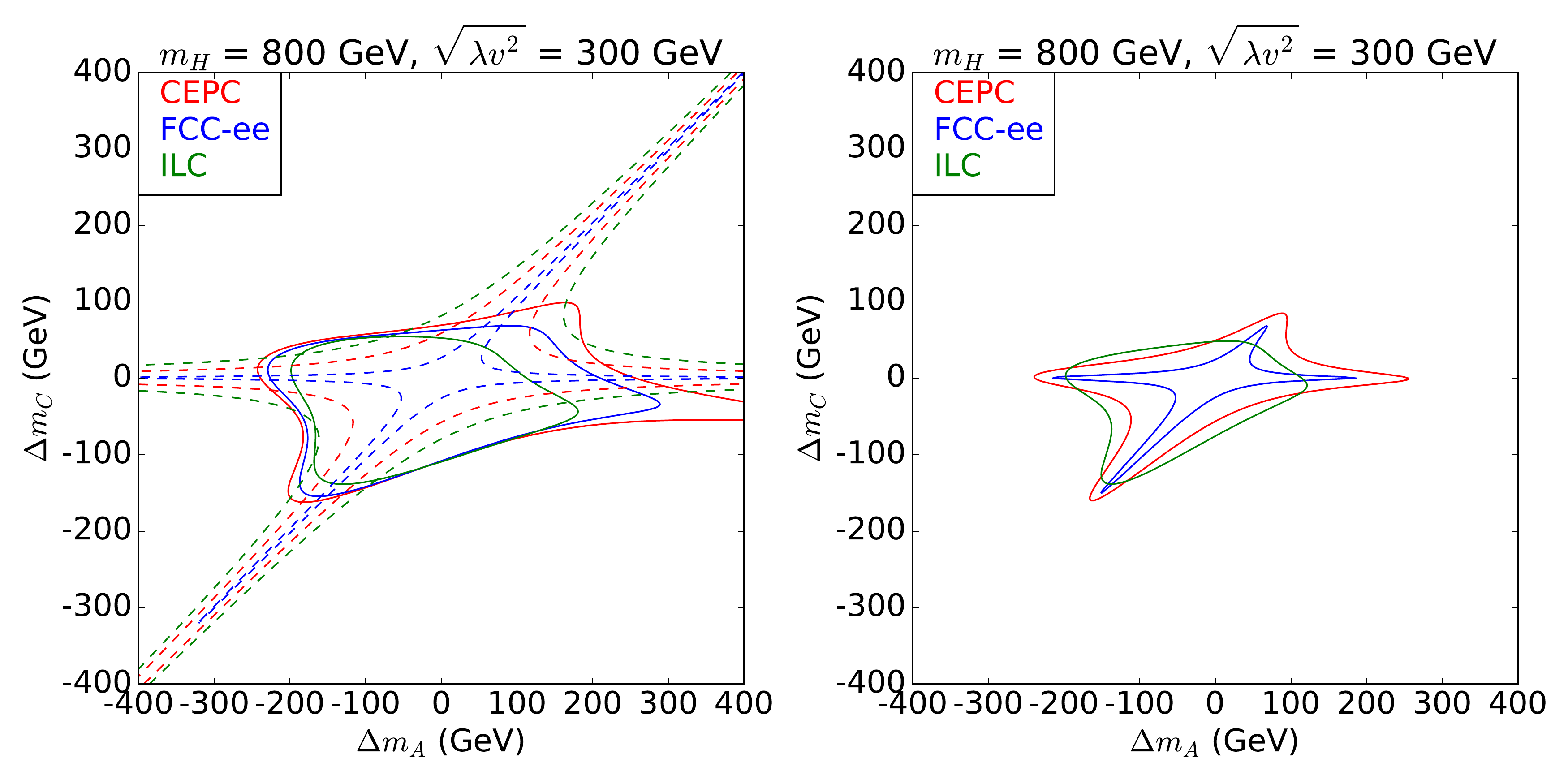}
\caption{Two-parameter fitting results at 95\% C.L. in the $\Delta m_A$-$\Delta m_C$ plane with CEPC (red), FCC-ee (blue) and ILC (green) precisions,  similar to~\autoref{fig:da-dc-cos}.   The left and right  panels are for Higgs/$Z$-pole results individually and combined, respectively. Here $m_H = 800 \gev, \lambvs = 300 \gev, \cosba=0$. }
\label{fig:ee-dmac}
\end{center}
\end{figure}

In this section, we present a brief comparison for the potential reach of different machines, including CEPC, FCC-ee, and ILC precision shown in~\autoref{tab:STU} for $Z$-pole precision and~\autoref{tab:mu_precision} for Higgs precision. In~\autoref{fig:ee-costanb}, we show the 95\% C.L. reach in the $\cos(\beta-\alpha)$-$\tan \beta$ plane for three different machines including both tree-level and loop effects, for benchmark points of $m_\Phi=800$ GeV (left panel), $m_\Phi=2000$ GeV (right panel),  and $\sqrt{\lambda v^2}=300$ GeV.
Dashed curves show the tree-level only results with CEPC precision as a comparison. The reach with Higgs precision is similar for CEPC and FCC-ee, while slightly better for ILC including center-of-mass energies of 250/350/500 GeV. The overall features are similar to those in~\autoref{fig:dmac_cepc}.

Finally, in~\autoref{fig:ee-dmac}, we show the comparison among three machines for Higgs and $Z$-pole precision constraints individually (left panel) and combined fitting results (right panel) in the $\Delta m_A$-$\Delta m_C$ plane, for benchmark point of $m_H=800$ GeV, $\cos(\beta-\alpha)=0$ and $\sqrt{\lambda v^2}=300$ GeV.   For the Higgs precisions, ILC has the best constraint because of the energy reach, while for the $Z$-pole precision, FCC-ee has the best performance because of the higher proposed luminosity at $Z$-pole.
For the combined fit, FCC-ee shows the best constraint, dominanted by the $Z$-pole effects.

\section{Summary and Conclusions}
\label{sec:conclu}

In this paper, we examined the impacts of the precision measurements of the SM parameters at the proposed $Z$-factories and Higgs factories on the extended Higgs sector.
We first summarized the anticipated accuracies on determining the EW observables at the $Z$-pole and the Higgs factories in~\autoref{sec:input}. Those expectations serve as the general guidances and inputs for the following studies for BSM Higgs sector.
We illustrated this by studying in great detail the well-motivated theory, the Type-II 2HDM.
Previous works focused on either just the tree-level deviations, or loop corrections under the alignment limit,  and with the assumption of degenerate masses of the heavy Higgs bosons.
In our analyses, we extended the existing results by including the tree-level and one-loop level effects of non-degenerate Higgs masses. The general formulation,  theoretical considerations and the existing constraints to the model parameters were presented in~\autoref{sec:2hdm}, see~\autoref{fig:STU_THDM_mamc}$-$\autoref{fig:LHC_heavyH_limits}.

The main results of the paper were presented in~\autoref{sec:results}, where we performed a global fit to the expected precision measurements in the full model-parameter space.
We first set up the global $\chi^2$-fitting framework. We then illustrated the simple case with degenerate heavy Higgs masses as in~\autoref{fig:dmac_cepc_ana} with the expected CEPC precision.  We found that  in the parameter space of $\cos(\beta-\alpha)$ and $\tan\beta$, the largest 95\% C.L. range of $|\cos(\beta-\alpha)|\lesssim 0.008$ could be achieved for $\tan\beta$ around 1, with smaller and larger values of $\tan\beta$ tightly constrained by $\kappa_{g,c}$ and $\kappa_{b,\tau}$, respectively. Comparing to the tree-level only results~\cite{Gu:2017ckc}, $\cos(\beta-\alpha)$ shifts to negative values for $\tan\beta >1$.   Smaller heavy Higgs masses and larger $\lambda v^2$ lead to larger loop corrections, as shown in~\autoref{fig:dmac_cepc}.

The limits on the heavy Higgs masses also depend on $\tan\beta$, $\lambda v^2$ and $\cos(\beta-\alpha)$, as shown in~\autoref{fig:mphitanb_cepc_lambda} and alternatively in~\autoref{fig:mphitanb_cepc_m12} varying $m_{12}^2$.  While the most relaxed limits can be obtained under the alignment limit with small $\lambda v^2$, deviation away from the alignment limit leads to much tighter constraints, especially for allowed range of $\tan\beta$.  The reach seen in the $m_\Phi$-$\tan\beta$ plane is complementary to direct non-SM Higgs search limits at the LHC and future $pp$ colliders, especially in the intermediate $\tan\beta$ region when the direct search limits are relaxed.

It is important to explore the extent to which the parametric deviations from the degenerate mass case can be probed by the precision measurements. \autoref{fig:la_dm1} showed the allowed deviation for $\Delta m_\Phi$ with the expected CEPC precision and~\autoref{fig:da-dc-ana} demonstrated the constraints from the individual decay channels of the SM Higgs boson. As shown in~\autoref{fig:dadc}, the Higgs precision measurements alone constrain $\Delta m_{A,C}$ to be less than about a few hundred GeV, with tighter constraints achieved for
small $m_H$, large $\lambda v^2$ and small/large values of $\tan\beta$.   $Z$-pole measurements, on the other hand, constrain the deviation from $m_{H^\pm}\sim m_{A,H}$.  We found that the expected accuracies at the $Z$-pole and at a Higgs factory are quite complementary in constraining mass splittings.
While $Z$-pole precision is more sensitive to the mass splittings between the charged Higgs and the neutral ones (either $m_H$ or $m_A$),  Higgs precision measurements in addition could impose an upper bound on the mass splitting between the neutral ones. Combining both Higgs and $Z$-pole precision measurements,  the mass splittings are constrained even further, as shown in~\autoref{fig:da-dc-cos}, especially when deviating from the alignment limit.   Furthermore, Higgs precision measurements are more sensitive to parameters like $\cos(\beta-\alpha)$, $\tan\beta$, $\sqrt{\lambda v^2}$ and the masses of heavy Higgs bosons.   We found that except for cancelations in some correlated parameter regions, the allowed ranges are typically
 \begin{equation}
\tan\beta \sim 0.2-5,\quad |\cos(\beta-\alpha)| < 0.008, \quad  |\Delta m_\Phi | < 200\ {\rm GeV}\,.
 \end{equation}

For the sake of illustration, we mostly presented our results using the CEPC precision on Higgs and $Z$-pole measurements.  The comparison among different proposed Higgs factories of CEPC, FCC-ee and ILC are shown in~\autoref{fig:ee-costanb} and~\autoref{fig:ee-dmac}.   While ILC with different center-of-mass energies has slightly better reach in Higgs precision fit, FCC-ee has slightly better reach in $Z$-pole precisions.

The precision measurements of the SM parameters at the proposed $Z$ and Higgs factories would significantly advance our understanding of the electroweak physics and shed lights on possible new physics beyond the SM, and could be complementary to the direct searches at the LHC and future hadron colliders.

\begin{acknowledgments}


We would like to thank Han Yuan and Huanian Zhang for collaboration at the early stage of this project.  We would also like to thank Liantao Wang and Manqi Ruan for valuable discussions.
NC is supported by the National Natural Science Foundation of China (under Grant No. 11575176) and Center for Future High Energy Physics (CFHEP).
TH is supported in part by the U.S.~Department of Energy under
grant No.~DE-FG02-95ER40896 and by the PITT PACC. SS is supported  by the Department of Energy under Grant No.~DE-FG02-13ER41976/DE-SC0009913.
WS were supported in part by the National Natural Science Foundation of China (NNSFC) under grant No.~11675242. YW is supported by the Natural Sciences and Engineering Research Council of Canada (NSERC). TH also acknowledges the hospitality of the Aspen Center for Physics, which is supported by National Science Foundation grant PHY-1607611.

 \end{acknowledgments}

\bibliographystyle{JHEP}
\bibliography{references}

\end{document}